\documentclass{LMCS}

\def\doi{8(4:17)2012}
\lmcsheading%
{\doi}
{1--44}
{}
{}
{Jan.~\phantom07, 2011}
{Nov.~23, 2012}
{}
 
\def\qedhere{}
\let\origparagraph\paragraph
\renewcommand{\paragraph}[1]{\origparagraph{\textbf{#1}}}

\newenvironment{definition}[1][]
{\begin{defi}\ifthenelse{\equal{#1}{}}{}{(\textsc{#1}).}\ignorespaces
}%
{\end{defi}\par\noindent\ignorespacesafterend}

\newenvironment{theorem}[1][]
{\begin{thm}\ifthenelse{\equal{#1}{}}{}{(\textsc{#1}).}\ignorespaces
}%
{\end{thm}\par\noindent\ignorespacesafterend}

\newenvironment{lemma}[1][]
{\begin{lem}\ifthenelse{\equal{#1}{}}{}{(\textsc{#1}).}\ignorespaces
}%
{\end{lem}\par\noindent\ignorespacesafterend}
\newenvironment{proposition}[1][]
{\begin{prop}\ifthenelse{\equal{#1}{}}{}{(\textsc{#1}).}\ignorespaces
}%
{\end{prop}\par\noindent\ignorespacesafterend}

\usepackage{enumerate}
\usepackage{wrapfig}

\usepackage{tikz}
\usetikzlibrary{positioning,chains,arrows,shapes,decorations,decorations.pathreplacing}
\usepackage{amsmath}
\DeclareMathOperator{\fraction}{frac}%
\DeclareMathOperator{\intpart}{intpart}%
\DeclareMathOperator{\digit}{bit}%
\definecolor{vred}{rgb}{.7,0,0}
\definecolor{vgreen}{rgb}{.1,.5,0}
\newcommand{\jupd}{\mathcal{A}}%

\usepackage{hyperref}

\usepackage[bbsets,setrelation,Dfprime]{math}
\usepackage[modernsign,substopindex,longmquant,mquantifiertype,mconnectiveformal,bracketinterpret,prefixinterpret,modifopindex,seqarrow,seqoptional,footnotecalculus,abbrseqcontext,shortterms,nosigmaterms,novarterms]{logic}

\usepackage[pretest,nocommandblocks,keywordfont=textsfs]{progreg}
\usepackage[bracketinterpret,prefixinterpret]{dL}

\usepackage{editingsupport}
\newcommand{\tweakp}[1]{}

\usepackage{prettyref}
\newcommand{\rref}[2][]{\prettyref{#2}}
\newrefformat{sec}{Section\,\ref{#1}}
\newrefformat{app}{Appendix\,\ref{#1}}
\newrefformat{def}{Def.\,\ref{#1}}
\newrefformat{thm}{Theorem\,\ref{#1}}
\newrefformat{prop}{Proposition\,\ref{#1}}
\newrefformat{lem}{Lemma\,\ref{#1}}
\newrefformat{ex}{Example\,\ref{#1}}
\newrefformat{tab}{Table\,\ref{#1}}
\newrefformat{fig}{Fig.\,\ref{#1}}
\newrefformat{calculus}{Fig.\,\ref{#1}}
\newrefformat{case}{case\,\ref{#1}}
\newrefformat{foot}{footnote\,\ref{#1}}

\newcommand{\oidV}[2][]{#1}

\newcommand{\FOQD}{\text{FOQD}\xspace}%
\let\FOD\FOQD

\newcommand*{\laforallplus}[4][]{\lforall[\laetype{#1}{\cup}\{#2\}]{#3}{#4}}%

\newcommand*{\hastype}[2]{#1\,{:}\,#2}
\DeclareMathOperator{\bvarop}{BV}%
\newcommand*{\bvar}[1]{\bvarop(#1)}%

\newcommand{\solutionf}{{y}}%
\newcommand*{\solutionfor}[2][]{\solutionf_{#1}}%
\newcommand*{\solutionupdate}[1]{\umod{f(\vec{s})}{\solutionfor[\vec{s}]{}(#1)}}%
\newcommand{\genDE}[1]{\theta}%
\newcommand{\ivr}{\chi}%
\newcommand{\inv}{\phi}%
\newcommand{\var}{\varphi}%

\newcommand{\bebecomes}{\mathrel{::=}}
\newcommand{\alternative}{~|~}

\def\lprogramsname{\operatorname{QHP}}
\def\linterpretationsname{\mathcal{S}}

\renewcommand{\qelim}[2][]{\ensuremath{\QE\ifthenelse{\equal{#2}{}}{}{(#2)}}}
\renewcommand{\lsequentimpl}[3][]{{#2}\ifx\blank#2\else\lseqinfers\fi{#3}}
\makeatletter
  \renewmethodimplementation{dL}{iportray}{1}[1]{%
    \Interpretation@state%
  }
\makeatother

  \renewcommand{\idomain}[2]{\iget[state]{#1}\ifthenelse{\equal{#2}{}}{}{(#2)}}
\newcommand{\stdI}{\dLint[state=\sigma,const=I]}
\newcommand{\I}{\stdI}
\newcommand{\It}{\dLint[state=\tau,const=I]}

\newcommand{\onew}[2][]{n_{#2}}%
\def\leftrule{l}
\def\rightrule{r}

  \newcommand{\DCCS}{\textit{DCCS}\xspace}
  \newcommand{\abrake}{b}%
  \newcommand{\amax}{a}%
  \newcommand{\cyct}{\varepsilon}%

\newcommand*{\oa}[2]{#1(#2)}%
\newcommand*{\dcseparate}[2]{\mathcal{M}(#1,#2)}%

\newcommand*{\dcseparatetf}[2]{%
   \oa{x}{#1}<\oa{x}{#2}\land\oa{v}{#1}\leq\oa{v}{#2}\land\oa{a}{#1}\leq\oa{a}{#2}}%
\newcommand*{\dcseparatets}[2]{%
  \oa{x}{#1}>\oa{x}{#2}\land\oa{v}{#1}\geq\oa{v}{#2}\land\oa{a}{#1}\geq\oa{a}{#2}}%
\newcommand{\dcinv}{\laforall[C]{i,j}{\dcseparate{i}{j}}}%
\newcommand{\dcnu}{\onew{}}%
\newcommand{\dcnup}{\pumod{\dcnu}{\pnew{C}}}%
\newcommand{\dcnusep}{\laforall[C]{i}{\dcseparate{i}{\dcnu}}}%
\newcommand{\dcevo}{\pevolve{\laforall[C]{i}{(\D[2]{\oa{x}{i}}=\oa{a}{i})}}}%

\newcommand{\dcsys}{\dcnup; \ptest{\dcnusep}; \dcevo}%

  \newcommand*{\dcaccelt}[1]{\umod{\oa{a}{#1}}{\piif{\SBforma{#1}}{\amax}{\,{-}\abrake}}}%
  \newcommand*{\dcaccelallt}{\pupdate{\laforall[C]{i}{\dcaccelt{i}}}}%
\newcommand*{\dcaccelseparatetf}[2]{%
   \oa{x}{#1}<\oa{x}{#2}\land\oa{v}{#1}^2<\oa{v}{#2}^2+2\abrake(\oa{x}{#2}-\oa{x}{#1}) \land \oa{v}{#1}\geq0\land\oa{v}{#2}\geq0}%
\newcommand*{\dcaccelseparatets}[2]{%
  \oa{x}{#1}>\oa{x}{#2}\land\oa{v}{#2}^2<\oa{v}{#1}^2+2\abrake(\oa{x}{#1}-\oa{x}{#2}) \land \oa{v}{#1}\geq0\land\oa{v}{#2}\geq0}%
  \newcommand*{\SBforma}[1]{\laforall[C]{j}{\SBform{#1}{j}}}%
  \newcommand{\SBform}[2]{\textit{far}(#1,#2)}
  \newcommand{\SBformt}[2]{\oa{x}{#2}>\oa{x}{#1} \limply 
    \oa{x}{#2}>\oa{x}{#1} + \frac{\oa{v}{#1}^2-\oa{v}{#2}^2}{2\abrake} + \left(\frac{\amax}{\abrake}+1\right)\left(\frac{\amax}{2}\cyct^2+\cyct\oa{v}{#1}\right)
  }%
\newcommand{\dcaccelevot}{\pupdate{\umod{\tau}{0}};~ \hevolvein{\laforall[C]{i}{(\D{\oa{x}{i}}=\oa{v}{i} \syssep \D{\oa{v}{i}}=\oa{a}{i}\syssep\D{\tau}=1}}{\oa{v}{i}\geq0 \land \tau\leq\cyct)}}%

\usepackage{ifpdf}
\ifpdf
\fi

\begin{document}

\title[A Complete Axiomatization of \QdL for Distributed Hybrid Systems]{A Complete Axiomatization of\\ Quantified Differential Dynamic Logic\\ for Distributed Hybrid Systems\rsuper*}

\author[A.~Platzer]{Andr\'e Platzer}
\address{Carnegie Mellon University, Computer Science Department, Pittsburgh, PA, USA}
\email{aplatzer@cs.cmu.edu}
\thanks{%
This material is based upon work supported by the National Science Foundation under
NSF CAREER Award CNS-1054246, NSF EXPEDITION CNS-0926181, and under Grant Nos.
CNS-1035800 and CNS-0931985, by the NASA grant NNG-05GF84H, and by the ONR award N00014-10-1-0188.
}


\titlecomment{{\lsuper*}An extended abstract has appeared at CSL'10 \cite{DBLP:conf/csl/Platzer10}.}

\begin{abstract}
  \noindent
  We address a fundamental mismatch between the combinations of dynamics that occur in cyber-physical systems and the limited kinds of dynamics supported in analysis.
  Modern applications combine communication, computation, and control.
  They may even form dynamic distributed networks, where neither structure nor dimension stay the same while the system follows hybrid dynamics, i.e., mixed discrete and continuous dynamics.
  
  We provide the logical foundations for closing this analytic gap.
  We develop a formal model for distributed hybrid systems. It combines quantified differential equations with quantified assignments and dynamic dimensionality-changes.
  We introduce a dynamic logic for verifying distributed hybrid systems and present a proof calculus for this logic.
  This is the first formal verification approach for distributed hybrid systems.
  We prove that our calculus is a sound and complete axiomatization of the behavior of distributed hybrid systems relative to quantified differential equations.
  In our calculus we have proven collision freedom in distributed car control even when an unbounded number of new cars may appear dynamically on the road.
\end{abstract}

\maketitle

\section{Introduction}
\irlabel{qelim|QE}
\newsavebox{\exbox}%
\sbox{\exbox}{\rotatebox[origin=c]{180}{$\exists$}}%

Many safety-critical computers are embedded in cyber-physical systems like cars \cite{HsuEskafiSachsVaraiya1991,SenguptaRSCDK06} and aircraft \cite{DowekMC05}.
How do we know that their designs will work as intended?
Most initial designs do not. And some deployed systems still do not.
Ensuring the correct functioning of cyber-physical systems is a central challenge in computer science, mathematics, and engineering, because it is the key to designing smart and reliable control.
Scientists and engineers need analytic tools to understand and predict the behavior of their systems.
As systems become ever more complex, it becomes prohibitively expensive or impossible to test all possible interactions and rule out unsafe behavior by simulation.
Formal verification techniques are used routinely to overcome this for finite systems.
But for cyber-physical systems, there is not even a foundation for verification that would cover all required behavior.

There is a fundamental mismatch between the actual dynamics of cyber-physical system applications and the limits imposed by current modeling and analysis.
Cyber-physical systems in automotive, aviation, railway, and power grids combine \emph{communication, computation, and control}.
Combining computation and control leads to \emph{hybrid systems} \cite{DBLP:conf/hybrid/AlurCHH92,DBLP:conf/hybrid/Branicky95,DBLP:conf/lics/Henzinger96,DBLP:journals/tac/BranickyBM98,Platzer10}, whose behavior involves both discrete and continuous dynamics originating, e.g., from discrete control decisions and differential equations of motion.
Combining communication and computation leads to \emph{distributed systems} \cite{Lynch,DBLP:conf/concur/AttieL01,AptdeBoerOlderog10}, whose dynamics are discrete transitions of system parts that communicate with each other.
They may form \emph{dynamic distributed systems}, where the structure of the system is not fixed but evolves over time and agents may appear or disappear during the system evolution.

\begin{wrapfigure}{r}{9cm}
  \centering
  \includegraphics[width=9cm]{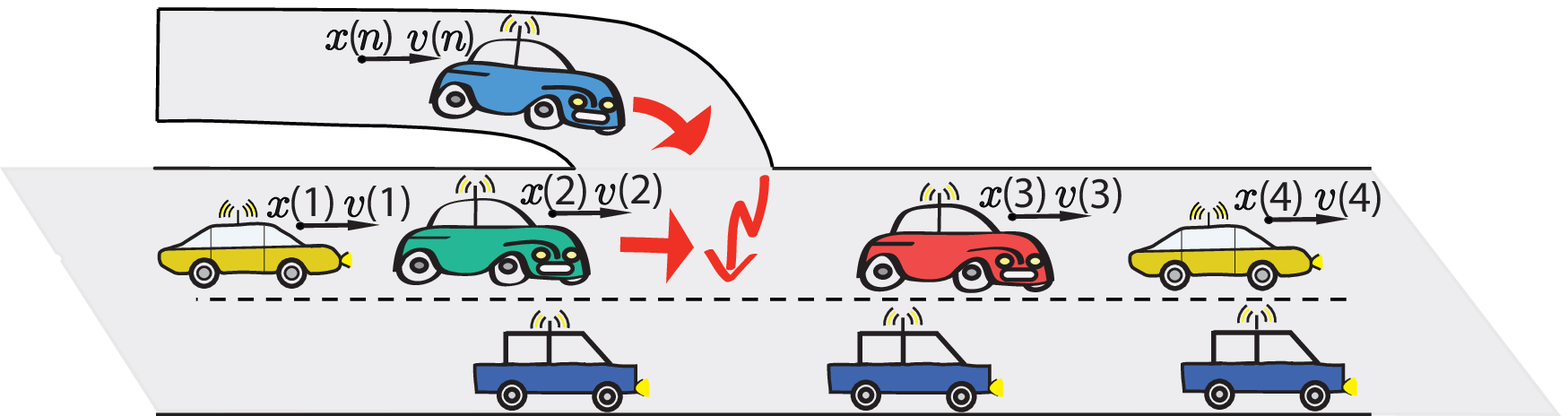}%
  \caption{Distributed car control.}
  \label{fig:distributed-car-control-new}
\end{wrapfigure}
Combinations of all three aspects (communication, computation, and control) are used in sophisticated applications, e.g., cooperative distributed car control \cite{HsuEskafiSachsVaraiya1991} and decentralized aircraft control \cite{PallottinoSFB06}.
Neither the structure nor dimension of the system stay the same, because new cars can appear on the street or leave it; see \rref{fig:distributed-car-control-new}.
These systems are \emph{(dynamic) distributed hybrid systems}, i.e., systems that combine the dynamics of distributed systems with the discrete and continuous dynamics of hybrid systems.
More generally, distributed hybrid systems are multi-agent hybrid systems that interact through remote communication or physical interaction.
They cannot be considered just as a distributed system (because, e.g., the continuous evolution of positions and velocities matters crucially for collision freedom in car control) nor just as a hybrid system (because the evolving system structure and appearance of new agents can make an otherwise collision-free system unsafe).
It is generally impossible to split the analysis of distributed hybrid systems soundly into an analysis of a distributed system (without continuous movement) and an analysis of a hybrid system (without structural changes or appearance), because all kinds of dynamics interact.
Just like hybrid systems are diffcult to analyze from a purely discrete or a purely continuous perspective \cite{DBLP:conf/lics/Henzinger96,DBLP:conf/lics/Platzer12b}.

Distributed hybrid systems have been considered to varying degrees in modeling languages \cite{DBLP:conf/hybrid/DeshpandeGV96,DBLP:conf/hybrid/Rounds04,DBLP:conf/hybrid/KratzSPL06,DBLP:conf/hybrid/MeseguerS06}.
In order to build these systems, however, scientists and engineers also need analytic tools to understand and predict their behavior. 
But formal verification and proof techniques do not yet support the required combination of dynamical effects---which is not surprising given the numerous sources of undecidability for distributed hybrid systems verification.

In this article, we provide the logical foundations to close this fundamental analytic gap.
We develop \emph{quantified hybrid programs} (\QHPs) as a formal model for distributed hybrid systems, which combine dynamical effects from multiple sources:
\emph{discrete transitions, continuous evolution, dimension changes, and structural dynamics}.
In order to account for changes in the dimension and for co-evolution of an unbounded and evolving number of participants, we generalize the notion of states from assignments for primitive system variables like $x$ to full first-order structures.
In a \QHP, function term $x(i)$ may denote the position of car $i$ of type $C$, the term $f(i)$ could be the car registered by communication as the car following car $i$, and the term $d(i,f(i))$ could denote the minimum safety distance negotiated between car $i$ and its follower $f(i)$.
The values of all these terms may evolve \emph{for all} $i$ as time progresses according to interacting laws of discrete and continuous dynamics, because all cars evolve simultaneously. They are also affected by changing the system dimension as new cars appear, disappear, or by reconfiguring the system structure dynamically, e.g., by remote communication or physical interaction.
The defining characteristic of \QHPs is that they allow \emph{quantified hybrid dynamics} in which variables like $i$ that occur in function arguments of the system dynamics are quantified over, such that the system co-evolves, e.g., \emph{for all} cars $i$ of type $C$.
This quantification is necessary to characterize the distributed hybrid systems dynamics with an unbounded and possibly evolving number of participants.
Quantification is also necessary to represent structural dynamics when the number of participants is not fixed.

There is a crucial difference between a primitive system variable~$x$ and a first-order function term~$x(i)$, where~$i$ is quantified over.
Hybrid dynamics of primitive system variables can model a concrete number of, say, four cars (putting scalability issues aside), but neither a parametric number of $n$ cars nor systems with a variable number of cars (a number $n$ that may change over time).
With first-order function symbols $x(i)$ and hybrid dynamics quantifying over all cars~$i$, a single \QHP can represent \emph{any} number of cars at once.
\QHPs can even represent (dis)appearance of cars by changing the domain that quantifiers range over dynamically at runtime.
\QHPs are thus a formal model for general (dynamic) distributed hybrid systems.

Verification of distributed hybrid systems is challenging.
We show that they have three independent sources of undecidability: discrete dynamics, continuous dynamics, and structural/dimensional dynamics.
As an analysis tool for distributed hybrid systems, we introduce a specification and verification logic for \QHPs that we call \emph{quantified differential dynamic logic} (\QdL).
\QdL provides dynamic logic \cite{DBLP:conf/focs/Pratt76,Harel_et_al_2000} modal operators~$\dbox{\alpha}{}$ and~$\ddiamond{\alpha}{}$ that refer to the states reachable by \QHP~$\alpha$ and can be placed in front of any formula.
Formula \m{\dbox{\alpha}{\phi}} expresses
that all states reachable by system~$\alpha$ satisfy formula~$\phi$, while \m{\ddiamond{\alpha}{\phi}} expresses that there is at least one reachable state satisfying $\phi$. 
These modalities can express necessary or possible properties of the transition behavior of \QHP~$\alpha$.
With its ability to specify and verify properties of (dynamic) distributed hybrid systems and quantified dynamics, \QdL is a major extension of prior work for static hybrid systems \cite{DBLP:journals/jar/Platzer08,DBLP:journals/logcom/Platzer10} and conventional discrete programs \cite{DBLP:conf/cade/BeckertP06,DBLP:conf/lpar/Rummer06}.

Our primary contributions are:
\begin{iteMize}{$\bullet$}
\item We introduce a \emph{formal system model and semantics} that succinctly captures the logical quintessence of (dynamic) distributed hybrid systems with joint discrete, continuous, structural, and dimension-changing dynamics.
\item We introduce a \emph{specification and verification logic} for (dynamic) distributed hybrid systems.
\item We present a \emph{proof calculus} for this logic, which, to the best of our knowledge, is the \emph{first verification approach} that can handle distributed hybrid systems with their hybrid dynamics and unbounded (and evolving) dimensions and structure.
\item We prove that this compositional calculus is a \emph{sound and complete axiomatization} of (dynamic) distributed hybrid systems relative to quantified differential equations.
\item We have used our proof calculus to verify \emph{collision freedom in a distributed car control system}, where an unbounded number of new cars may appear dynamically on the road.
\end{iteMize}
In particular, we extend our previous extended abstract \cite{DBLP:conf/csl/Platzer10} by 28 pages worth of
\begin{iteMize}{$\bullet$}
\item soundness and relative completeness proofs
\item new results on ineffective fragments
\item more detailed explanations and more examples
\item new derived proof rules 
\item new formal proofs illustrating the interaction of quantifiers, first-order function symbols, and quantified system dynamics in detail
\item a proof of collision avoidance in a simple distributed car control system, and a new result about a more advanced distributed car control system.
\end{iteMize}
This work constitutes the logical foundation for analysis of distributed hybrid systems.
Since distributed hybrid control is the key to control numerous advanced systems, analytic approaches have significant potential for applications. 
With a theorem prover based on our approach, we have verified collision avoidance in a distributed car control system, which is out of scope for other approaches.
The approach presented here has been used subsequently for verifying distributed adaptive cruise control systems for highways \cite{DBLP:conf/fm/LoosPN11} and distributed air traffic control \cite{DBLP:conf/hybrid/Platzer11}.

Our verification approach for distributed hybrid systems is a fundamental extension compared to previous approaches.
In much the same way as first-order logic increases the expressive power over propositional logic (quantifiers and function symbols are required to express properties of unbounded structures), \QdL increases the expressive power over its predecessors (because first-order functions and quantifiers in the dynamics of \QHPs are required to characterize systems with unbounded and changing dimensions).
\section{Related Work} \label{sec:Related}

Multi-party distributed control has been suggested for car control \cite{HsuEskafiSachsVaraiya1991} and air traffic control \cite{DowekMC05}.
Due to limits in verification technology, no formal analysis of the distributed hybrid dynamics has been possible for these systems yet.
Analysis results include discrete message handling \cite{HsuEskafiSachsVaraiya1991} or collision avoidance for two participants \cite{DowekMC05}.
In distributed car control and air traffic control systems, appearance of new participants is a major unsolved challenge for formal verification.

Ad-hoc informal arguments have been used to discuss distribution effects away, e.g., assuming that at most 4 cars are close to one another.
These arguments are treacherous, though.
They are very case-specific and do not lend themselves to formal verification within one proof system because they need arguments outside the proof system to work.
In distributed car control, for instance, it might, at first sight, be convincing to suspect that it would be enough to consider every possible constellation of, say, four cars.
This breaks down at second thought, though, because, without a formal proof, there is no reason to believe that a locally consistent and safe system would be globally safe and consistent.
Consider an example for the situation in \rref{fig:distributed-car-control-new}, for instance. 
Even if hybrid systems verification techniques could show that local patterns consisting of the four cars $\{1,2,\dcnu,3\}$ are safe and that local patterns consisting of the four cars $\{2,\dcnu,3,4\}$ are safe, the full system consisting of all cars $\{1,2,\dcnu,3,4\}$ still does not have to be safe.
For example, the local pattern $\{1,2,\dcnu,3\}$ could be safe, because it will ask car $\dcnu$ to change lanes and ask car $2$ to keep speed and car $3$ to speed up.
But the pattern $\{2,\dcnu,3,4\}$ could be safe, because it will ask car $\dcnu$ to change lanes but, instead, ask car $2$ to slow down and car $3$ to keep speed.
Those two locally safe patterns still lead to a globally incompatible maneuver choice resulting in a crash, because both cars $2$ and $3$ would be forced to keep the speed (for they would otherwise collide with car $1$ or $4$, respectively) and, henceforth, collide with car $\dcnu$ during its lane change.
More generally, independent actions in different parts of a system may still end up interacting by rippling effects.
It is, thus, crucial to understand and verify the emergent behavior resulting from local control principles.
The full distributed hybrid systems dynamics needs to be considered and we cannot generally hope to prove meaningful properties by simply ignoring part of the dynamics.

The importance of understanding dynamic / reconfigurable distributed hybrid systems was recognized in modeling languages SHIFT \cite{DBLP:conf/hybrid/DeshpandeGV96} and R-Charon \cite{DBLP:conf/hybrid/KratzSPL06} before.
They focused on simulation and compilation \cite{DBLP:conf/hybrid/DeshpandeGV96} or the development of a semantics \cite{DBLP:conf/hybrid/KratzSPL06}, so that no verification is possible yet.
For stochastic simulation, see \cite{DBLP:conf/hybrid/MeseguerS06}, where soundness has not been proven, because ensuring coverage is difficult by a random simulation.
See \cite{DBLP:conf/hybrid/ZulianiPC10} for a discussion of statistical evidence that can be obtained for randomized discrete-time hybrid systems by fair (i.i.d. sampled) simulation.
This technique neither covers distributed hybrid systems nor continuous-time hybrid systems nor nondeterministic dynamics, all of which we cover in this article.

For distributed hybrid systems, even giving a formal semantics is very challenging \cite{DBLP:conf/hybrid/ChaochenJR95,DBLP:conf/hybrid/Rounds04,DBLP:conf/hybrid/KratzSPL06,DBLP:journals/jlp/BeekMRRS06}!
Zhou et al.\ \cite{DBLP:conf/hybrid/ChaochenJR95} gave a semantics for a hybrid version of CSP in the Extended Duration Calculus \cite{DBLP:conf/hybrid/ChaochenRH92}.
Rounds \cite{DBLP:conf/hybrid/Rounds04} gave a semantics in a rich set theory for a spatial logic for a hybrid version of the $\pi$-calculus.
In the hybrid $\pi$-calculus, processes interact with a continuously changing environment, but cannot themselves evolve continuously, which would be crucial to capture the physical movement of traffic agents.
From the semantics alone, no verification is possible in these approaches, except perhaps by manual reasoning in the semantics.

Other process-algebraic approaches, like $\chi$ \cite{DBLP:journals/jlp/BeekMRRS06}, have been developed for modeling and simulation purposes.
Verification is still limited to small fragments that can be translated directly to other verification tools like PHAVer or UPPAAL, which have fixed dimensions and restricted dynamics (thus no distributed hybrid systems).

Our approach is completely different.
It is based on first-order structures and dynamic logic. We focus on developing a logic that supports distributed hybrid dynamics directly and that is amenable to automated theorem proving in the logic itself.

For a detailed discussion of verification approaches for static real-time and hybrid systems, we refer to previous work \cite{DBLP:journals/jar/Platzer08,DBLP:journals/logcom/Platzer10,Platzer08,Platzer10}.
Our previous work and other verification approaches for static hybrid systems cannot verify distributed hybrid systems.
Distributed hybrid systems may have an unbounded and changing number of components/participants, which cannot be represented with any fixed finite number of dimensions of the state space.
In distributed car control, for instance, there is no prior limit on the number of cars on the street.
Even when there is a limit, explicit replication of the system, say, 100 times, does not yield a scalable verification approach, because most hybrid systems verification approaches scale exponentially in the number of participants or worse.

Approaches for distributed systems \cite{DBLP:conf/concur/AttieL01} do not cover hybrid systems, because the addition of differential equations to distributed systems is even more challenging than the addition of differential equations to discrete dynamics has been when forming hybrid systems.
There is not even a bound on the number of differential equations that would need to be added to faithfully hybridize a distributed system.

In summary, previous approaches to distributed hybrid systems are limited to modeling, simulation, or the definition of a semantics. 
No formal verification technique was known for distributed hybrid systems before.

\section{Syntax} \label{sec:QdL-syntax}
As a formal logic for specifying and verifying correctness properties of distributed hybrid systems, we introduce \emph{quantified differential dynamic logic} (\QdL).
\QdL combines dynamic logic for reasoning about all (\m{\dbox{\alpha}{\phi}}) or some (\m{\ddiamond{\alpha}{\phi}}) system runs of a system~$\alpha$ \cite{DBLP:conf/focs/Pratt76,Harel_et_al_2000} with many-sorted first-order logic for reasoning about all (\m{\lforall[C]{i}{\phi}}) or some (\m{\lexists[C]{i}{\phi}}) objects of a sort $C$, e.g., the sort of all cars.
The most important defining characteristic of \QdL is that~$\alpha$ can be a distributed hybrid system, because the \QdL system model of \emph{quantified hybrid programs} (\QHP) supports quantified operations that affect \emph{all} objects of a sort $C$ at once.
If $C$ is the sort of cars, the quantified assignment \m{\pupdate{\lforall[C]{i}{\pumod{a(i)}{a(i)+1}}}} increases the respective accelerations $a(i)$ of \emph{all cars} $i$ at once by a single instantaneous discrete jump.
It can be used to model simultaneous discrete changes in multiple agents at once.
Discrete changes where only some of the cars change their acceleration, others do not, are easy to model with quantified assignments by masking.
The quantified differential equation \m{\hevolve{\lforall[C]{i}{\D{v(i)}=a(i)}}} represents a continuous evolution of the respective velocities $v(i)$ of \emph{all cars}~$i$ at the same time according to their acceleration by their respective differential equations \m{\hevolve{\D{v(i)}=a(i)}}.
Again, continuous evolutions where only some of the cars evolve, others remain stopped, are easy to model with quantified differential equations by masking.
These quantified assignments and quantified differential equation systems of \QHPs are crucial for representing distributed hybrid systems where an unbounded number of objects co-evolve simultaneously, because no finite set of classical assignments and classical differential equations could represent that.
Note that, because of the close semantical relationship, we use the same quantifier notation \m{\pupdate{\lforall[C]{i}}}for quantified operations in programs and for quantifiers in logical formulas, instead of a separate notation $\Pi_{i:C}$ for parallel products in programs.

Interaction by communication can be modeled by (possibly quantified) discrete assignments to share data between agents $i$ and $j$ in \QHPs.
Physical interaction, instead, may be modeled either by (possibly quantified) discrete assignments when an agent $i$ activates a response in agent $j$ by an instantaneous discrete action (e.g., pushing a physical button) or by a (possibly quantified) differential equation involving multiple agents $i$ and $j$ when they come into physical contact and act jointly over a (nonzero) period of time (e.g., both agents jointly lifting and pulling on a rigid object).
Observe that the cyber structure of the system reconfigures dynamically when discrete communication topologies change, whereas the physical structure reconfigures dynamically when agents engage in physical contact.
\QHPs for the latter case may involve structural changes in the quantified differential equation.

We model the appearance of new participants in the distributed hybrid system, e.g., new cars entering the road, by a \QHP \m{\pumod{\onew{}}{\pnew{C}}}. It creates a new object of type $C$, thereby extending the range of all subsequent quantified assignments or quantified differential equations ranging over created objects of type $C$.
With quantifiers and function terms, $\pnew{}$ can be handled in an entirely modular way.
In order to reduce the conceptual complexity, we first focus on the syntax and semantics of \QdL and postpone the discussion of actual existence and creation until \rref{sec:objectcreation}.
We will see that actual existence and creation are completely modular extensions.

The model of \QHPs is of independent interest as a formal model for distributed hybrid systems.
Inside a \QHP, logical formulas can occur in state tests for conditional execution.
We thus explain logical formulas, terms, and sorts first.
Conversely, however, a \QHP~$\alpha$ occurs inside the modalities (\m{\dbox{\alpha}{}} and \m{\ddiamond{\alpha}{}}) of \QdL formulas, which state properties of the behavior of~$\alpha$.
Hence, \QHPs may occur inside \QdL formulas yet formulas may occur inside \QHPs.
The subsequent definitions of \QdL and \QHP are thus to be understood by simultaneous induction.
It is easier to start with sorts, terms, and logical formulas first and then explain the \QHP model subsequently.

\subsection{Quantified Differential Dynamic Logic} \label{sec:QdL}
We introduce quantified differential dynamic logic (\QdL), which is the first formal logic for specifying and verifying correctness properties of distributed hybrid systems.
\QdL is a combination of many-sorted first-order logic with dynamic logic, generalized to a system model (\QHPs) for distributed hybrid systems.

\paragraph{Sorts}
\QdL supports a (finite) number of object sorts, e.g., the sort of all cars and that of all aircraft.
For continuous quantities of distributed hybrid systems like positions or velocities, we add the sort $\reals$ of real numbers.
It would be easy to add subtyping of sorts;  see previous work \cite{DBLP:conf/cade/BeckertP06} for details.
We refrain from doing so, because that just obscures the logical essence of our approach.

The primary purpose of the sorts is to distinguish different kinds of objects in multi-agent hybrid systems in which different kinds of agents occur, e.g., cars of sort $C$, traffic lights of sort $T$, lanes of sort $L$, and aircraft of sort $A$.

\paragraph{Terms}
\QdL terms are built from a set of (sorted) function and variable symbols as in many-sorted first-order logic.
In particular, each function symbol~$f$ has a fixed type \m{C_1\times\dots\times C_n\to D} for some $n\in\naturals$ and some sorts \m{D,C_1,\dots,C_n} such that $f$ only accepts argument terms~\m{\theta_1,\dots,\theta_n} of the respective sorts \m{C_1,\dots,C_n} and then \m{f(\theta_1,\dots,\theta_n)} is a term of sort $D$.
We use these function symbols to represent the state of the system or other parameters.
In a car control scenario like that in \rref{fig:distributed-car-control-new}, for instance, we could use function symbol $x$ to represent the positions of cars, i.e., the term $x(i)$ could represent the position of car $i$ and $x(j)$ the position of car $j$.
Similarly, the term $v(i)$ could represent the velocity of car $i$ and $a(i)$ its acceleration.
These terms have sort $\reals$, whereas a term $l(i)$ that represents the car in front of car $i$ has sort $C$.

Unlike in first-order logic, the interpretation of function symbols can change when transitioning from one state to the other while following the dynamics of a distributed hybrid system.
The value of position $x(i)$ will change over time as car $i$ drives down the street.
The value of $x(i)$ would also change if the argument term $i$ changes its value and now refers to a different car than before.
Even objects may appear or disappear as the distributed hybrid system evolves.
We use function symbol $\laexisting{\cdot}$ to distinguish between objects $i$ that actually exist (\m{\laexisting{i}=1}) and those that have not been created yet or exist no longer (\m{\laexisting{i}=0}), depending on the value of $\laexisting{i}$, which may also change its interpretation from state to state.
We use $0,1,+,-,\cdot$ with the usual notation and fixed semantics for nonlinear real arithmetic.
Divisions can be added when guarding against divisions by zero \cite{DBLP:journals/jar/Platzer08}.
For \m{n\geq0} we abbreviate \m{f(s_1,\dots,s_n)} by \m{f(\vec{s})} using vectorial notation and we use \m{\vec{s}=\vec{t}} for component-wise equality.

\paragraph{Formulas}
The formulas of \QdL are defined as in first-order dynamic logic plus many-sorted first-order logic .
\begin{definition}[\QdL formulas]
The formulas of \QdL are defined by the following grammar ($\phi,\psi$ are formulas, $\theta_1,\theta_2$ are terms of the same sort, $i$ is a variable of sort $C$, and $\alpha$ is a \QHP as defined in \rref{sec:QHP}):
\begin{equation*}
  \begin{array}{@{}l@{}}
  \phi,\psi ~\bebecomes~ %
  \theta_1=\theta_2 \alternative
  \theta_1\geq\theta_2 \alternative
  \lnot \phi \alternative
  \phi \land \psi \alternative
  \lforall[C]{i}{\phi} \alternative 
  \lexists[C]{i}{\phi} \alternative
  \dbox{\alpha}{\phi} \alternative 
  \ddiamond{\alpha}{\phi}
  \end{array}
\end{equation*}
\end{definition}

We use standard abbreviations to define $\leq,>,<,\lor,\limply$.
Sorts \m{C\neq\reals} have no ordering and only \m{\theta_1=\theta_2} is allowed, not \m{\theta_1\geq\theta_2}.
For sort $\reals$, we abbreviate \m{\lforall[\reals]{x}{\phi}} by \m{\lforall{x}{\phi}} and \m{\lexists[\reals]{x}{\phi}} by \m{\lexists{x}{\phi}}.
In the following, all formulas and terms have to be well-typed.
For instance, \m{x(i)=l(i)} is no formula if $x$ has type \m{C\to\reals} and $l$ has type \m{C\to C} for a sort \m{C\neq\reals} or if $i$ has a sort \m{D\neq C}.
\QdL formula \m{\dbox{\alpha}{\phi}} expresses that \emph{all states} reachable by \QHP~$\alpha$ satisfy formula~$\phi$. Likewise, \m{\ddiamond{\alpha}{\phi}} expresses
that \emph{there is at least one state} reachable by~$\alpha$ for
which~$\phi$ holds.

For short notation, we allow \emph{conditional terms} of the form \m{\piif{\phi}{\theta_1}{\theta_2}} (where $\theta_1$ and $\theta_2$ have the same sort).
This term evaluates to $\theta_1$ if the formula $\phi$ is true and to $\theta_2$ otherwise.
We generally consider formulas with conditional terms as abbreviations, e.g.,
\m{\mapply{\psi}{\piif{\phi}{\theta_1}{\theta_2}}} abbreviates
\m{(\phi \limply \mapply{\psi}{\theta_1})  \land  (\lnot\phi \limply \mapply{\psi}{\theta_2})}.
Conditional terms can be understood as an additional operator for terms and formulas as well.
\paragraph{Example}
{%
  \let\laforall\lforall%
A major challenge in distributed car control systems \cite{HsuEskafiSachsVaraiya1991} is that they do not follow fixed, static setups.
Instead, new situations can arise dynamically that change structure and dimension of the system whenever new cars appear on the road from on-ramps or leave it; see \rref{fig:distributed-car-control-new}.
As a running example, we model a \emph{distributed car control system} \DCCS.
First, we consider desirable \QdL properties of the system \DCCS for which we will later develop a series of increasingly more realistic models as \QHPs.

If $i$ is a term of type $C$ (for cars), let $x(i)$ denote the position of car $i$, $v(i)$ its current velocity, and $a(i)$ its current acceleration; see \rref{fig:distributed-car-control-new}.
A car control system is collision-free at a state if all cars are at different positions, i.e., \m{\lforall[C]{i{\neq}j}{\oa{x}{i}{\neq}\oa{x}{j}}}.
Without a quantifier we could not describe that all cars on a highway are in a collision-free state, because there is a large number of cars on the highway and we may not know how many.
The car control system is globally collision-free if it will always stay collision-free.
The following \QdL formula expresses that the system \DCCS controls cars in a way that is always collision-free:
\begin{equation}
  {(\dcinv)\,} \limply {\,\dbox{\DCCS}{~\laforall[C]{i{\neq}j}{\oa{x}{i}{\neq}\oa{x}{j}}}}
  \label{eq:distributed-car-control-new}
\end{equation}
It says that cars following the distributed hybrid systems dynamics of \DCCS are always collision-free (postcondition), provided that \DCCS starts in an initial state satisfying a formula \m{\dcseparate{i}{j}} for all cars $i,j$ (precondition).
In particular, the modality \m{\dbox{\DCCS}{}} expresses that all states reachable by following the distributed hybrid system \DCCS satisfy the postcondition \m{\laforall[C]{i{\neq}j}{\oa{x}{i}{\neq}\oa{x}{j}}}.
The simple-most choice for the formula \m{\dcseparate{i}{j}} in the precondition is a formula that characterizes a simple compatibility condition: for different cars $i\neq j$, the car that is further down the road (i.e., with greater position) neither moves slower nor accelerates slower than the other car, i.e.:
\begin{align}
  \dcseparate{i}{j} ~\mequiv~
  i \neq j &\limply \big((\dcseparatetf{i}{j})\qquad
  \notag
  \\&~\lor(\dcseparatets{i}{j})\big)%
  \label{eq:distributed-car-control-separate}
\end{align}
Even though this monotonicity condition is not the only safe choice for \m{\dcseparate{i}{j}}, some precondition like \m{\dcinv} is necessary, because car control is unsafe if the cars start with incompatible velocities or acceleration choices initially.
In fact, we may suspect that a corresponding condition like this may have to hold all the time for the system to remain safe.
The car controllers will thus have to make sure they maintain \m{\dcinv} always.
And formal verification will have to make sure that formula \rref{eq:distributed-car-control-new} is actually valid for the appropriate choices of \DCCS.

How do we design the distributed hybrid system \DCCS that satisfies the \QdL formula \rref{eq:distributed-car-control-new}?
What is an appropriate model for distributed hybrid systems?
How can we then prove that \rref{eq:distributed-car-control-new} is true?
Next, we introduce \QHPs as a general model for distributed hybrid systems and then discuss possible choices of \QHPs for \DCCS.
}%
The reader should note that more sophisticated combinations of nested quantifiers and modalities are possible with \QdL as well.

\subsection{Quantified Hybrid Programs} \label{sec:QHP}
As a formal model for distributed hybrid systems, we introduce \emph{quantified hybrid programs}~(\QHPs).
These are regular programs from dynamic logic \cite{Harel_et_al_2000} to which we add quantified assignments and quantified differential equation systems for \emph{distributed} hybrid dynamics.
From these quantified assignments and quantified differential equations, \QHPs are built like a Kleene algebra with tests \cite{DBLP:journals/toplas/Kozen97}.
\begin{definition}[Quantified hybrid programs]
\QHPs are defined by the following grammar ($\alpha,\beta$ are \QHPs, $i$ a variable of sort $C$, $f$ is a function symbol, $\vec{s}$ is a vector of terms with sorts compatible to the arguments of $f$, $\theta$ is a term with sort compatible to the result of $f$, and $\ivr$ is a formula of many-sorted first-order logic):
\begin{equation*}
  \begin{array}{@{}l@{}}
  \alpha,\beta ~\bebecomes~
  \pupdate{\lforall[C]{i}{\pumod{f(\vec{s})}{\theta}}}
  \alternative
  \hevolvein{\lforall[C]{i}{\D{f(\vec{s})}=\theta}}{\ivr}
  \alternative
  \ptest{\ivr}
  \alternative
  \alpha\cup\beta
  \alternative
  \alpha;\beta
  \alternative
  \prepeat{\alpha}
  \end{array}
\end{equation*}
\end{definition}

In order to simplify technical difficulties, we impose regularity assumptions on $f(\vec{s})$ in quantified assignments and quantified differential equations.
We assume $\vec{s}$ to be either a vector of length 0 or that the mapping from the quantified variable $i$ to $\vec{s}$ is \emph{injective}.
That is, each value of $\vec{s}$ can be exhibited by at most one choice of $i$.
A system is injective, e.g., when at least one component of $\vec{s}$ is the quantified variable $i$.
These assumptions can be relaxed, but are sufficient for our purposes; see \rref{sec:QdL-semantics} for a discussion of injectivity.
For quantified differential equations, we further assume that $f$ is an $\reals$-valued function symbol so that derivatives can be defined.

\paragraph{Quantified State Change}
The effect of \dfn[assignment!quantified]{quantified assignment} 
\m{\pupdate{\lforall[C]{i}{\pumod{f(\vec{s})}{\theta}}}} is an instantaneous discrete jump assigning~$\theta$ to~$f(\vec{s})$ simultaneously for all objects $i$ of sort $C$.
Hence all $f(\vec{s})$ that are affected by \m{\pupdate{\lforall[C]{i}{\pumod{f(\vec{s})}{\theta}}}} will change their value to the respective~$\theta$ simultaneously for all choices of~$i$ in a single discrete instant of time.
Usually,~$i$ occurs in term~$\theta$, but does not have to.
The effect of \dfn[differential~equation!quantified]{quantified differential equation}
\m{\hevolvein{\lforall[C]{i}{\D{f(\vec{s})}=\theta}}{\ivr}} is a continuous evolution where, for all objects $i$ of sort $C$, all differential equations \m{\D{f(\vec{s})}=\theta} hold at the same time and formula~$\ivr$ holds throughout the evolution (the state always remains in the region described by~$\ivr$, i.e., the evolution stops at any arbitrary time before it leaves~$\ivr$).
Again,~$i$ usually occurs in term~$\theta$.
For the trivial evolution domain restriction~\m{\ivr\mequiv\ltrue}, which is always satisfied, we also write \m{\hevolve{\lforall[C]{i}{\D{f(\vec{s})}=\theta}}} instead of \m{\hevolvein{\lforall[C]{i}{\D{f(\vec{s})}=\theta}}{\ltrue}}.

The dynamics of \QHPs changes the interpretation of terms over time:
\m{\D{f(\vec{s})}} is intended to denote the derivative of the interpretation of the term \m{f(\vec{s})} over time during continuous evolution, not the derivative of \m{f(\vec{s})} by its argument $\vec{s}$.
For \m{\D{f(\vec{s})}} to be defined, we assume $f$ is an $\reals$-valued function symbol.
Although our approach can be extended, we assume that~$f$ does not occur in~$\vec{s}$.
The most common choice of $\vec{s}$ in quantified assignments and quantified differential equations is just $i$.
Other choices are possible for $\vec{s}$, e.g., \m{\vec{s}=(i,f(i))} in \m{\pupdate{\lforall[C]{i}{\pumod{d(i,f(i))}{\frac{1}{2}a(i)+\frac{1}{2}a(f(i))}}}}.
The latter \QHP could be used to model that, for each car $i$, the average acceleration of a car $i$ and its follower $f(i)$ is assigned to a data field $d(i,f(i))$ that car $i$ and its follower use to determine their safe distance.

Time itself is not special but implicit. If a clock variable $t$ is needed in a \QHP, it can be axiomatized by \m{\hevolve{\D{t}=1}}, which is equivalent to
\m{\hevolve{\lforall[C]{i}{\D{t}=1}}} where $i$ does not occur in $t$.
For such \dfn[vacuous!quantifier]{vacuous quantification} ($i$ does not occur anywhere), we may omit $\lforall[C]{i}{}$from assignments and differential equations, which are then classical assignments and ordinary differential equations.
Similarly, we may omit vectors $\vec{s}$ of length 0.

\paragraph{Regular Programs}
The \dfn{test} action~\m{\ptest{\ivr}} is used to define conditions. Its effect is that of a \textit{no-op} if the formula~$\ivr$ is true in the current state; otherwise, like \textit{abort}, it allows no transitions.
That is, if the test succeeds because formula~$\ivr$ holds in the current state, then the state does not change, and the system execution continues normally.
If the test fails because formula~$\ivr$ does not hold in the current state, then the system execution cannot continue, is cut off and not considered any further.

The nondeterministic choice~\m{\pchoice{\alpha}{\beta}}, sequential composition~\m{\alpha;\beta}, and non\-de\-ter\-min\-is\-tic repetition~\m{\prepeat{\alpha}} of programs are as in regular expressions but generalized to a semantics in distributed hybrid systems.
\dfn[nondeterministic!choice]{Nondeterministic choice} \m{\pchoice{\alpha}{\beta}} is used to express behavioral alternatives between the transitions of~$\alpha$ and~$\beta$.
That is, the \QHP~\m{\pchoice{\alpha}{\beta}} can choose nondeterministically to follow the transitions of \QHP~$\alpha$, or, instead, to follow the transitions of \QHP~$\beta$.
The \dfn[composition!sequential]{sequential composition}~\m{\alpha;\beta} says that the \QHP~$\beta$ starts executing after \QHP~$\alpha$ has finished ($\beta$ never starts if~$\alpha$ does not terminate).
In~\m{\alpha;\beta}, the transitions of~$\alpha$ take effect first, until~$\alpha$ terminates (if it does), and then~$\beta$ continues.
Observe that, like repetitions, continuous evolutions within~$\alpha$ can take more or less time, which causes uncountable nondeterminism.
This nondeterminism is inherent in distributed hybrid systems, because they can operate in so many different ways, which is as such reflected in \QHPs.
\dfn[nondeterministic!repetition]{Nondeterministic repetition}~\m{\prepeat{\alpha}} is used to express that the \QHP~$\alpha$ repeats any number of times, including zero times.
When following~\m{\prepeat{\alpha}}, the transitions of \QHP~$\alpha$ can be repeated over and over again, any nondeterministic number of times (\m{{\geq}0}).

\QHPs (with their semantics and our proof rules) can be extended to systems of quantified differential equations, systems of simultaneous assignments to multiple functions $f,g$, and statements with multiple quantifiers (\m{\lforall[C]{i}{\lforall[D]{j}{\dots}}}).
This includes the quantified differential equation system \m{\hevolve{\lforall[C]{i}{(\D{x(i)}=v(i)\syssep\D{v(i)}=a(i))}}}, which we can understand as a second-order quantified differential equation \m{\hevolve{\lforall[C]{i}{(\D[2]{x(i)}=a(i))}}} or as a vectorial first-order quantified differential equation \m{\hevolve{\lforall[C]{i}{\D{\vec{z}(i)}=\vec{\theta}}}} with \m{\vec{z}(i)=(x(i),v(i))} and \m{\vec{\theta}=(v(i),a(i))}; see \cite{DBLP:journals/jar/Platzer08} for details on how to handle vectorial differential equations.
It is similarly simple to extend our approach to quantified assignments with multiple function symbols like \m{\pupdate{\lforall[C]{i}{(\pumod{a(i)}{a(i)+1}\syssep\pumod{t(i)}{0})}}}, which is a vectorial extension that can be handled like parallel updates in programs \cite{DBLP:conf/cade/BeckertP06}.
Our approach can also be extended to multiple quantifiers like in the quantified differential equation \m{\hevolve{\lforall[C]{i}{\lforall[D]{j}{\D{f(i,j)}=a(i)-d(i,j)}}}}
or the quantified assignment \m{\pupdate{\lforall[C]{i}{\lforall[D]{j}{\pumod{d(i,j)}{d(i,j)+a(i)+1}}}}}.
These quantifier blocks correspond to \m{\lforall[\vec{C}]{\vec{i}}{}}with a vectorial variable $\vec{i}$ and a vectorial sort $\vec{C}$.
Since these simple vectorial extensions \cite{DBLP:journals/jar/Platzer08,DBLP:conf/cade/BeckertP06} are a diversion from the logical essence of our approach, we simplify notation and do not consider these cases formally.

\paragraph{Example}
{%
  \let\laforall\lforall%
Continuous movement of position $x(i)$ of car $i$ with acceleration $a(i)$ is expressed by differential equation \m{\hevolve{\D[2]{x(i)}=a(i)}}, which corresponds to the first-order differential equation system \m{\hevolve{\D{x(i)}=v(i)\syssep\D{v(i)}=a(i)}} where $v(i)$ is the velocity of car $i$.
Simultaneous movement of all cars with their respective accelerations $a(i)$ is expressed by the quantified differential equation
\m{\dcevo} where quantifier $\lforall[C]{i}$ranges over all cars, such that all cars co-evolve along their respective differential equations at the same time.

In addition to continuous dynamics, cars have discrete control.
In the following \QHP, discrete and continuous dynamics interact (repeatedly because of the~$\prepeat{}$ repetition operator):
\begin{equation}
  \renewcommand*{\dcaccelt}[1]{({\oa{a}{#1}}\,{\mathrel{{:}{=}}}\,{\piif{\SBforma{#1}}{\amax}{\,{-}\abrake}})}%
  \big(\dcaccelallt;\,\,\,
   \dcevo \prepeat{\big)}
  \label{eq:distributed-car-control-accel}
\end{equation}
First, all cars $i$ control their acceleration $a(i)$.
Each car $i$ chooses maximum acceleration \m{\amax\geq0} for \m{a(i)} if its distance to all other cars $j$ is far enough (some condition \m{\textit{far}(i,j)}).
Otherwise, $i$ chooses full braking \m{-\abrake<0}.
After all accelerations have been set, all cars move continuously along \m{\dcevo}.
Accelerations may change repeatedly, because the repetition operator $\prepeat{}$ can repeat the  \QHP when the continuous evolution stops at any time.
}%

Note that the presence of the function argument $i$ in $x(i), v(i), a(i)$ is a decisive difference when comparing the \QHP in \rref{eq:distributed-car-control-accel} to hybrid systems and when comparing the \QdL formula in \rref{eq:distributed-car-control-new} to  hybrid systems properties.
In hybrid systems, we are limited to using variables $x,v,a$ of a single car.
If we want to add a second car to a hybrid system model, new state variables $y,w,c$, new dynamics \m{\hevolve{\D{y}=w\syssep\D{w}=c}}, and new control need to be added for the second car.
We can keep on adding any fixed finite number of state variables that way, but we need to know exactly how many cars there are on the street.
This does not work when we want to model and verify situations with arbitrarily many cars or in distributed car control scenarios like \rref{fig:distributed-car-control-new}, where new cars appear or disappear during the evolution of the system.
A quantified differential equation like \m{\hevolve{\lforall[C]{i}{(\D{x(i)}=v(i)\syssep\D{v(i)}=a(i))}}}, for example, cannot be expressed in hybrid systems, because we do not know how many cars $i$ ranges over.
If $i$ did range over exactly 3 cars, called 1, 2, and 3, we could replace it by
\[\hevolve{\D{x(1)}=v(1)\syssep\D{v(1)}=a(1)\syssep\D{x(2)}=v(2)\syssep\D{v(2)}=a(2)\syssep\D{x(3)}=v(3)\syssep\D{v(3)}=a(3)}\]
and change notation to obtain primitive state variables $x_1,v_1,a_1,x_2,v_2,a_2,x_3,v_3,a_3$ in an ordinary differential equation system
\[\hevolve{\D{x_1}=v_1\syssep\D{v_1}=a_1\syssep\D{x_2}=v_2\syssep\D{v_2}=a_2\syssep\D{x_3}=v_3\syssep\D{v_3}=a_3}\]
But this replacement does not work unless we know exactly how many cars are in the system.
Even for systems with a fixed known but large number of participants, such flat representations as (non-distributed) hybrid systems are inefficient, because the system dimension is exponential in the number of participants and all reasoning needs to be repeated for each participant, or even for each pair of participants (collision freedom requires each pair of cars to remain safely separated).
This is why we benefit from studying distributed hybrid systems.

{%
  \let\laforall\lforall%
  \renewcommand*{\dcaccelt}[1]{\umod{\oa{a}{#1}}{(\piif{\SBforma{#1}}{\amax}{\piif{\oa{v}{i}>0}{\,{-}\abrake}{0}}})}%
\renewcommand{\dcevo}{\hevolvein{\laforall[C]{i}{(\D{\oa{x}{i}}=\oa{v}{i}\syssep\D{\oa{v}{i}}=\oa{a}{i}}}{\oa{v}{i}\geq0)}}%
One remaining issue with \QHP \rref{eq:distributed-car-control-accel} is that cars could still move backwards by braking long enough.
But this does not capture braking.
In order to say that cars can accelerate or brake but may never move backwards, we refine \QHP \rref{eq:distributed-car-control-accel} to the following \QHP in which the evolution domain of the quantified differential equation is restricted (by $\&$) to stay in the region \m{\oa{v}{i}\geq0} where each car $i$ has a nonnegative velocity:
\begin{multline*}
  \big(\dcaccelallt;\\
   \dcevo \prepeat{\big)}
\end{multline*}
Observe that this controller is also smarter about the acceleration choices of cars than that in \rref{eq:distributed-car-control-accel}.
It will choose $0$ for $a(i)$ if car $i$ does not move (\m{v(i)=0}) but car $i$ cannot accelerate safely either, because not all cars $j$ are far enough away.

\paragraph{System Structure}
The \emph{communication model} that \QdL supports is that of shared variable communication.
Suppose a car $i$ has direct control over the acceleration of car $j$. Then, when $i$ decides to brake, it could directly change the acceleration of car $j$ as well using the \QHP \m{\pupdate{\pumod{a(j)}{a(j)-2}}}.
In most system designs, control variables of other agents are not directly accessible but communication has to be used instead.
In \QdL, communication can be implemented by assigning to shared variables (delays in communication are easy to model by combining assignments with differential equations).
Suppose $s(i)$ is the data field that car $i$ queries periodically to track how much distance it is supposed to maintain relative to its leader car. 
Then the \QHP \m{\pupdate{\lforall[C]{i}{\pumod{s(f(i))}{s(f(i))+10}}}} would cause each car $i$ to tell its respective follower car $f(i)$ to increase the safety distance $s(i)$ by 10, e.g., when the road conditions are slippery.

Shared (first-order) variables are sufficient to model \emph{discrete structural dynamics}, e.g., of changing communication links.
If, for example, the car $f(i)$ following car $i$ has left the street, car $i$ may update its communication link to reflect this change in the structure of the system by running the \QHP \m{\pupdate{\pumod{f(i)}{f(f(i))}}} that updates the follower of $i$ to the follower of $f(i)$, i.e., the follower of the follower of $i$.
Other discrete structural changes in the system and communication patterns as well as all data structures can be modeled easily, since a complete object-oriented programming language \cite{DBLP:conf/cade/BeckertP06} can be defined in \QdL.
Shared (first-order) variables are sufficient to model \emph{continuous structural dynamics}, since structural changes in the continuous dynamics can be modeled by quantified differential equations that change their connectivity, i.e., which parts of the quantified differential equation depend on which other parts.
For example, in \QHP
\m{\hevolve{\lforall[C]{i}{(\D[2]{x(i)}=a(i)+c(i,f(i))a(f(i)))}}}
the connectivity term $c(i,f(i))$ models whether or not the follower $f(i)$ of car $i$ has physical bumper-to-bumper contact with car $i$, such that the acceleration $a(f(i))$ of car $f(i)$ also pushes car $i$ forwards, not just car $f(i)$.
The change of $c(i,f(i))$ from zero to non-zero represents a structural change in the physical dynamics structurally, because it structurally changes the effect of the continuous dynamics.

These examples illustrate how the discrete dynamics, continuous dynamics, and discrete and continuous structural dynamics of distributed hybrid systems with an arbitrary parametric number of participants can be modeled as a \QHP.
We defer the explanation of dimensional dynamics, i.e., dynamic appearance and disappearance of agents, to \rref{sec:objectcreation}.
}

\section{Semantics} \label{sec:QdL-semantics}
The \QdL semantics is a \emph{\oidV[constant]{varying} domain Kripke semantics \cite{Fitting_Mendelsohn_1999} with first-order structures as states} that associate total functions of appropriate type with function symbols.
In constant domain, all states share the same domain for quantifiers.
In particular, we choose to represent object creation not by changing the domain of states, but by changing the interpretation of the createdness flag $\laexisting{i}$ of the object denoted by $i$.
With $\laexisting{i}$, object creation is definable in a modular way (as we elaborate in \rref{sec:objectcreation}).

\paragraph{States}
A \emph{state} $\iget[state]{\I}$ 
associates an (infinite) set $\idomain{\I}{C}$ of objects with each sort $C$, and it
associates a function $\iget[state]{\I}(f)$ of appropriate type with each function symbol $f$, including $\laexisting{\cdot}$.
We assume $\laexisting{\cdot}$ to have (unbounded but) finite support, i.e., each state only has a finite number of positions $i$ at which \m{\laexisting{i}=1}.
This makes sense in practice, because there is a varying and possibly large but still finite numbers of participants (e.g., cars).
For simplicity, $\iget[state]{\I}$ also associates a value $\iget[state]{\I}(i)$ of appropriate type with each variable $i$.
The domain of $\reals$ and the interpretation of $0,1,+,-,\cdot$ is that of real arithmetic.
We assume \emph{constant domain} for each sort $C$: all states $\iget[state]{\I},\iget[state]{\It}$ share the same domains \m{\idomain{\I}{C}=\idomain{\It}{C}} for~$C$.
Sorts $C\neq D$ are disjoint: \m{\idomain{\I}{C} \cap \idomain{\I}{D} = \emptyset}.
\oidV{Object identifiers \m{\onew{C}\neq\onew[']{D}} are interpreted distinctly \m{\iget[state]{\I}(\onew{C}) \neq \iget[state]{\I}(\onew[']{D})} and ....}%
The set of all states is denoted by \m{\linterpretations{\Sigma}{V}}.
The state
\m{\iget[state]{\imodif[state]{\I}{i}{e}}} agrees with~$\iget[state]{\I}$ except for the interpretation of variable~$i$, which is changed to~\m{e \ignore{\idomain{\I}{C}}}.

\paragraph{Formulas} \label{sec:QdL-valuation}
We use $\ivaluation{\I}{\theta}$ to denote the value of term~$\theta$ at $\iname[state]{\I}$~$\iget[state]{\I}$, which is defined as in first-order logic.
Especially, \m{\ivaluation{\imodif[state]{\I}{i}{e}}{\theta}} denotes the value of $\theta$ in state \m{\iget[state]{\imodif[state]{\I}{i}{e}}}, i.e., in state $\iget[state]{\I}$ with $i$ interpreted as $e$.
Further, \m{\iaccess[\alpha]{\I} \subseteq \linterpretations{\Sigma}{V} \times \linterpretations{\Sigma}{V}} denotes the state transition relation of \QHP~$\alpha$, which we define below.
\begin{definition}[Semantics of {\QdL}]
    \newcommand{\Id}{\imodif[state]{\I}{i}{e}}%
  The \dfn{interpretation} \m{\imodels{\I}{\phi}} of \QdL formula~$\phi$ with respect to $\iname[state]{\I}~\iget[state]{\I}$ is defined inductively as:
  \begin{enumerate}[(1)]
  \item $\imodels{\I}{(\theta_1=\theta_2)}$
    iff $\ivaluation{\I}{\theta_1} = \ivaluation{\I}{\theta_2}$;
    accordingly for $\geq$ (greater or equal).
  \item $\imodels{\I}{\phi \land \psi}$ iff
    $\imodels{\I}{\phi}$ and $\imodels{\I}{\psi}$;
    accordingly for $\lnot$ (not).
  \item $\imodels{\I}{\lforall[C]{i}{\phi}}$
    iff
    $\imodels{\Id}{\phi}$
    for all objects \m{e\in\idomain{\I}{C}}.
    \index{$\lforall{}{}$}
  \item $\imodels{\I}{\lexists[C]{i}{\phi}}$
    iff
    $\imodels{\Id}{\phi}$
    for some object \m{e\in\idomain{\I}{C}}.
    \index{$\lexists{}{}$}
  \item
    $\imodels{\I}{\dbox{\alpha}{\phi}}$
      iff
      $\imodels{\It}{\phi}$
      for all states~$\iget[state]{\It}$ with
        $\related{\iaccess[\alpha]{\I}}{\iget[state]{\I}}{\iget[state]{\It}}$.
      \index{$\dbox{\alpha}{}$}
     \item $\imodels{\I}{\ddiamond{\alpha}{\phi}}$
       iff
       $\imodels{\It}{\phi}
       \mexistsr[\untweak{\iname[state]{\I}}]{\iget[state]{\It}\tweak{\,}\untweak{~}\textrm{with}~
         \related{\iaccess[\alpha]{\I}}{\iget[state]{\I}}{\iget[state]{\It}}}$.
       \index{$\ddiamond{\alpha}{}$}
     \end{enumerate}
\end{definition}
We say that $\phi$ is true at $\iportray{\I}$ if \m{\imodels{\I}{\phi}}.
\QdL formula $\phi$ is \dfn{valid}, written \m{\entails\phi}, iff \m{\imodels{\I}{\phi}} for all $\iportray{\I}$.

\paragraph{Programs} \label{sec:QdL-QHP-transition}
\QHPs have a compositional semantics.
The semantics of a \QHP is its reachability relation.
{\newcommand{\Ii}{\imodif[state]{\I}{i}{e}}%
   \newcommand{\ws}{\sigma}%
    \newcommand{\Ifz}[1][\zeta]{\iconcat[state=\varphi(#1)]{\stdI}}%
    \newcommand{\Ifzi}[1][\zeta]{\imodif[state]{\Ifz[#1]}{i}{e}}%
\begin{definition}[Transition semantics of {\QHP}]
   The \dfn[transition~relation]{transition relation, \m{\iaccess[\alpha]{\I} \subseteq \linterpretations{\Sigma}{V} \times \linterpretations{\Sigma}{V}}}, of \QHP~$\alpha$
    specifies which $\iname[state]{\I}$ $\iget[state]{\It} \in \linterpretations{\Sigma}{V}$ is reachable from $\iget[state]{\I} \in \linterpretations{\Sigma}{V}$ by running \QHP $\alpha$.
    It is defined inductively:
    \begin{enumerate}[(1)]
    \item \label{case:QdL-QHP-transition-assign}
      $\relateds{\iaccess[\pupdate{\lforall[C]{i}{\pumod{f(\vec{s})}{\theta}}}]{\I}}{\iget[state]{\I}}{\iget[state]{\It}}$
      iff $\iname[state]{\I}$~$\iget[state]{\It}$ is identical to~$\iget[state]{\I}$ except that
      at each position \m{\vec{o} \ignore{\in \iget[state]{\I}(\vec{S})}} of $f$:
      if \m{\ivaluation{\Ii}{\vec{s}} = \vec{o}} for some object \m{e\in\idomain{\I}{C}}, then
      \m{\iget[state]{\It}(f)\big(\ivaluation{\Ii}{\vec{s}}\big) = \ivaluation{\Ii}{\theta}}.
      If there are multiple objects $e$ giving the same position \m{\ivaluation{\Ii}{\vec{s}} = \vec{o}}, then all of the resulting states $\iget[state]{\It}$ are reachable.
      \oidV{If any $\onew{D}$ occurs in \m{\pupdate{\lforall[C]{i}{\pumod{f(\vec{s})}{\theta}}}}, then, in addition, the domain is updated to \m{\idomain{\It}{C}\eqdef\idomain{\I}{C}\cup\{\iget[const]{\I}(\onew{C})\}}.}
      
    \item \label{case:QdL-QHP-transition-evolve}
      $\relateds{\iaccess[\hevolvein{\lforall[C]{i}{\D{f(\vec{s})}=\theta}}{\ivr}]{\I}}{\iget[state]{\I}}{\iget[state]{\It}}$
      iff
      there is a\ignore{ (\emph{flow})} function
      ${{\varphi}{:}{\interval{[0,r]}\to\linterpretations{\Sigma}{V}}}$
      for some \m{r\geq0} with
      $\varphi(0)=\iget[state]{\I}$ and $\varphi(r)=\iget[state]{\It}$
      satisfying the following conditions.
      At each time \m{t \in \interval{[0,r]}}, state $\iget[state]{\Ifz[t]}$ is identical to $\iget[state]{\I}$, except that
      at each position \m{\vec{o} \ignore{\in \iget[state]{\I}(\vec{S})}} of $f$:
        if \m{\ivaluation{\Ii}{\vec{s}} = \vec{o}} for some object \m{e\in\idomain{\I}{C}}, then, at each time \m{\zeta \in \interval{[0,r]}}:
      \begin{iteMize}{$\bullet$}
        \item All differential equations hold and corresponding derivatives exist (trivial for \m{r=0}):
        \[
        \D[t]{\,(\ivaluation{\Ifzi[t]}{f(\vec{s})})} (\zeta) = (\ivaluation{\Ifzi[\zeta]}{\theta})
        \]
      \item The evolution domain is respected:
      \m{\imodels{\Ifzi}{\ivr}}.
      \end{iteMize}
      If there are multiple objects $e$ giving the same position \m{\ivaluation{\Ii}{\vec{s}} = \vec{o}}, then all of the resulting states $\iget[state]{\It}$ are reachable.
    \item $\iaccess[\ptest{\ivr}]{\I} =
      \{(\iget[state]{\I},\iget[state]{\I}) {\with}  \imodels{\I}{\ivr}\}$
    \item $\iaccess[\pchoice{\alpha}{\beta}]{\I} =
      \iaccess[\alpha]{\I} \cup \iaccess[\beta]{\I}$
      \index{$\pchoice{}{}$}
    \item $\iaccess[{\alpha};{\beta}]{\I} =
      \iaccess[\beta]{\I} \compose \iaccess[\alpha]{\I} =
      \{(\iget[state]{\I},\iget[state]{\It}) {\with}  \related{\iaccess[\alpha]{\I}}{\iget[state]{\I}}{z} ~\text{and}~ \related{\iaccess[\beta]{\I}}{z}{\iget[state]{\It}}$ $\text{for a}~\iname[state]{\I}~z\}$
    \item
      $\relateds{\iaccess[\prepeat{\alpha}]{\I}}{\iget[state]{\I}}{\iget[state]{\It}}$
      iff there is an \m{n \in \mathbb{N}} with \m{n\geq0} and there are states $\iget[state]{\I}=\ws_0,
      \ldots, \ws_n=\iget[state]{\It}$ such that $
      \related{\iaccess[\alpha]{\I}}{\ws_i}{\ws_{i+1}}$ for all \m{0
        \leq i < n}.\index{$\prepeat{}$}
    \end{enumerate}
\end{definition}
The semantics is \emph{explicit change}:
nothing changes unless an assignment or differential equation specifies how.
In cases~\ref{case:QdL-QHP-transition-assign}--\ref{case:QdL-QHP-transition-evolve}, only $f$ changes and only at positions of the form \m{\ivaluation{\Ii}{\vec{s}}} for some interpretation \m{e\in\idomain{\I}{C}} of $i$.
If there are multiple such $e$ that affect the same position $\vec{o}$, any of those changes can take effect by a nondeterministic choice.
\QHP \m{\hupdate{\lforall[C]{i}{\umod{x}{a(i)}}}} may change $x$ to \emph{any} $a(i)$.
Hence,
\m{\dbox{\hupdate{\lforall[C]{i}{\umod{x}{a(i)}}}}{\mapply{\phi}{x}} \mequiv \lforall[C]{i}{\mapply{\phi}{a(i)}}},
because that modality considers \emph{all} possibilities of changing $x$ to \emph{any} $a(i)$.
In contrast,
\m{\ddiamond{\hupdate{\lforall[C]{i}{\umod{x}{a(i)}}}}{\mapply{\phi}{x}} \mequiv \lexists[C]{i}{\mapply{\phi}{a(i)}}},
because that modality considers \emph{some} possibility of changing $x$ to \emph{any} $a(i)$.
Similarly, $x$ can evolve along \m{\hevolve{\lforall[C]{i}{\D{x}=a(i)}}} with any of the slopes $a(i)$. But evolutions cannot start with slope $a(c)$ and then switch to a different slope $a(d)$ later.
Any choice for the quantified variable $i$ is possible but $i$ remains unchanged during each evolution.

We call a quantified assignment \m{\pupdate{\lforall[C]{i}{\pumod{f(\vec{s})}{\theta}}}}  or a quantified differential equation \m{\hevolvein{\lforall[C]{i}{\D{f(\vec{s})}=\theta}}{\ivr}} \dfn[injective!QHP]{injective} iff there is at most one $e$ satisfying cases~\ref{case:QdL-QHP-transition-assign}--\ref{case:QdL-QHP-transition-evolve}.
For injective quantified assignments and injective quantified differential equations, conditions~\ref{case:QdL-QHP-transition-assign}--\ref{case:QdL-QHP-transition-evolve} can be simplified as follows:
\begin{enumerate}[  ($1'$) ]
\item
      $\relateds{\iaccess[\pupdate{\lforall[C]{i}{\pumod{f(\vec{s})}{\theta}}}]{\I}}{\iget[state]{\I}}{\iget[state]{\It}}$
      iff $\iname[state]{\I}$~$\iget[state]{\It}$ is identical to~$\iget[state]{\I}$ except that
      for each \m{e\in\idomain{\I}{C}}:
      \m{\iget[state]{\It}(f)\big(\ivaluation{\Ii}{\vec{s}}\big) = \ivaluation{\Ii}{\theta}}.
\item
      $\relateds{\iaccess[\hevolvein{\lforall[C]{i}{\D{f(\vec{s})}=\theta}}{\ivr}]{\I}}{\iget[state]{\I}}{\iget[state]{\It}}$
      iff
      there is a\ignore{ (\emph{flow})} function
      ${{\varphi}{:}{\interval{[0,r]}\to\linterpretations{\Sigma}{V}}}$
      for some \m{r\geq0} with
      $\varphi(0)=\iget[state]{\I}$ and $\varphi(r)=\iget[state]{\It}$
      such that
      for each \m{e\in\idomain{\I}{C}} and each time \m{\zeta \in \interval{[0,r]}}:
      \begin{iteMize}{$\bullet$}
        \item All differential equations hold and corresponding derivatives exist (trivial for \m{r=0}):
        \[
        \D[t]{\,(\ivaluation{\Ifzi[t]}{f(\vec{s})})} (\zeta) = (\ivaluation{\Ifzi[\zeta]}{\theta})
        \]
      \item The evolution domain is respected:
      \m{\imodels{\Ifzi}{\ivr}}.
      \end{iteMize}
\end{enumerate}
We call quantified assignments and quantified differential equations \dfn[schematic!QHP]{schematic} iff $\vec{s}$ is $i$ (thus injective) and the only arguments to function symbols in $\theta$ are $i$.
Schematic quantified differential equations like \m{\hevolvein{\lforall[C]{i}{\D{f(i)}=a(i)}}{\ivr}} are very common, because distributed hybrid systems often have a family of similar differential equations replicated for multiple participants $i$. Their synchronization often comes from discrete communication on top of their continuous dynamics. Physically coupled differential equations are possible as well.
They correspond to continuous physical interactions, e.g., if a car bumps into another car from the side, it radically changes the structure of the differential equations that determine its movement.
Either case can be represented in \QHPs, even if the schematic case is more common.

Cases~\ref{case:QdL-QHP-transition-assign}--\ref{case:QdL-QHP-transition-evolve} can be defined accordingly for vectorial extensions.
These vectorial extensions are simple, just notationally cumbersome.
For quantified assignments to multiple function symbols like in
\m{\pupdate{\lforall[C]{i}{(\pumod{f(\vec{s})}{\theta}\syssep\pumod{g(\vec{t})}{\vartheta})}}}
all changes to $f$ and $g$ according to \rref{case:QdL-QHP-transition-assign} are performed simultaneously when transitioning from state $\iget[state]{\I}$ to $\iget[state]{\It}$ \cite{DBLP:journals/jar/Platzer08,DBLP:journals/logcom/Platzer10,DBLP:conf/lpar/Rummer06}.
The only difference to the sequential composition
\m{(\pupdate{\lforall[C]{i}{\pumod{f(\vec{s})}{\theta}}});~(\pupdate{\lforall[C]{i}{\pumod{g(\vec{t})}{\vartheta}}})} is that in the quantified assignment to multiple functions, the change is simultaneous, hence $\vec{t}$ and $\vartheta$ are evaluated in the original state $\iget[state]{\I}$, not in the intermediate state that is reached after $f$ has already been modified by \m{\pupdate{\lforall[C]{i}{\pumod{f(\vec{s})}{\theta}}}}.
For quantified differential equation systems with multiple function symbols like in
\m{\hevolvein{\lforall[C]{i}{(D{f(\vec{s})}=\theta\syssep\D{g(\vec{t})}=\vartheta}}{\ivr)}} the changes to $f$ and $g$ according to \rref{case:QdL-QHP-transition-evolve} are again simultaneous and all differential equations of the differential equation system need to hold at the same time.
Multiple quantifiers like \m{\lforall[C]{i}{\lforall[D]{j}{}}}in the quantified differential equation and quantified assignment are vectorial, i.e., ``for some object \m{e\in\idomain{\I}{C}}'' in cases~\ref{case:QdL-QHP-transition-assign}--\ref{case:QdL-QHP-transition-evolve} is replaced by ``for some object \m{e\in\idomain{\I}{C}} and some object \m{c\in\idomain{\I}{D}}'', which are for $i$ and $j$, respectively.
That is, we replace \m{\iget[state]{\imodif[state]{\I}{i}{e}}} with     
\m{\iget[state]{\imodif[state]{\imodif[state]{\I}{i}{e}}{j}{c}}}
and \m{\iget[state]{\imodif[state]{\iconcat[state=\varphi(t)]{\stdI}}{i}{e}}} with \m{\iget[state]{\imodif[state]{\imodif[state]{\iconcat[state=\varphi(t)]{\stdI}}{i}{e}}{j}{c}}}
as well as \m{\iget[state]{\imodif[state]{\iconcat[state=\varphi(\zeta)]{\stdI}}{i}{e}}} with \m{\iget[state]{\imodif[state]{\imodif[state]{\iconcat[state=\varphi(\zeta)]{\stdI}}{i}{e}}{j}{c}}}
in cases~\ref{case:QdL-QHP-transition-assign}--\ref{case:QdL-QHP-transition-evolve}.

Note that existence/uniqueness theorems for solutions of differential equations \cite{Walter:ODE} carry over to quantified differential equations.
In particular, existence/uniqueness of solutions by Picard-Lindel\"of / Cauchy-Lipschitz theorem \cite[Theorem~10.VI]{Walter:ODE} and by Peano theorem \cite[Theorem~10.IX]{Walter:ODE} carry over to \rref{case:QdL-QHP-transition-evolve} of the semantics $\iaccess[\alpha]{\I}$ if it only affects a finite subdomain of \m{\idomain{\I}{C}}, because the quantifier then corresponds to a finite set of classical differential equations. (The number of differential equations may still change dynamically over time, though, so that the quantified differential equation system \emph{cannot} be replaced with an unquantified differential equation system in the \QHP).
For infinite \m{\idomain{\I}{C}}, the theorems carry over to schematic \m{\hevolvein{\lforall[C]{i}{\D{f(i)}=\theta}}{\ivr}}, which give an (infinite) set of disconnected classical differential equations.
In all these cases, Picard-Lindel\"of's theorem implies that the solution is unique, when terms are continuously differentiable (on the open domain where divisors are non-zero).
For an overview of results about general infinite-dimensional differential equations, see \cite{Bogachev95}.

}%

\section{Actual Existence and Object Creation} \label{sec:objectcreation}

Up to now, we have been neglecting the effects of object creation and just pretended that the domain of objects would never change.
In this section, we consider object creation and distinguish objects that actually exist physically from those that have not been created yet (or are not physically present in the part of the world reflected in the model).
We will see that this distinction does not require any change of \QdL.
It is just a conceptual change of our understanding.

\paragraph{Actual Existence}
For the \QdL semantics, we chose constant domain semantics, i.e., all states share the same domains. Thus quantifiers range over all possible objects (\emph{possibilist quantification} in constant domain semantics) not just over active existing objects (\emph{actualist quantification} in varying domain semantics) \cite{Fitting_Mendelsohn_1999}.
In order to distinguish between \emph{actual objects} that exist in a state, because they have already been created and can now actively take part in its evolution, versus \emph{possible objects} that still passively await creation, we use function symbol $\laexisting{\cdot}$.
Function symbol $\laexisting{\cdot}$ is similar to existence predicates in first-order modal logic \cite{Fitting_Mendelsohn_1999}, except that its value can be assigned to in \QHPs.

\paragraph{Object Creation}
For a term $i$ of type \m{C\neq\reals}, we use \m{\laexisting{i}=1} to represent that the object denoted by $i$ has been created and actually exists. We use \m{\laexisting{i}=0} to represent that $i$ has not been created or does not exist any longer.
Object creation amounts to changing the interpretation of $\laexisting{i}$. For an object denoted by $i$ that has not been created (\m{\lnaexisting{i}}), object creation corresponds to the state change caused by assignment \m{\pupdate{\pumod{\laexisting{i}}{1}}}.
With quantified assignments and function symbols, \dfn[object!creation]{object creation}
is definable by a \QHP:
\begin{align}
\pupdate{\pumod{\onew{}}{\pnew{C}}}
&\,\mequiv\,
(\pupdate{\lforall[C]{j}{\pumod{\onew{}}{j}}});~ \ptest{(\lnaexisting{\onew{}})};~ \paexisting{\onew{}}
\label{eq:new}
\end{align}
This \QHP assigns an arbitrary $j$ of type $C$ to $\onew{}$ (\m{\pupdate{\lforall[C]{j}{\pumod{\onew{}}{j}}}}) that did not exist before (subsequent test $\ptest{\lnaexisting{\onew{}}}$) and adjusts existence (\m{\paexisting{\onew{}}}).
\emph{Disappearance} of object $i$ corresponds to \m{\pupdate{\pumod{\laexisting{i}}{0}}}.
Our choice of constant domain semantics avoids semantic subtleties of varying domains about the meaning of free variables denoting non-existent objects as in free logics \cite{Fitting_Mendelsohn_1999}.
Denotation is standard in \QdL. Terms may just denote objects that have not been activated yet.
This is even useful to initialize new objects (e.g., \m{\pupdate{\pumod{x(\onew{})}{8}}}) before activation (\m{\paexisting{\onew{}}}).

\paragraph{Actualist Quantifiers}
We define abbreviations for \emph{actualist quantifiers} in formulas, quantified assignments, and quantified differential equations that range only over previously \emph{created objects}, similar to relativization in modal logic \cite{Fitting_Mendelsohn_1999} by masking:
\begin{align*}
  \laforall[C]{i}{\phi} &\mequiv \lforall[C]{i}{(\lpaexisting{i}\limply\phi)}
  \\
  \laexists[C]{i}{\phi} &\mequiv \lexists[C]{i}{(\lpaexisting{i}\land\phi)}
  \\
  \pupdate{\laforall[C]{i}{\umod{f(\vec{s})}{\theta}}} &\mequiv 
  \pupdate{\lforall[C]{i}{\umod{f(\vec{s})}{(\piif{\lpaexisting{i}}{\theta}{f(\vec{s})})}}}
  \\
  \renewcommand{\hevolvein}[2]{#1}
  \hevolvein{\laforall[C]{i}{\D{f(\vec{s})}={\theta}}}{\ivr} &\mequiv 
  \renewcommand{\hevolvein}[2]{#1}
  \hevolvein{\lforall[C]{i}{\D{f(\vec{s})}={(\piif{\lpaexisting{i}}{\theta}{0})}}}{\ivr}
  ~\mequiv~
  \renewcommand{\hevolvein}[2]{#1}
  \hevolvein{\lforall[C]{i}{\D{f(\vec{s})}=\laexisting{i} \theta}}{\ivr}
\end{align*}
The first two cases define quantifiers for actually existing objects.
The last two cases define quantified state change for actually existing objects using conditional terms that choose effect $\theta$ if $\lpaexisting{i}$ and choose no effect, retaining the old value $f(\vec{s})$ or evolving with slope 0, if $\lnaexisting{i}$.
The conditional terms can be avoided as indicated in the last column of the last row (similarly for quantified assignments).
In all cases, the notation $\laetype{C}$ signifies that the quantifier domain is restricted to actually existing objects of type $C$.
Hence, \m{\lforall[C]{i}{}}ranges over all objects of sort $C$, existent or not, whereas \m{\laforall[C]{i}{}}ranges only over those objects of sort $C$ that actually exist in the current state.

We generally assume that \QHPs involve only quantified assignments and differential equations that are restricted to created objects, because real systems only affect objects that are physically present, not those that will be created later.
We still treat actualist quantification over $\laetype{C}$ as a defined notion, in order to simplify the semantics and proof calculus by separating object creation from quantified state change rules in a modular way.

If only finitely many objects have been created in the initial state (say 0), then it is easy to see that only finitely many new objects will be created with finitely many such \QHP transitions, because each quantified state change for $\laetype{C}$ only ranges over a finite domain then.
Recall that we assume $\laexisting{\cdot}$ to have \emph{(unbounded but) finite support}, i.e., each state only has a finite number of positions $i$ at which \m{\laexisting{i}=1}.
This makes sense in practice, because there is a varying and possibly large but still finite numbers of participants (e.g., cars).

\paragraph{Example}
The car control examples in \rref{sec:QdL-syntax} were unaware of the distinction between actual existing and possible objects.
Car control, of course, only affects created cars that are physically present, not the possible cars that have not been built yet or that are not present yet.
To reflect this, the dynamics and properties, we only need to replace each occurrence of \m{\lforall[C]{i}{}}with \m{\laforall[C]{i}{}} in the car control examples of \rref{sec:QdL-syntax}.
For instance the \QdL formula \rref{eq:distributed-car-control-new} will be restricted to actual cars by adding $\laetype{C}$ as follows:
\begin{equation}
  {(\dcinv)\,} \limply {\,\dbox{\DCCS}{~\laforall[C]{i{\neq}j}{\oa{x}{i}{\neq}\oa{x}{j}}}}
  \label{eq:distributed-car-control-new-ae}
\end{equation}
In the precondition, we only demand that all cars that actually exist (\m{\laforall[C]{i,j}{\dots}}) start from compatible positions with compatible velocities and accelerations, because we do not care about non-existent cars.
In the postcondition, we only guarantee that all existing cars are at different positions, because we cannot really say what happens with cars that do not yet exist and that are beyond our control.
The controller and dynamics in the \QHP \DCCS can be restricted to actual cars in the same way, e.g., in the following variant of \rref{eq:distributed-car-control-accel}:
\begin{equation}
  \renewcommand*{\dcaccelt}[1]{({\oa{a}{#1}}\,{\mathrel{{:}{=}}}\,{\piif{\SBforma{#1}}{\amax}{\,{-}\abrake}})}%
  \big(\dcaccelallt;\,\,\,
   \dcevo \prepeat{\big)}
  \label{eq:distributed-car-control-accel-ae}
\end{equation}

Except conceptually, this restriction to created cars does not really affect the specification (nor its verification).
This gets much more involved as soon as we create new objects at runtime or let them disappear again.
When we create a new car that joins the system, or when a new car appears from an on-ramp (\rref{fig:distributed-car-control-new}), then one more set of positions $x(n)$, velocities $v(n)$, and accelerations $a(n)$ comes out of nowhere and starts evolving along with the distributed car control dynamics.
That new car $n$ has not even been considered in the dynamics before it has been created.
A real system cannot control what is not part of the system yet and thus must deal with new agents dynamically whenever they arrive.

A fairly challenging feature of distributed car control, thus, is that new cars may appear dynamically from on-ramps (\rref{fig:distributed-car-control-new}) changing the set of active objects dynamically at runtime.
To model this, we consider the following \QHP:
\begin{equation}
  \DCCS ~\mequiv~
  \prepeat{(\dcsys)}
  \label{eq:distributed-car-control-new-model}
\end{equation}
Before following the continuous dynamics, this \QHP creates a new car $\dcnu$ at an arbitrary position $x(\dcnu)$ satisfying compatibility condition \m{\dcseparate{i}{\dcnu}} with respect to all other created cars $i$.
Hence \DCCS allows new cars to appear, but not drop right out of the sky in front of a fast car or run at the speed of light only 2 meters away.
When cars appear into the horizon from on-ramps, this condition captures that a car is only allowed to join the lane (``appear'' into the model world) if it cannot cause a crash with other existing cars (\rref{fig:distributed-car-control-new}).
Unboundedly many cars may appear during the operation of \DCCS and change the system dimension arbitrarily, because of the repetition operator $\prepeat{}$.

\DCCS is simple but shows how properties of distributed hybrid systems can be expressed in \QdL.
Joint dynamics of multiple components corresponds to compositions of quantified differential equation systems, quantified assignments, and object (dis)appearance.
Structural dynamics corresponds to assignments to function terms.
Say, $f(i)$ is the car registered by communication as the car following car $i$. Then a term $d(i,f(i))$, which denotes the minimum safety distance negotiated between car $i$ and its follower, is a crucial part of the system dynamics.
Restructuring the system in response to lane change corresponds to assigning a new value to $f(i)$, which impacts the value of $d(i,f(i))$ in the system dynamics.

\section{Proof Calculus} \label{sec:QdL-calculus}

\begin{figure}[p]
  \def\globalrule{g}%
  \tabcolsep=0pt%
  \renewcommand{\linferenceRuleFootnoteSeparation}{\,}%
\renewcommand{\linferenceRuleNameSeparation}{\,}%
  \renewcommand{\linferPremissSeparation}{\quad}%
  \linferenceRulevskipamount=0.8em%
  \newdimen\linferenceRulehskipamount%
  \linferenceRulehskipamount=4pt%
  \linferenceRulehskipamount=12pt%
  \newdimen\lcalculuscollectionvskipamount%
  \lcalculuscollectionvskipamount=0.1em%
  \newcommand{\dupdate}[2]{\dmodality{\pupdate{#1}}{#2}}%
  \newcommand{\dupdatevar}[1]{\dmodality{\updatevar{#1}}}%
  \begin{calculuscollections}{\textwidth}
  \begin{calculuscollection}[prefix=D,reset]
    \begin{calculus}
      \cinferenceRule[choiceb|${[\cup]}$]{branch}
      {\linferenceRule[sequent]
        {\lsequent[s]{}{\dbox{\alpha}{\phi}
              \land \dbox{\beta}{\phi}}
        }
        {\lsequent[s]{}{
              \dbox{\pchoice{\alpha}{\beta}}{\phi}}}
      }{}
      \cinferenceRule[choiced|$\langle\cup\rangle$]{branch}
      {\linferenceRule[sequent]
        {\lsequent[s]{}{\ddiamond{\alpha}{\phi}
              \lor \ddiamond{\beta}{\phi}}
        }
        {\lsequent[s]{}{
              \ddiamond{\pchoice{\alpha}{\beta}}{\phi}}}
      }{}
    \end{calculus}
    \hspace{\linferenceRulehskipamount}%
    \begin{calculus}
      \cinferenceRule[composeb|${[{;}]}$]{composition}
      {\linferenceRule[sequent]
        {\lsequent[s]{}{
              \dbox{\alpha}{\dbox{\beta}{\phi}}}}
        {\lsequent[s]{}{
              \dbox{\alpha;\beta}{\phi}}}
      }{}
      \cinferenceRule[composed|$\langle{;}\rangle$]{composition}
      {\linferenceRule[sequent]
        {\lsequent[s]{}{
              \ddiamond{\alpha}{\ddiamond{\beta}{\phi}}}}
        {\lsequent[s]{}{
              \ddiamond{\alpha;\beta}{\phi}}}
      }{}
    \end{calculus}
    \hspace{\linferenceRulehskipamount}%
    \hspace{\linferenceRulehskipamount}%
    \begin{calculus}
      \cinferenceRule[testb|${[?]}$]{test}
      {\linferenceRule[sequent]
        {\lsequent[s]{}{\ivr \limply \psi}}
        {\lsequent[s]{}{\dbox{\ptest{\ivr}}{\psi}}}
      }{}
      \cinferenceRule[testd|$\langle?\rangle$]{test}
      {\linferenceRule[sequent]
        {\lsequent[s]{}{\ivr \land \psi}}
        {\lsequent[s]{}{\ddiamond{\ptest{\ivr}}{\psi}}}
      }{}
    \end{calculus}
  \end{calculuscollection}%
  \\[\lcalculuscollectionvskipamount]
  \begin{calculuscollection}[prefix=D]
    \begin{calculus}
      \cinferenceRule[evolveb|${[']}$]{evolve}
      {\linferenceRule[sequent]
        {\lsequent[s]{}{
              \lforall{t{\geq}0}{\big(
                (\lforall{0{\leq}\tilde{t}{\leq}t}{\dbox{\pupdate{\lforall[C]{i}{\solutionupdate{\tilde{t}}}}}{\ivr}})
                \limply
                \dbox{\pupdate{\lforall[C]{i}{\solutionupdate{t}}}}{\phi}
              \big)}
            }}
        {\lsequent[s]{}{
              \dbox{\hevolvein{\lforall[C]{i}{\D{f(\vec{s})}=\theta}}{\ivr}}{\phi}}}
      }{$t, \tilde{t}$ are new logical variables and \m{\solutionfor[\vec{s}]{}(t)} the simultaneous solutions of the (injective) differential equations \m{\hevolve{\lforall[C]{i}{\D{f(\vec{s})}=\theta}}} with~\m{f(\vec{s})} as symbolic initial values. \label{foot:evolveb}}
      \cinferenceRule[evolved|$\langle '\rangle$]{evolve}
      {\linferenceRule[sequent]
        {\lsequent[s]{}{
              \lexists{t{\geq}0}{\big(
                (\lforall{0{\leq}\tilde{t}{\leq}t}{\ddiamond{\pupdate{\lforall[C]{i}{\solutionupdate{\tilde{t}}}}}{\ivr}})
                \land
                \ddiamond{\pupdate{\lforall[C]{i}{\solutionupdate{t}}}}{\phi}
              \big)}
            }}
        {\lsequent[s]{}{
              \ddiamond{\hevolvein{\lforall[C]{i}{\D{f(\vec{s})}=\theta}}{\ivr}}{\phi}}}
      }{\ldito}
    \end{calculus}%
  \end{calculuscollection}%
  \\[\lcalculuscollectionvskipamount]
  \begin{calculuscollection}[prefix=D]
    \begin{calculus}
      \cinferenceRule[assignb|${[:=]}$]{apply (possibly quantified) update}
      {\linferenceRule[sequent]
        {\lsequent[s]{}{
            \piif{\lexists[C]{i}{\vec{s}=\dbox{\jupd}{\vec{u}}}}
            {\lforall[C]{i}{(\vec{s}=\dbox{\jupd}{\vec{u}} \limply \mapply{\phi}{\theta})}}
            {\mapply{\phi}{f(\dbox{\jupd}{\vec{u}})}}
        }}
        {\lsequent[s]{}{\mapply{\phi}{\dbox{\pupdate{\lforall[C]{i}{\umod{f(\vec{s})}{\theta}}}}{f(\vec{u})}}}}
      }{The occurrence of \m{f(\vec{u})} in \m{\mapply{\phi}{f(\vec{u})}} is not in scope of a modality (admissible substitution) and we abbreviate assignment \m{\hupdate{\lforall[C]{i}{\umod{f(\vec{s})}{\theta}}}} by $\jupd$, which is assumed to be injective.}
      \irlabel{upapply|$[:=]$}%
      \cinferenceRule[assignd|$\langle:=\rangle$]{apply (possibly quantified) update}
      {\linferenceRule[sequent]
        {\lsequent[s]{}{
            \piif{\lexists[C]{i}{\vec{s}=\ddiamond{\jupd}{\vec{u}}}}
            {\lexists[C]{i}{(\vec{s}=\ddiamond{\jupd}{\vec{u}} \land \mapply{\phi}{\theta})}}
            {\mapply{\phi}{f(\ddiamond{\jupd}{\vec{u}})}}
        }}
        {\lsequent[s]{}{\mapply{\phi}{\ddiamond{\pupdate{\lforall[C]{i}{\umod{f(\vec{s})}{\theta}}}}{f(\vec{u})}}}}
      }{\ldito}
    \end{calculus}
  \end{calculuscollection}
  \\[\lcalculuscollectionvskipamount]
  \begin{calculuscollection}[prefix=D]
    \hspace{\linferenceRulehskipamount}%
    \begin{calculus}
      \cinferenceRule[upskip|${[:]}$]{update skip}
      {\let\dmodality\dbox
        \linferenceRule[sequent]
        {\lsequent[s]{}{\mapply{\mascriptor}{\dmodality{\pupdate{\lforall[C]{i}{\umod{f(\vec{s})}{\theta}}}}{\vec{u}}}}}
        {\lsequent[s]{}{\dmodality{\pupdate{\lforall[C]{i}{\umod{f(\vec{s})}{\theta}}}}{\mapply{\mascriptor}{\vec{u}}}}}
      }{\m{f\neq \mascriptor} and the quantified assignment \m{\hupdate{\lforall[C]{i}{\umod{f(\vec{s})}{\theta}}}} is injective. The same rule applies for \m{\ddiamond{\hupdate{\lforall[C]{i}{\umod{f(\vec{s})}{\theta}}}}{}} instead of  \m{\dbox{\hupdate{\lforall[C]{i}{\umod{f(\vec{s})}{\theta}}}}{}}.}
      \oidV[{
      \cinferenceRule[newex|\usebox{\exbox}]{new existence pool}
      {\linferenceRule[sequent]
        {\lsequent[s]{}{\ltrue}}
        {\lsequent[s]{}{\lexists[C]{\onew{}}{\lnaexisting{\onew{}}}}}
      }{}%
      }]{}
    \end{calculus}
    \hspace{\linferenceRulehskipamount}%
    \begin{calculus}
      \cinferenceRule[assignrb|${[{:}{*}]}$]{random assignment}
      {\linferenceRule[sequent]
        {\lsequent[s]{}{\lforall[C]{j}{\mapply{\phi}{\theta}}}}
        {\lsequent[s]{}{\dbox{\pupdate{\lforall[C]{j}{\pumod{\onew{}}{\theta}}}}{\mapply{\phi}{\onew{}}}}}
      }{}
    \end{calculus}
    \hspace{\linferenceRulehskipamount}%
    \begin{calculus}
      \cinferenceRule[assignrd|${{\langle}{:}{*}{\rangle}}$]{random assignment}
      {\linferenceRule[sequent]
        {\lsequent[s]{}{\lexists[C]{j}{\mapply{\phi}{\theta}}}}
        {\lsequent[s]{}{\ddiamond{\pupdate{\lforall[C]{j}{\pumod{\onew{}}{\theta}}}}{\mapply{\phi}{\onew{}}}}}
      }{}
    \end{calculus}
  \end{calculuscollection}
  \\[\lcalculuscollectionvskipamount]
  \begin{calculuscollection}[prefix=F,context=L]
    \begin{calculus}
      \cinferenceRule[existsr|$\exists$\rightrule]{$\lexists{}{}$ right}
      {\linferenceRule[sequent]
        {\lsequent{}{\mapply{\phi}{\theta}, \lexists[C]{x}{\mapply{\phi}{x}}}}
        {\lsequent{}{\lexists[C]{x}{\mapply{\phi}{x}}}}
      }{$\theta$ is an arbitrary term of sort $C$, often a new logical variable $X$.}
      \irlabel{existsrinst|$\exists$\rightrule}
      \cinferenceRule[alll|$\forall$\leftrule]{$\lforall{}{}$ left instantiation}
      {\linferenceRule[sequent]
        {\lsequent{\mapply{\phi}{\theta},\lforall[C]{x}{\mapply{\phi}{x}}}{}}
        {\lsequent{\lforall[C]{x}{\mapply{\phi}{x}}}{}}
      }{\ldito}
      \irlabel{alllinst|$\forall$\leftrule}
    \end{calculus}
    \hspace{\linferenceRulehskipamount}\hspace{0.48cm}
    \begin{calculus}
      \cinferenceRule[allr|$\forall$\rightrule]{$\lforall{}{}$ right}
      {\linferenceRule[sequent]
        {\lsequent{}{\mapply{\phi}{\skolem{f}(X_1,\sdots,X_n)}}}
        {\lsequent{}{\lforall[C]{x}{\mapply{\phi}{x}}}}
      }{$\skolem{f}$ is a new (Skolem) function of appropriate type and~$X_1,\sdots,X_n$ are all free logical variables of~\m{\lforall{x}{\mapply{\phi}{x}}}.}
      \cinferenceRule[existsl|$\exists$\leftrule]{$\lexists{}{}$ left}
      {\linferenceRule[sequent]
        {\lsequent{\mapply{\phi}{\skolem{f}(X_1,\sdots,X_n)}}{}}
        {\lsequent{\lexists[C]{x}{\mapply{\phi}{x}}}{}}
      }{\ldito}
    \end{calculus}
  \end{calculuscollection}
  \vspace*{-\baselineskip}
  \\[\lcalculuscollectionvskipamount]
  \begin{calculuscollection}[prefix=F,context=L]
    \begin{calculus}[context=l]
      \cinferenceRule[iallr|i$\forall$]{$\lforall{}{}$ right inverse}
      {\linferenceRule[sequent]
        {\lsequent{}{\qelim{
          \lforall{X,Y}{(\piif{\vec{s}=\vec{t}}{\lsequent[f]{\mapply{\Phi}{X}}{\mapply{\Psi}{X}}}{\lsequent[f]{\mapply{\Phi}{X}}{\mapply{\Psi}{Y}}})}
        }}}
        {\lsequent{\mapply{\Phi}{f(\vec{s})}}{\mapply{\Psi}{f(\vec{t})}}}
      }{\m{X,Y} are new logical variables of sort $\reals$.
        \qelim{} needs to be applicable to the formula in the premise.}
    \end{calculus}%
    \hfill%
    \begin{calculus}[context=l]
      \cinferenceRule[iexistsr|i$\exists$]{$\lexists{}{}$ right inverse}
      {\linferenceRule[sequent]
        {\lsequent{}{\qelim{
          \lexists{X}{\landfold_i (\lsequent[f]{\Phi_i}{\Psi_i})}
        }}}
        {\lsequent{\Phi_1}{\Psi_1} ~~ \dots ~~ \lsequent{\Phi_n}{\Psi_n}}
      }{among all branches, the free (existential) logical variable~$X$ of sort $\reals$ only occurs in the branches~\m{\lsequent{\Phi_i}{\Psi_i}}.
        \qelim{} needs to be defined for the formula in the premise, especially, no Skolem dependencies on~$X$ occur.}
    \end{calculus}
  \end{calculuscollection}
  \\[\lcalculuscollectionvskipamount]
  \begin{calculuscollection}[prefix=G,context=L]
    \begin{calculus}
      \cinferenceRule[genb|${[]}gen$]{$\ddiamond{}{}/\dbox{}{}$ generalisation}
      {\linferenceRule[sequent]
        {\lsequent[\globalrule]{\phi}{\psi}}
        {\lsequent{\dbox{\alpha}{\phi}}{\dbox{\alpha}{\psi}}}
      }{}
    \end{calculus}
    \hspace{\linferenceRulehskipamount}\hspace{0.57cm}%
    \begin{calculus}
      \cinferenceRule[gend|${\langle\rangle}gen$]{$\ddiamond{}{}/\dbox{}{}$ generalisation}
      {\linferenceRule[sequent]
        {\lsequent[\globalrule]{\phi}{\psi}}
        {\lsequent{\ddiamond{\alpha}{\phi}}{\ddiamond{\alpha}{\psi}}}
      }{}
    \end{calculus}
    \hspace{\linferenceRulehskipamount}\hspace{0.15cm}%
    \begin{calculus}
      \cinferenceRule[invind|$ind$]{inductive invariant}
      {\linferenceRule[sequent]
        {\lsequent[g]{\inv}{\dbox{\alpha}{\inv}}}
        {\lsequent{\inv}{\dbox{\prepeat{\alpha}}{\inv}}}
      }{}
    \end{calculus}
    \hspace{\linferenceRulehskipamount}%
    \begin{calculus}
      \cinferenceRule[con|$con$]{loop convergence right}
      {\linferenceRule[sequent]
        {\lsequent[g]{}{v>0 \land {\var(v)}\limply{\ddiamond{\alpha}{\var(v-1)}}}}
        {\lsequent{\lexists{v}{\var(v)}}{
            \ddiamond{\prepeat{\alpha}}{\lexists{v{\leq}0}{\var(v)}}}}
      }{logical variable~$v$ does not occur in~$\alpha$.}
    \end{calculus}
  \end{calculuscollection}
  \index{$\lnot$}\index{$\land$}\index{$\lor$}\index{$\limply$}%
  \index{$\lforall{}{}$}\index{$\lexists{}{}$}%
  \index{$;$}%
  \index{calculus!of~QdL_@of~\QdL|textbf}%
  \end{calculuscollections}
  \caption{Rule schemata of the proof calculus for quantified differential dynamic logic.}
  \label{calculus:QdL}
\end{figure}

In \rref{calculus:QdL}, we present a proof calculus for \QdL formulas.
The basic principle behind the proof rules is that they transform a \QHP into structurally simpler logical formulas by symbolic decomposition\index{symbolic~decomposition}.
For our purposes, it is sufficient to understand the sequent notation  informally, just for a systematic proof structure.
With finite sets of formulas for the \dfn{antecedent}~$\Gamma$ and \dfn{succedent}~$\Delta$, \dfn{sequent} \m{\lsequent{\Gamma}{\Delta}} is an abbreviation for the formula
\m{\landfold_{\phi \in \Gamma} \phi \,\limply\, \lorfold_{\psi \in \Delta} \psi}.
Our calculus uses standard proof rules for propositional logic with the cut rule; see \rref{calculus:propositional}.
The proof rules are used backwards from the \emph{conclusion} (goal below horizontal bar) to the \emph{premises} (subgoals above bar).

In the \QdL calculus, we use substitutions that take effect within formulas and programs (defined as usual).
Only admissible substitutions are applicable, however, which is crucial for soundness.
  An application of a substitution~$\sigma$ is \dfn[substitution!admissible]{admissible} if no
  replaced term~$\theta$ occurs in the scope of a quantifier or modality binding a
  symbol in~$\theta$ or in its replacement~$\applysubst{\sigma}{\theta}$.
  A modality \dfn[symbol!bound]{binds} a symbol~$f$ iff it contains an assignment to~$f$ (like \m{\pupdate{\lforall[C]{i}{\umod{f(\vec{s})}{\theta}}}}) or a differential equation containing an~$\D{f(\vec{s})}$ (like \m{\hevolve{\lforall[C]{i}{\D{f(\vec{s})}=\theta}}}).
The substitutions in \rref{calculus:QdL} that insert a term~\m{\theta} into~\m{\mapply{\phi}{\theta}} also have to be admissible for the proof rules to be applicable.
We explain the \QdL proof rules in the sequel.

\begin{figure}[tbp]
  \def\globalrule{g}%
  \tabcolsep=0pt%
\renewcommand{\linferenceRuleNameSeparation}{}%
  \renewcommand{\linferPremissSeparation}{\quad}%
  \linferenceRulevskipamount=0.9em%
  \newdimen\linferenceRulehskipamount%
  \linferenceRulehskipamount=9pt%
  \newdimen\lcalculuscollectionvskipamount%
  \lcalculuscollectionvskipamount=0.1em%
  \newcommand{\dupdate}[2]{\dmodality{\pupdate{#1}}{#2}}%
  \newcommand{\dupdatevar}[1]{\dmodality{\updatevar{#1}}}%
  \begin{calculuscollections}{\textwidth}
  \begin{calculuscollection}[prefix=P,reset,context=L]
    \begin{calculus}
      \cinferenceRule[notr|$\lnot$\rightrule]{$\lnot$ right}
      {\linferenceRule[sequent]
        {\lsequent{\phi}{}}
        {\lsequent{}{\lnot \phi}}
      }{}
      \cinferenceRule[notl|$\lnot$\leftrule]{$\lnot$ left}
      {\linferenceRule[sequent]
        {\lsequent{}{\phi}}
        {\lsequent{\lnot \phi}{}}
      }{}
    \end{calculus}
    \hspace{\linferenceRulehskipamount}%
    \begin{calculus}
      \cinferenceRule[orr|$\lor$\rightrule]{$\lor$ right}
      {\linferenceRule[sequent]
        {\lsequent{}{\phi, \psi}}
        {\lsequent{}{\phi \lor \psi}}
      }{}
      \cinferenceRule[orl|$\lor$\leftrule]{$\lor$ left}
      {\linferenceRule[sequent]
        {\lsequent{\phi}{}
          & \lsequent{\psi}{}}
        {\lsequent{\phi \lor \psi}{}}
      }{}
    \end{calculus}
    \hspace{\linferenceRulehskipamount}%
    \begin{calculus}
      \cinferenceRule[andr|$\land$\rightrule]{$\land$ right}
      {\linferenceRule[sequent]
        {\lsequent{}{\phi}
          & \lsequent{}{\psi}}
        {\lsequent{}{\phi \land \psi}}
      }{}
      \cinferenceRule[andl|$\land$\leftrule]{$\land$ left}
      {\linferenceRule[sequent]
        {\lsequent{\phi , \psi}{}}
        {\lsequent{\phi \land \psi}{}}
      }{}
    \end{calculus}
    \hspace{\linferenceRulehskipamount}%
    \begin{calculus}
      \cinferenceRule[implyr|$\limply$\rightrule]{$\limply$ right}
      {\linferenceRule[sequent]
        {\lsequent{\phi}{\psi}}
        {\lsequent{}{(\phi \limply \psi)}}
      }{}
      \cinferenceRule[implyl|$\limply$\leftrule]{$\limply$ left}
      {\linferenceRule[sequent]
        {\lsequent{}{\phi}
          & \lsequent{\psi}{}}
        {\lsequent{(\phi \limply \psi)}{}}
      }{}
    \end{calculus}
  \end{calculuscollection}
  \\[\lcalculuscollectionvskipamount]
  \begin{calculuscollection}[prefix=P,context=L]
    \begin{calculus}
      \cinferenceRule[axiom|$ax$]{axiom}
      {\linferenceRule[sequent]
        {}
        {\lsequent{\phi}{\phi}}
      }{}
    \end{calculus}
    \hspace{\linferenceRulehskipamount}
    \begin{calculus}
      \cinferenceRule[cut|$cut$]{cut}
      {\linferenceRule[sequent]
        {\lsequent{}{\phi}
        &\lsequent{\phi}{}}
        {\lsequent{}{}}
      }{}
    \end{calculus}
  \end{calculuscollection}
  \end{calculuscollections}
  \caption{Propositional rule schemata}
  \label{calculus:propositional}
\end{figure}

\paragraph{Regular Rules}
The first proof rules in \rref{calculus:QdL} axiomatize sequential compositions (\irref{composeb+composed}), nondeterministic choices (\irref{choiceb+choiced}), and tests (\irref{testb+testd}) of regular programs as in dynamic logic \cite{Harel_et_al_2000}.
Like most other rules in \rref{calculus:QdL}, these rules do not contain sequent symbol$\lsequent{\,}{}$, i.e., they can be applied to any subformula.
These rules represent (directed) equivalences: conclusion and premise are equivalent. 
The equivalences are directed in the sense that we only use them to replace occurrences of the conclusion with the premise (which is structurally simpler), not the other way around.

Nondeterministic choices split into their alternatives (\irref{choiceb+choiced}).
For rule \irref{choiceb}: If all $\alpha$ transitions lead to states satisfying~$\phi$ (i.e., \m{\dbox{\alpha}{\phi}} holds) and all $\beta$ transitions lead to states satisfying $\phi$ (i.e., \m{\dbox{\beta}{\phi}} holds), then, all transitions of \QHP \m{\pchoice{\alpha}{\beta}}, which choose between following $\alpha$ and following $\beta$, also lead to states satisfying $\phi$ (i.e., \m{\dbox{\pchoice{\alpha}{\beta}}{\phi}} holds).
Dually for rule \irref{choiced}, if there is an $\alpha$ transition to a $\phi$ state (\m{\ddiamond{\alpha}{\phi}}) or a $\beta$-transition to a $\phi$ state (\m{\ddiamond{\beta}{\phi}}), then, in either case, there is a transition of \m{\pchoice{\alpha}{\beta}} to $\phi$ (\m{\ddiamond{\pchoice{\alpha}{\beta}}{\phi}} holds), because \m{\pchoice{\alpha}{\beta}} can choose which of those transitions to follow.
A general principle behind the \QdL proof rules is most noticeable in \irref{choiceb+choiced}: these proof rules symbolically decompose the reasoning into two separate parts and analyze the fragments~$\alpha$ and~$\beta$ separately, which makes the problem tractable and is good for scalability.
For these symbolic structural decompositions\index{symbolic~decomposition}, it is very helpful that \QdL is a full logic that is closed under all logical operators, including disjunction and conjunction, for then the premises in \irref{choiceb+choiced} are \QdL formulas again (unlike in Hoare logic \cite{DBLP:journals/cacm/Hoare69}).

Sequential compositions are proven using nested modalities (\irref{composeb+composed}).
For rule \irref{composeb}: If after all $\alpha$-transitions, all $\beta$-transitions lead to states satisfying~$\phi$ (i.e., \m{\dbox{\alpha}{\dbox{\beta}{\phi}}} holds), then also all transitions of the sequential composition \m{\alpha;\beta} lead to states satisfying $\phi$ (i.e., \m{\dbox{\alpha;\beta}{\phi}} holds).
The dual rule \irref{composed} uses the fact that if there is an $\alpha$-transition, after which there is a $\beta$-transition leading to $\phi$ (i.e., \m{\ddiamond{\alpha}{\ddiamond{\beta}{\phi}}}), then there is a transition of \m{\alpha;\beta} leading to $\phi$ (that is, \m{\ddiamond{\alpha;\beta}{\phi}}), because the transitions of \m{\alpha;\beta} are just those that first do any $\alpha$-transition, followed by any $\beta$-transition (\rref{sec:QdL-semantics}).
Again, it is crucial that \QdL is a full logic that considers reachability statements as modal operators, which can be nested, for then the premises in \irref{composeb+composed} are \QdL formulas again (unlike in Hoare logic \cite{DBLP:journals/cacm/Hoare69}).

Tests are proven by assuming (with an implication in rule \irref{testb})  or showing (with a conjunction in rule \irref{testd}) that the test succeeds, because test~$\ptest{\ivr}$ can only make a transition when condition~$\ivr$ actually holds true (\rref{sec:QdL-semantics}).
Thus, for \QdL formula \m{\ddiamond{\ptest{\ivr}}{\phi}}, rule \irref{testd} is used to prove that formula $\ivr$ holds true (otherwise there is no transition and thus the reachability property is false) and that formula $\phi$ holds after the resulting no-op.
Dually, rule \irref{testb} for \QdL formula \m{\dbox{\ptest{\ivr}}{\phi}} assumes that formula $\ivr$ holds true (otherwise there is no transition and thus nothing to show) and shows that $\phi$ holds after the resulting no-op.

\paragraph{Quantified Differential Equations}
Rules \irref{evolveb+evolved} handle continuous evolutions for quantified differential equations with first-order definable solutions.
Given a solution for the quantified differential equation system with symbolic initial values~$f(\vec{s})$, continuous evolution along differential equations can be replaced with a quantified assignment \m{\pupdate{\lforall[C]{i}{\solutionupdate{t}}}} corresponding to the simultaneous solution (of the differential equations \m{\hevolve{\lforall[C]{i}{\D{f(\vec{s})}=\theta}}} with~\m{f(\vec{s})} as symbolic initial values)
and an additional quantifier for the evolution time~$t$.
In rule \irref{evolveb}, postcondition $\phi$ needs to hold \emph{for all} evolution durations \m{t\geq0}.
In rule \irref{evolved}, it needs to hold after \emph{some} duration \m{t\geq0}.
The constraint on~$\ivr$ restricts the continuous evolution such that its solution~\m{\solutionupdate{\tilde{t}}} remains in the evolution domain region~$\ivr$ at all intermediate times~\m{\tilde{t}\leq t}.
This constraint simplifies to~$\ltrue$ if~$\ivr$ is~$\ltrue$.

For schematic cases like \m{\hevolve{\lforall[C]{i}{\D{f(i)}=a(i)}}}, first-order definable solutions can be obtained by adding argument $i$ to first-order definable solutions of the deparametrized version \m{\hevolve{\D{f}=a}}.
For example, the following proof step uses rule \irref{evolveb} to turn a quantified differential equation system into a quantified assignment with an extra quantifier for the duration $t$ of the evolution.
{
   \def\arraystretch{1.3}%
    \begin{sequentdeduction}[array]
          \linfer[evolveb]
           {\lsequent{\lforall{i{\neq}j}{x(i){\neq}x(j)}}  {\lforall{t{\geq}0}{\dbox{\hupdate{\lforall{i}{\humod{x(i)}{{-}{\frac{b}{2}}t^2+v(i)t + x(i)}}}}{\,\lforall{j{\neq}k}{x(j){\neq}x(k)}}}}}
         {\lsequent{\lforall{i{\neq}j}{x(i){\neq}x(j)}} {\dbox{\hevolve{\lforall{i}{\D{x(i)}=v(i)\syssep\D{v(i)}=-b}}}{\,\lforall{j{\neq}k}{x(j){\neq}x(k)}}}}
  \end{sequentdeduction}
}%
The quantified assignment \m{\hupdate{\lforall{i}{\humod{x(i)}{{-}{\frac{b}{2}}t^2+v(i)t + x(i)}}}} solving the above quantified differential equation system can be obtained easily from the solution \m{\hupdate{\humod{x}{{-}{\frac{b}{2}}t^2+vt + x}}} of the deparametrized differential equation system \m{\hevolve{\D{x}=v\syssep\D{v}=-b}}, just by adding the parameter $i$ back in and checking whether this gives a solution.

We only present proof rules for first-order definable solutions of quantified differential equations here.
We refer to previous work \cite{DBLP:journals/logcom/Platzer10} for induction techniques that handle differential equations without solving them and that work for nondeterministic differential equations with disturbances.
We have shown recently that these differential induction techniques extend to quantified differential equations using \emph{quantified differential invariants} \cite{DBLP:conf/hybrid/Platzer11}.

\paragraph{Quantified Assignments}
Rules \irref{assignb+assignd+upskip} handle quantified assignments.
Rule \irref{upskip} characterizes the fact that quantified assignments to $f$ have no effect on all other operators $\mascriptor\neq f$ (including other function symbols, $\land$, $\piif{}{}{}$), so that $\mascriptor$ will not be affected by the quantified assignment and can be skipped over.
The argument $\vec{u}$ may still be affected by the quantified assignment, hence \irref{upskip} prefixes $\vec{u}$ (component-wise) by \m{\pupdate{\lforall[C]{i}{\umod{f(\vec{s})}{\theta}}}}.
Hence, the \irref{upskip} rule maps a quantified assignment over all arguments homomorphically.
For example, if $\mascriptor$ is an operator taking two arguments and is not the function symbol $f$, then rule \irref{upskip} derives the proof step
\begin{sequentdeduction}
        \linfer[upskip]
        {\lsequent[s]{}{\mapply{\mascriptor}{\dbox{\pupdate{\lforall[C]{i}{\umod{f(\vec{s})}{\theta}}}}{u_1},\dbox{\pupdate{\lforall[C]{i}{\umod{f(\vec{s})}{\theta}}}}{u_2}}}}
        {\lsequent[s]{}{\dbox{\pupdate{\lforall[C]{i}{\umod{f(\vec{s})}{\theta}}}}{\mapply{\mascriptor}{u_1,u_2}}}}
\end{sequentdeduction}

Rules \irref{assignb+assignd} characterize how a quantified assignment to $f$ affects the value of a term $f(\vec{u})$ (these rules are equivalent for the injective case, i.e., a match for at most one $i$).
Their effect depends on whether the quantified assignment \m{\pupdate{\lforall[C]{i}{\umod{f(\vec{s})}{\theta}}}} \dfn[match]{matches} \m{f(\vec{u})}, i.e., there is a choice for $i$ such that \m{f(\vec{u})} is affected by the assignment, because $\vec{u}$ is of the form $\vec{s}$ for some $i$.
Whether it matches or not cannot always be decided statically, because it may depend on the particular interpretations.
Hence, the premises of rules \irref{assignb+assignd} make a case distinction on matching by yielding an \keywordfont{if-then-else} formula.
The formula \m{\piif{\phi}{\phi_1}{\phi_2}} is short notation for
\m{(\phi \limply \phi_1)  \land  (\lnot\phi \limply \phi_2)}.
If the quantified assignment does not match (\keywordfont{else} part), the occurrence of $f$ in \m{\mapply{\phi}{f(\vec{u})}} will be left unchanged, because $f$ is not changed at position $\vec{u}$.
If it matches (\keywordfont{then} part), the premise uses the term $\theta$ assigned to \m{f(\vec{s})} instead of \m{f(\vec{u})}, either for all possible \m{\hastype{i}{C}} that match \m{f(\vec{u})} in case of \irref{assignb}, or for some of those \m{\hastype{i}{C}} in case of \irref{assignd}.
The universal and existential quantifiers pick the same unique $i$, because the quantified assignment needs to be injective for \irref{assignb+assignd}.
In all cases, the original quantified assignment \m{\pupdate{\lforall[C]{i}{\umod{f(\vec{s})}{\theta}}}}, which we abbreviate by $\jupd$, will be applied to \m{\vec{u}} in the premise, because the value of argument \m{\vec{u}} may also be affected by \m{\jupd}, recursively.

A special case of \irref{assignb} applies to the schematic case where $\vec{s}$ is of the form $i$, which matches trivially:
\[
      \linfer[assignb]
        {\lsequent[s]{}{
            \lforall[C]{i}{(i=\dbox{\pupdate{\lforall[C]{i}{\umod{f(i)}{\theta}}}}{u} \limply \mapply{\phi}{\theta})}
        }}
        {\lsequent[s]{}{\mapply{\phi}{\dbox{\pupdate{\lforall[C]{i}{\umod{f(i)}{\theta}}}}{f(u)}}}}
\]
If $f$ does not occur in $u$, then \irref{upskip} simplifies this proof step further:
\[
      \linfer[assignb+upskip]
        {\lsequent[s]{}{
            \lforall[C]{i}{(i=u \limply \mapply{\phi}{\theta})}
        }}
        {\lsequent[s]{}{\mapply{\phi}{\dbox{\pupdate{\lforall[C]{i}{\umod{f(i)}{\theta}}}}{f(u)}}}}
\]
Recall that $\subst[\theta]{i}{u}$ is the term $\theta$ with $i$ replaced by $u$.
Standard first-order reasoning simplifies the above to a derived rule that we again denote by \irref{assignb} (where $f$ does not occur in $u$)
\[
      \linfer[assignb]
        {\lsequent[s]{}{
            \mapply{\phi}{\subst[\theta]{i}{u}}
        }}
        {\lsequent[s]{}{\mapply{\phi}{\dbox{\pupdate{\lforall[C]{i}{\umod{f(i)}{\theta}}}}{f(u)}}}}
\]
Together with \irref{upskip} to propagate the change to both arguments of $\neq$, this derived rule proves, for example, the following proof step:
\begin{sequentdeduction}[array]
  \linfer[assignb+upskip]
                   {\lsequent{\lforall{i{\neq}j}{x(i){\neq}x(j)}} {\lforall{j{\neq}k}{({-}{\frac{b}{2}}\skolem{s}^2+v(j)\skolem{s} + x(j) \neq {-}{\frac{b}{2}}\skolem{s}^2+v(k)\skolem{s} + x(k))}}}
               {\lsequent{\lforall{i{\neq}j}{x(i){\neq}x(j)}} {\lforall{j{\neq}k}{\dbox{\hupdate{\lforall{i}{\humod{x(i)}{{-}{\frac{b}{2}}\skolem{s}^2+v(i)\skolem{s} + x(i)}}}}{\,x(j){\neq}x(k)}}}}
\end{sequentdeduction}

Rules \irref{assignb+assignd+upskip} also apply for assignments without quantifiers, which correspond to vacuous quantification $\lforall[C]{i}{}$where $i$ does not occur anywhere.
The following rule, for example, is a special case of \irref{assignb}
\begin{sequentdeduction}[array]
  \linfer[assignb]
        {\lsequent{}{
            \piif{s=\dbox{\pupdate{\umod{f(s)}{\theta}}}{k}}
            {\mapply{\phi}{\theta}}
            {\mapply{\phi}{f(\dbox{\pupdate{\umod{f(s)}{\theta}}}{k})}}
        }}
        {\lsequent{}{\mapply{\phi}{\dbox{\pupdate{\umod{f(s)}{\theta}}}{f(k)}}}}
\end{sequentdeduction}
If $f$ does not occur in term $k$, then this special case of \irref{assignb} simplifies further to
\begin{sequentdeduction}[array]
  \linfer[assignb]
        {\lsequent{}{
            \piif{s=k}
            {\mapply{\phi}{\theta}}
            {\mapply{\phi}{f(k)}}
        }}
        {\lsequent{}{\mapply{\phi}{\dbox{\pupdate{\umod{f(s)}{\theta}}}{f(k)}}}}
\end{sequentdeduction}
Note that the \keywordfont{if-then-else} case distinction is necessary in general, because the effect of the (vacuously quantified) assignment depends on whether $s=k$ holds, which may depend on what value $k$ has at the moment.
Rules \irref{assignrb+assignrd} reduce nondeterministic assignments to universal or existential quantification. %
For the handling of other general nondeterministic assignments and nondeterministic differential equations, also see previous work \cite{DBLP:journals/logcom/Platzer10}.

It is easy, just notationally cumbersome, to extend rules \irref{assignb+assignd+upskip} to vectorial extensions including systems of quantified assignments to multiple function symbols like \m{\pupdate{\lforall[C]{i}{(\pumod{a(i)}{a(i)+1}\syssep\pumod{t(i)}{0})}}} following the ideas of parallel updates \cite{DBLP:conf/cade/BeckertP06,DBLP:conf/lpar/Rummer06}.
With those, it is also easy to extend rules \irref{evolveb+evolved} to quantified differential equation systems like \m{\hevolve{\lforall[C]{i}{(\D{x(i)}=v(i)\syssep\D{v(i)}=a(i))}}} where the solution is a system of quantified assignments.

\paragraph{Object Creation}
Given our definition of $\pnew{C}$ as a \QHP from \rref{sec:objectcreation}, object creation can be proven by the other proof rules in \rref{calculus:QdL}.
With this definition of $\pnew{C}$, we obtain, for example, the following derived rule using \irref{composeb+assignrb+testb}
\[
      {\linferenceRule[sequent]
        {\lsequent[s]{}{\lforall[C]{\onew{}}{(\lnaexisting{\onew{}} \limply \dbox{\umod{\laexisting{\onew{}}}{\laetrue}}{\phi})}}}
        {\lsequent[s]{}{\dbox{\umod{\onew{}}{\pnew{C}}}{\phi}}}
      }{}
\]
In addition, axiom \irref{newex} expresses that, for sort $C\neq\reals$, there always is a new object $\onew{}$ that has not been created yet ($\lnaexisting{\onew{}}$), because domains are infinite.
This is the only place where we are using the assumption about infinite domains.
The primary purpose is to simplify technicalities that would arise if object creation could run out of objects and may thus fail if, e.g., no more cars can be created.
If this resource limitation is intended in a particular system, it can be modeled easily using patterns like \m{\pchoice{\pumod{\onew{}}{\pnew{C}}}{\textit{fail}}}.

\paragraph{Quantifiers}
For quantifiers, we cannot just use standard rules~\cite{Fitting96a}, because these are for uninterpreted first-order logic and work by instantiating quantifiers, eagerly as in ground tableaux or lazily by unification as in free variable tableaux~\cite{Fitting96a}.
\QdL is based on first-order logic interpreted over the reals~\cite{tarski_decisionalgebra51,DBLP:journals/jsc/CollinsH91}.
A formula like~\m{\lexists[\reals]{a}{\lforall[\reals]{x}{(x^2+a>0)}}} cannot be proven by the instantiation rules for the quantifiers but it is still valid for reals.
Thus, for handling quantifiers over the reals, we would like to use the standard decision procedure for first-order real arithmetic (i.e., real-closed fields) instead, which is quantifier elimination~\cite{tarski_decisionalgebra51,DBLP:journals/jsc/CollinsH91}.
\begin{definition}[Quantifier elimination] \label{def:qelim}
  A first-order theory admits
  \dfn[quantifier~elimination]{quantifier elimination} if, with each formula~$\phi$, a quantifier-free formula $\qelim{\phi}$\indexn[_QE]{\QE}\index{QE~(quantifier~elimination)@$\QE$~(quantifier~elimination)} can be associated effectively that is equivalent (i.e., $\phi\lbisubjunct\qelim{\phi}$ is valid) and has no additional free variables\index{symbol!free} or function symbols. The operation $\qelim{}$ is further assumed to evaluate formulas without variables,
  yielding a decision procedure for closed formulas\index{formula!closed}\index{closed!formula|see{formula, closed}} of this theory (i.e., formulas without free variables)\index{free~variable}.
\end{definition}
Unfortunately, we cannot use quantifier elimination of the theory of real-closed fields~\cite{tarski_decisionalgebra51,DBLP:journals/jsc/CollinsH91} either, because it cannot be applied to \QdL formulas with modalities, since these are quantified reachability statements.
Even in discrete dynamic logic, quantifiers plus modalities make validity $\Pi^1_1$-complete \cite[Theorem 13.1]{Harel_et_al_2000}.
\qelim{} cannot handle sorts \m{C\neq\reals}.

Instead, our \QdL proof rules combine quantifier handling of many-sorted logic based on instantiation with theory reasoning by \qelim{} for the theory of reals.
Figure~\ref{calculus:QdL} shows proof rules for quantifiers that combine with decision procedures for real-closed fields.
Classical instantiation is sound for sort $\reals$, just incomplete.
For example, rule \irref{existsr} can solve the following arithmetic by instantiation:
\begin{sequentdeduction}[array]
  \linfer[existsr]
    {\lsequent{a>0}{(a+1)^2>a,\lexists{x}{x^2>a}}}
    {\lsequent{a>0}{\lexists{x}{x^2>a}}}
\end{sequentdeduction}

Rules \irref{existsr} and \irref{alll} instantiate $x$ with arbitrary terms $\theta$, including a new free variable $X$, in which case \irref{existsr} and \irref{alll} become the usual $\gamma$-rules of free-variable proof calculi \cite{Fitting96a,Fitting_Mendelsohn_1999}:
\begin{center}
    \begin{calculus}[context=L]
      \dinferenceRule[existsrgamma|$\exists$\rightrule]{$\lexists{}{}$ right}
      {\linferenceRule[sequent]
        {\lsequent{}{\mapply{\phi}{X}, \lexists[C]{x}{\mapply{\phi}{x}}}}
        {\lsequent{}{\lexists[C]{x}{\mapply{\phi}{x}}}}
      }{}%
    \end{calculus}
    \hspace{\linferenceRulehskipamount}\qquad
    \begin{calculus}[context=L]
      \dinferenceRule[alllgamma|$\forall$\leftrule]{$\lforall{}{}$ left instantiation}
      {\linferenceRule[sequent]
        {\lsequent{\mapply{\phi}{X},\lforall[C]{x}{\mapply{\phi}{x}}}{}}
        {\lsequent{\lforall[C]{x}{\mapply{\phi}{x}}}{}}
      }{}%
\end{calculus}
\end{center}
Rules \irref{allr} and \irref{existsl} correspond to the liberalized $\delta^+$-rule~\cite{DBLP:journals/jar/HahnleS94} that is a refinement of the classical $\delta$-rule of free-variable tableaux \cite{Fitting96a}.
As in our previous work~\cite{DBLP:journals/jar/Platzer08}, rules \irref{iallr} and \irref{iexistsr} reintroduce and eliminate quantifiers over $\reals$ once \qelim{} is applicable, because the remaining constraints are first-order real arithmetical in the respective variables.
In particular, the quantifier rules can be used to postpone quantifier elimination until the remaining constraints are first-order, where the quantifier can be reintroduced by \irref{iallr} and \irref{iexistsr} \cite{DBLP:journals/jar/Platzer08}.

Unlike in previous work, however, functions and different argument vectors can occur in \QdL.
If the argument vectors $\vec{s}$ and $\vec{t}$ in \irref{iallr} have the same value, the same variable $X$ can be reintroduced for $f(\vec{s})$ and $f(\vec{t})$, otherwise different variables $X\neq Y$ have to be used.
Whether $\vec{s}$ and $\vec{t}$ have the same value cannot always be decided statically, so rule \irref{iallr} makes a case distinction by an \keywordfont{if-then-else}.
Rule \irref{iallr} works accordingly for multiple occurrences of $f(\vec{s}), f(\vec{t}), f(\vec{u})$ and so on in arbitrary positions in the formula, where more variables $X,Y,Z$ are introduced to quantify over.
It is easy to turn rule \irref{iallr} into a rule that successively substitutes one term $f(\vec{s})$ by a fresh variable $X$ everywhere at a time instead of handling all $f(\vec{s}),f(\vec{t}),f(\vec{u})$ at once.

Rule \irref{iexistsr} can reintroduce an existential quantifier for a free (existential) logical variable $X$ and merges all proof branches containing $X$, because $X$ has to satisfy all branches simultaneously.
It thus has multiple conclusions.
Rule \irref{iexistsr} reintroduces an existential quantifier and performs quantifier elimination for a free (existential) logical variable $X$ that has been introduced by \irref{existsr+alll} before by choosing a fresh variable $X$ for $\theta$.
We use the same rule \irref{iexistsr} as in previous work and refer to that work \cite{DBLP:journals/jar/Platzer08} for further explanations of merging.

Rules \irref{iallr} and \irref{iexistsr} require that quantifier elimination (\qelim{}) is applicable to the resulting formula.
If the resulting formulas still have occurrences of the quantified variables in the scope of modalities, then \qelim{} is not applicable and rules \irref{iallr} and \irref{iexistsr} have to be postponed until the modalities have been dealt with by other proof rules from \rref{calculus:QdL}.
Even for first-order formulas, we cannot just apply classical quantifier elimination in real-closed fields \cite{tarski_decisionalgebra51}, because the first-order theory of real-closed fields does not include function symbols.
For example, \m{\lforall{i}{(a(i)\geq0)}} is a formula of first-order real arithmetic \emph{augmented with function symbols}, hence quantifier elimination in real-closed fields due to Tarski \cite{tarski_decisionalgebra51} is not applicable.
It cannot even be expressed in quantifier-free form, because its truth-value depends on the value of function $a$ at unboundedly many positions.
This makes sense. 
\qelim{} is a decision procedure for first-order real arithmetic.
But first-order logic (even without arithmetic) is only semidecidable, so we cannot handle it by \qelim{} and need to rely on the instantiation rules \irref{allr+alll+existsr+existsl}, which are complete for first-order logic.
Nevertheless, from previous work \cite{DBLP:journals/jar/Platzer08}, we obtain the following result on how to lift \qelim{} to the presence of function symbols:
\begin{lemma}[Quantifier elimination lifting \cite{DBLP:journals/jar/Platzer08}] \label{lem:qelim-lift}
  Quantifier elimination can be lifted to instances of formulas of first-order theories that admit quantifier elimination,
  i.e., to formulas that result from the base theory by substitution.
  \index{quantifier~elimination!lifting}
  \index{lifting|see{quantifier~elimination, lifting}}
\end{lemma}
For example, \m{\lforall{y}{(a(i)<y^2)}} is a formula of first-order real arithmetic augmented with function symbols such that quantifier elimination in real-closed fields due to Tarski \cite{tarski_decisionalgebra51} is not (directly) applicable.
By \rref{lem:qelim-lift}, however, \qelim{} can be lifted to this formula, because it is an instance of \m{\lforall{y}{(Z<y^2)}}, for $Z$ replaced with $a(i)$.
Hence,
\[\qelim{\lforall{y}{(a(i)<y^2)}} \mequiv \subst[(\qelim{\lforall{y}{(Z<y^2)}})]{Z}{a(i)} \mequiv \subst[(Z<0)]{Z}{a(i)} \mequiv a(i)<0\]

\paragraph{Global Rules}
The proof rules in the last block of \rref{calculus:QdL} depend on the truth of their premises in all states,
thus the context $\Gamma,\Delta$ cannot be used in the premise, because it may be specific to the current state.
The rules are given in a form that best displays their underlying logical principles.
The general pattern for applying these rules to prove that the succedent of their conclusion holds is to prove that both their premise and the antecedent of their conclusion hold.
In particular, the antecedent can be thought of as holding in the current state, whereas the premise can be thought of as holding in all states because the context $\Gamma,\Delta$ is gone.

Rules \irref{genb+gend} are G\"odel generalization rules and can be used to strengthen postconditions: antecedent~$\dbox{\alpha}{\phi}$ is sufficient for proving succedent~$\dbox{\alpha}{\psi}$ when postcondition~$\phi$ entails~$\psi$ in all states, as shown in the premise of \irref{genb}.
Clearly, for rule \irref{genb}, if all states reachable by~$\alpha$ satisfy~$\phi$ (antecedent \m{\dbox{\alpha}{\phi}}) and~$\phi$ implies~$\psi$ in all states (premise \m{\lsequent{\phi}{\psi}}), then~$\psi$ also holds in all states reachable by~$\alpha$ (succedent \m{\dbox{\alpha}{\psi}}).
Similarly, for rule \irref{gend}, if some state reachable by~$\alpha$ satisfies~$\phi$ (antecedent \m{\ddiamond{\alpha}{\phi}}) and~$\phi$ implies~$\psi$ in all states (premise \m{\lsequent{\phi}{\psi}}), then~$\psi$ also holds in some state reachable by~$\alpha$ (succedent \m{\ddiamond{\alpha}{\psi}}).

Rule \irref{invind} is an induction schema for loops with \dfn[invariant!inductive]{inductive invariant}~$\inv$ \cite{Harel_et_al_2000,DBLP:journals/jar/Platzer08}.
Rule \irref{invind} says that~$\inv$ holds after any number of repetitions of~$\alpha$ if it holds initially (antecedent) and, for all states, invariant~$\inv$ remains true after one iteration of~$\alpha$ (premise).
If~$\inv$ is true after executing~$\alpha$ whenever~$\inv$ has been true before (premise), then, if~$\inv$ holds in the beginning,~$\inv$ will continue to hold, no matter how often we repeat~$\alpha$ in \m{\dbox{\prepeat{\alpha}}{\inv}}.

Similarly, \irref{con} generalizes Harel's convergence rule~\cite{Harel_et_al_2000} to the hybrid case with decreasing \dfn{variant}~$\varphi$ \cite{DBLP:journals/jar/Platzer08}.
Rule \irref{con} expresses that the variant~$\var(v)$ holds for some real number~\m{v\leq0} after repeating~$\alpha$ sufficiently often (succedent) if~$\var(v)$ holds for some real number at all in the beginning (antecedent) and, by premise,~$\var(v)$ can decrease after every execution of~$\alpha$ by~1 (or another positive real constant).
This rule can be used to show positive progress (by~1) with respect to~$\var(v)$ by executing~$\alpha$.

\paragraph{Example}
{\def\prem{\lforall{i{\neq}j}{x(i){\neq}x(j)}}%
As a simple example illustrating how the \QdL proof calculus works, we consider the \QdL derivation in \rref{fig:verification-example} for a simple \QdL formula.
The \QdL formula that we consider here follows the pattern of the running example formula in \rref{eq:distributed-car-control-new}.
But we simplify the formula to consider just one case and postpone the discussion of the full system to \rref{sec:distributed-car-control-verification}.
Here we consider the \QdL formula:
\begin{equation}
  {\prem} \limply {\dbox{\hevolve{\lforall{i}{(\D{x(i)}=v(i)\syssep\D{v(i)}=-b)}}}{\,\lforall{j{\neq}k}{x(j){\neq}x(k)}}}
  \label{eq:distributed-car-control-discovery}
\end{equation}
The difference of the simpler \QdL formula \rref{eq:distributed-car-control-discovery} compared to the full \QdL formula \rref{eq:distributed-car-control-new} is that the simpler formula considers only the case of the \QHP dynamics where all cars are braking.
Certainly, if the system would not be safe when all cars are braking (which is one possible behavior of \DCCS), then it would not be safe always.
The other difference is that \rref{eq:distributed-car-control-discovery} has a weaker assumption in the precondition.
It only assumes that cars start from different positions (\m{\prem}), not that they respect the compatibility constraint \m{\let\laforall\lforall \dcinv}.
In fact, we are using the derivation in \rref{fig:verification-example} to find out how we need to choose \m{\dcseparate{i}{j}} to ensure collision freedom, because \m{\dcseparate{i}{j}} needs to imply at least that all cars would remain safe when braking.

The derivation in \rref{fig:verification-example} can be used to find out  under which circumstances the \QdL formula \rref{eq:distributed-car-control-discovery}, from which we start the derivation at the bottom of \rref{fig:verification-example}, is true.
Formula \rref{eq:distributed-car-control-discovery} claims that cars would never crash if they start at different positions (\m{\prem}) and all cars brake by following the dynamics \m{\hevolve{\lforall{i}{\D[2]{x(i)}=-b}}}.
Since braking is the safest operation for cars, we might think that car control would always be safe in this most conservative scenario.
But that is not the case.
If the cars start with incompatible velocities and distances, then not even braking can prevent a crash.
The premise discovered by the \QdL derivation in \rref{fig:verification-example} reveals that collisions will only be prevented by braking if the initial velocities and positions satisfy the monotonicity condition \m{\dcseparate{j}{k}} that we have already shown in \rref{eq:distributed-car-control-separate}.

\begin{figure}[tbh]
     \def\arraystretch{1.3}%
      \begin{sequentdeduction}[array]
          \linfer[evolveb]
            {\linfer[allr]
              {\linfer[implyr]
                {\linfer[upskip]
                {\linfer[assignb]
                  {\linfer[allr]
                    {\linfer[iallr]
                      {\linfer[qelim]
                        {\linfer[iallr]
                          {\lsequent{} {\qelim{\lforall{X,Y,V,W}{(\skolem{j}\neq\skolem{k} \land X\neq Y \limply X{\leq}Y{\land}V{\leq}W \lor X{\geq}Y{\land}V{\geq}W)}}}}
                          {\lsequent{\prem} {(\skolem{j}{\neq}\skolem{k}\limply 
                            {(x(\skolem{j}){\leq}x(\skolem{k}){\land}v(\skolem{j}){\leq}v(\skolem{k}) \lor x(\skolem{j}){\geq}x(\skolem{k}){\land}v(\skolem{j}){\geq}v(\skolem{k}))})
                          }}
                        }%
                        {\lsequent{\prem} {\qelim{\lforall{\skolem{s}{\geq}0}{(\skolem{j}{\neq}\skolem{k} \limply
                          {{-}{\frac{b}{2}}\skolem{s}^2+v(\skolem{j})\skolem{s} + x(\skolem{j}) \neq {-}{\frac{b}{2}}\skolem{s}^2+v(\skolem{k})\skolem{s} + x(\skolem{k})})}}
                        }}
                      }%
                      {\lsequent{\prem, \skolem{s}{\geq}0} {(\skolem{j}{\neq}\skolem{k} \limply
                        {{-}{\frac{b}{2}}\skolem{s}^2+v(\skolem{j})\skolem{s} + x(\skolem{j}) \neq {-}{\frac{b}{2}}\skolem{s}^2+v(\skolem{k})\skolem{s} + x(\skolem{k})})
                      }}
                    }%
                   {\lsequent{\prem, \skolem{s}{\geq}0} {\lforall{j{\neq}k}{({-}{\frac{b}{2}}\skolem{s}^2+v(j)\skolem{s} + x(j) \neq {-}{\frac{b}{2}}\skolem{s}^2+v(k)\skolem{s} + x(k))}}}
                 }%
               {\lsequent{\prem, \skolem{s}{\geq}0} {\lforall{j{\neq}k}{\dbox{\hupdate{\lforall{i}{\humod{x(i)}{{-}{\frac{b}{2}}\skolem{s}^2+v(i)\skolem{s} + x(i)}}}}{\,x(j){\neq}x(k)}}}}
             }%
               {\lsequent{\prem, \skolem{s}{\geq}0} {\dbox{\hupdate{\lforall{i}{\humod{x(i)}{{-}{\frac{b}{2}}\skolem{s}^2+v(i)\skolem{s} + x(i)}}}}{\,\lforall{j{\neq}k}{x(j){\neq}x(k)}}}}
             }%
            {\lsequent{\prem} {(\skolem{s}{\geq}0 \limply \dbox{\hupdate{\lforall{i}{\humod{x(i)}{{-}{\frac{b}{2}}\skolem{s}^2+v(i)\skolem{s} + x(i)}}}}{\,\lforall{j{\neq}k}{x(j){\neq}x(k)}})}}
           }%
           {\lsequent{\prem}  {\lforall{t{\geq}0}{\dbox{\hupdate{\lforall{i}{\humod{x(i)}{{-}{\frac{b}{2}}t^2+v(i)t + x(i)}}}}{\,\lforall{j{\neq}k}{x(j){\neq}x(k)}}}}}
         }%
         {\lsequent{\prem} {\dbox{\hevolve{\lforall{i}{(\D{x(i)}=v(i)\syssep\D{v(i)}=-b)}}}{\,\lforall{j{\neq}k}{x(j){\neq}x(k)}}}}
      \end{sequentdeduction}
  \caption{Example of a \QdL derivation to prove collision-freedom of simple car control.}
  \label{fig:verification-example}
\end{figure}
The proof in \rref{fig:verification-example} starts with the conjecture at the bottom (goal).
The proof uses rule \irref{evolveb} to turn the quantified differential equation system into a quantified assignment with an extra quantifier for the duration $t$ of the evolution.
The quantified differential equation system is easy to solve.
The quantified assignment \m{\hupdate{\lforall{i}{\humod{x(i)}{{-}{\frac{b}{2}}t^2+v(i)t + x(i)}}}} solving it can be obtained easily from the solution \m{\hupdate{\humod{x}{{-}{\frac{b}{2}}t^2+vt + x}}} of the deparametrized differential equation system \m{\hevolve{\D{x}=v\syssep\D{v}=-b}}, just by adding the parameter $i$ back in and checking that the resulting terms solve the quantified differential equation.
Now the top-most logical operator in the succedent is the quantifier \m{\forall{t}}.
Even though it is a quantifier over a real variable, we cannot use the decision procedure of quantifier elimination for real-closed fields \cite{tarski_decisionalgebra51} to handle it, because we do not have a formula of first-order real arithmetic, but still a \QdL formula with a modality expressing a property of all reachable states.
Instead, we use rule \irref{allr} to postpone quantifier elimination and turn variable $t$ into a Skolem function $\skolem{s}$.
This Skolem function has no arguments, because no free (existential) logical variables occur in the formula \cite{DBLP:journals/jar/Platzer08}.
After that, we use the standard propositional sequent rule \irref{implyr} to normalize implications in the succedent into sequent form by moving their left-hand side to the antecedent.

The resulting quantified assignment to $x(i)$ (for all $i$) takes effect on the postcondition \m{\lforall{j{\neq}k}{x(j)\neq x(k)}} by skipping over the quantifier \m{\forall{j{\neq}k}} with rule \irref{upskip} and then affecting $x(j)$ and $x(k)$ subsequently by rule \irref{assignb} (and another application of \irref{upskip} to skip over $\neq$, which is not shown in \rref{fig:verification-example}).

At this point (the top-most use of rule \irref{allr} in \rref{fig:verification-example}), we already have a first-order formula and it may seem as if we could apply \irref{iallr} directly instead of \irref{allr}.
This would not work, however, because quantifier elimination works from inside out and will have to eliminate the inner quantifier \m{\forall{j{\neq}k}} before the outer quantifier \m{\forall{\skolem{s}}}.
Yet, the resulting formula is not an instance of first-order real arithmetic (not even when using \rref{lem:qelim-lift}), because there are dependencies on the quantified variables $j,k$ in function arguments of the resulting formula:
\[
\lforall{\skolem{s}{\geq}0}{\lforall{j{\neq}k}
                        {\left({-}{\frac{b}{2}}\skolem{s}^2+v(j)\skolem{s} + x(j) \neq {-}{\frac{b}{2}}\skolem{s}^2+v(k)\skolem{s} + x(k)\right)}}
\]
Instead, the proof in \rref{fig:verification-example} uses rule \irref{allr} to turn the quantified variables $j,k$ into Skolem functions, which, for simplicity, we again denote by $\skolem{j}$ and $\skolem{k}$.
Subsequently, we can use rule \irref{iallr} to reintroduce a quantifier for the Skolem function $\skolem{s}$.
Rule \irref{iallr} does not produce an if-then-else, because $\skolem{s}$ has no arguments.
This time, the formula is still not in first-order real arithmetic, because function symbols like $v(j)$ occur.
However, it is an instance ($v(\skolem{j})$ for $V$ and $x(\skolem{j})$ for $X$ and $v(\skolem{k})$ for $W$ and $x(\skolem{k})$ for $Y$) of the following formula of first-order real arithmetic:
\begin{equation}
\lforall{\skolem{s}{\geq}0}{\left(\skolem{j}{\neq}\skolem{k} \limply
                          {{-}{\frac{b}{2}}\skolem{s}^2+V\skolem{s} + X \neq {-}{\frac{b}{2}}\skolem{s}^2+W\skolem{s} + Y}\right)}
\label{eq:qelim-lift-ex}
\end{equation}
and thus quantifier elimination can be lifted by \rref{lem:qelim-lift}.
The result of quantifier elimination is an instance (with the same instantiation as above) of the result of applying \qelim{} to \rref{eq:qelim-lift-ex}.
To improve traceability, we show the application of \qelim{} as a separate proof step (indicated by \irref{qelim}).

Finally (the top-most rule), we use rule \irref{iallr} to finish the deduction.
We still cannot yet use rule \irref{iallr} for $\skolem{j},\skolem{k}$, but we can use rule \irref{iallr} for the (non-Skolem) function symbols $x$ and $v$.
This time, the use of rule \irref{iallr} is more involved than before, because the functions $x$ and $v$ have arguments.
When using rule \irref{iallr} on
\[
\lsequent{\prem} {(\skolem{j}{\neq}\skolem{k}\limply 
                            {(x(\skolem{j}){\leq}x(\skolem{k}){\land}v(\skolem{j}){\leq}v(\skolem{k}) \lor x(\skolem{j}){\geq}x(\skolem{k}){\land}v(\skolem{j}){\geq}v(\skolem{k}))})}
\]
we formally obtain
\begin{align*}
\qelim{}\big(\lforall{X,Y,V,W}{\big(}
\keywordfont{if}\,j=k\,\keywordfont{then}\,&\\&
{\skolem{j}\neq\skolem{k} \land X\neq X \limply X{\leq}X{\land}V{\leq}V \lor X{\geq}X{\land}V{\geq}V}
\\\keywordfont{else}\,\\&
{\skolem{j}\neq\skolem{k} \land X\neq Y \limply X{\leq}Y{\land}V{\leq}W \lor X{\geq}Y{\land}V{\geq}W}
\big)\big)
\end{align*}
Since the condition \m{\keywordfont{if}\,j=k} contradicts the assumption \m{j\neq k}, this formula simplifies to:
\[
\lsequent{} {\qelim{\lforall{X,Y,V,W}{(\skolem{j}\neq\skolem{k} \land X\neq Y \limply X{\leq}Y{\land}V{\leq}W \lor X{\geq}Y{\land}V{\geq}W)}}}
\]
Simplifications like those arise often and can be exploited for automated theorem proving.
Applying \qelim{} in the above formula yields $\lfalse$, so the derivation in \rref{fig:verification-example} does not result in a closed proof.
This is good news, however, because the conjecture at the bottom of \rref{fig:verification-example} is not true under all interpretations.
The constraints at the top of \rref{fig:verification-example} can be used to construct the constraints required for safety, which coincide with \m{\dcseparate{j}{k}} from \rref{eq:distributed-car-control-separate}.
}

\paragraph{Derived Rules}
Several useful rules can be derived from the \QdL rules in \rref{calculus:QdL} to shortcut common reasoning cases.
For instance, the following derived rules characterize the effect of creating objects of type $C$ on actualist quantifiers over type $\laetype{C}$ (where {$\onew{}$ is of type $C$}):
\begin{center}
  \let\dmodality\dbox
  \tabcolsep=1pt%
  \newdimen\linferenceRulehskipamount%
  \linferenceRulehskipamount=5mm%
    \begin{calculus}
      \cinferenceRule[upnewall|$\nu\forall$]{$\pnew{C}$ update on $\forall$ quantifier}
      {\linferenceRule[sequent]
        {\lsequent[s]{}{\daexisting{\onew{}}{\mapply{\phi}{\onew{}}}
            \land \laforall[C]{i}{\daexisting{\onew{}}{\mapply{\phi}{i}}}}}
        {\lsequent[s]{}{\daexisting{\onew{}}{\laforall[C]{i}{\mapply{\phi}{i}}}}}
      }{}%
    \end{calculus}
    \hspace{\linferenceRulehskipamount}
    \begin{calculus}
      \cinferenceRule[upnewexists|$\nu\exists$]{$\pnew{C}$ update on $\exists$ quantifier}
      {\linferenceRule[sequent]
        {\lsequent[s]{}{\daexisting{\onew{}}{\mapply{\phi}{\onew{}}}
            \lor \laexists[C]{i}{\daexisting{\onew{}}{\mapply{\phi}{i}}}}}
        {\lsequent[s]{}{\daexisting{\onew{}}{\laexists[C]{i}{\mapply{\phi}{i}}}}}
      }{}%
    \end{calculus}
\end{center}
They commute the effect $\let\dmodality\dbox \daexisting{\onew{}}{}$ of object creation with quantification, retaining the effect on the new object explicitly.
Rule \irref{upnewall} states that the new object denoted by $\onew{}$---which may not have been created before---needs to satisfy $\mapply{\phi}{\onew{}}$ too in order for \m{\laforall[C]{i}{\mapply{\phi}{i}}} to hold after \m{\paexisting{\onew{}}} ensures that $\onew{}$ is created.
Dually, rule \irref{upnewexists} states that created object $\onew{}$ is an alternative choice for $i$, in addition to the previous domain of $\laetype{C}$.

A similar derived rule \irref{upnewu} states that, after creating an object of type $C$, this created object will be affected by actualist quantified assignments ranging over $\laetype{C}$, so that commuting has to take care of the effect on the new object explicitly.
\begin{center}
  \let\dmodality\dbox
  \tabcolsep=1pt%
  \newdimen\linferenceRulehskipamount%
  \linferenceRulehskipamount=5mm%
    \begin{calculus}
      \cinferenceRule[upnewu|$\nu A$]{$\pnew{C}$ update on quantified update}
      {\linferenceRule[sequent]
        {\lsequent[s]{}{\dmodality{\pupdate{\laforallplus[C]{\onew{}}{i}{\umod{f(\vec{s})}{\theta}}}}{\daexisting{\onew{}}{\phi}}}}
        {\lsequent[s]{}{\daexisting{\onew{}}{\dmodality{\pupdate{\laforall[C]{i}{\umod{f(\vec{s})}{\theta}}}}{\phi}}}}
      }{}%
  \end{calculus}
\end{center}
For this situation where $\onew{}$ is adjoined to the range of quantification ($\onew{}$ might even have been in the range before, so the union is not necessarily disjoint), we use the following mnemonic abbreviation in the premise of \irref{upnewu}:
\begin{align*}
  \pupdate{\laforallplus[C]{\onew{}}{i}{\umod{f(\vec{s})}{\theta}}}
  &~\mequiv~
  \pupdate{\lforall[C]{i}{(\umod{f(\vec{s})}{\piif{i=\onew{}\lor\laexisting{i}}{\theta}{f(\vec{s})}})}}
\end{align*}
Note that we cannot simply apply the assignment to $\onew{}$ separately before \m{\pupdate{\laforall[C]{i}{\umod{f(\vec{s})}{\theta}}}} as in \m{\pupdate{\pumod{i}{\onew{}}};~\pupdate{\umod{f(\vec{s})}{\theta}};~\pupdate{\laforall[C]{i}{\umod{f(\vec{s})}{\theta}}}}, because that would change $f$ twice if $\onew{}$ already existed initially.

\section{Soundness} \label{sec:Sound}

We have presented a proof calculus for \QdL in \rref{sec:QdL-calculus}.
One of the most important questions about it is whether we can rely on the proofs and know that every \QdL formula proven in the \QdL calculus is really a valid formula.
That is, the question is whether the \QdL calculus is sound.
An unsound calculus would be disastrous, because we could use it to ``prove'' counterfactual properties.
We need to make sure that the proof calculus fits to the semantics of \QdL.
Indeed it does.
\begin{theorem}[Soundness] \label{thm:QdL-sound}
  The \QdL calculus is \emph{sound}: every \QdL formula that can be proven in the \QdL calculus is valid, i.e., true in all states.
\end{theorem}
\proof
The calculus is sound if each rule instance is sound.
Some of the rules of the \QdL calculus\ignore{except \irref{allr},\irref{existsl} and \irref{iexistsr} and \irref{genb+gend+invind+con}} are even \dfn[sound!locally]{locally sound}, i.e., their conclusion is true at state~$\iportray{\I}$ if all its premises are true at~$\iportray{\I}$, which implies soundness.
The proofs for the propositional rules, and regular rules \irref{composeb+composed+choiceb+choiced+testb+testd} are as usual \cite{Platzer10}.
We refer to previous work \cite{DBLP:journals/jar/Platzer08,Platzer10} for the soundness proofs for \irref{existsr+alll+allr+existsl+iexistsr}, which are more involved.
\begin{desCription}
 \item\noindent{\hskip-12 pt\irref{iallr}\ }\ Rule \irref{iallr} is locally sound.
  For this, we assume that the premise holds, i.e., we assume
  \m{\imodels{\I}{\qelim{\lforall{X,Y}{(\piif{\vec{s}=\vec{t}}{\lsequent[f]{\mapply{\Phi}{X}}{\mapply{\Psi}{X}}}{\lsequent[f]{\mapply{\Phi}{X}}{\mapply{\Psi}{Y}}})}}}}.
  Since \qelim{} yields an equivalence, we conclude
  \m{\imodels{\I}{\lforall{X,Y}{(\piif{\vec{s}=\vec{t}}{\lsequent[f]{\mapply{\Phi}{X}}{\mapply{\Psi}{X}}}{\lsequent[f]{\mapply{\Phi}{X}}{\mapply{\Psi}{Y}}})}}}.
  This is equivalent to \m{\imodels{\I}{\piif{\vec{s}=\vec{t}}{\lforall{X}{(\lsequent[f]{\mapply{\Phi}{X}}{\mapply{\Psi}{X}})}}{\lforall{X,Y}{(\lsequent[f]{\mapply{\Phi}{X}}{\mapply{\Psi}{Y}})}}}}, because the fresh variables $X,Y$ do not occur in $\vec{s}$ or $\vec{t}$.
  Then we assume the antecedent of the conclusion is true, i.e., \m{\imodels{\I}{\mapply{\Phi}{f(\vec{s})}}}.
  We conclude that the succedent of the conclusion is true, \m{\imodels{\I}{\mapply{\Psi}{f(\vec{t})}}},
  by choosing \m{\ivaluation{\I}{f(\vec{s})}} for~$X$ and \m{\ivaluation{\I}{f(\vec{t})}} for~$Y$ in the premise.
  If \m{\imodels{\I}{\lnot(\vec{s}=\vec{t})}} then \m{\imodels{\I}{\mapply{\Psi}{f(\vec{t})}}} follows directly from the premise.
  If, otherwise, \m{\imodels{\I}{\vec{s}=\vec{t}}}, then \m{\imodels{\I}{\mapply{\Psi}{f(\vec{t})}}} also follows, because the choice \m{\ivaluation{\I}{f(\vec{s})}} for~$X$ is identical to the choice \m{\ivaluation{\I}{f(\vec{t})}} for $Y$ in the premise.
  By admissibility of substitutions, any variables occurring in terms $\vec{s}$ and $\vec{t}$ are free at all occurrences of \m{f(\vec{s})} and \m{f(\vec{t})}, hence their value is the same in all occurrences.

 \item\noindent{\hskip-12 pt\irref{assignd}\ }\ 
 {\newcommand{\Ie}{\imodif[state]{\I}{i}{e}}%
  \newcommand{\Ied}{\imodif[state]{\Ie}{z}{d}}%
  \newcommand{\Id}{\imodif[state]{\I}{z}{d}}%
  Rule \irref{assignd} is locally sound for injective
  \m{\pupdate{\lforall[C]{i}{\umod{f(\vec{s})}{\theta}}}}, which we abbreviate as $\jupd$.
  Injective $\jupd$ give a deterministic transition.
  We assume that the premise holds
  \m{\imodels{\I}{\piif{\lexists[C]{i}{\vec{s}=\ddiamond{\jupd}{\vec{u}}}}
            {\lexists[C]{i}{(\vec{s}=\ddiamond{\jupd}{\vec{u}} \land \mapply{\phi}{\theta})}}
            {\mapply{\phi}{f(\ddiamond{\jupd}{\vec{u}})}}}}.
  We now show that
  \m{\imodels{\I}{\mapply{\phi}{\ddiamond{\pupdate{\lforall[C]{i}{\umod{f(\vec{s})}{\theta}}}}{f(\vec{u})}}}}.
  First assume that, with a fresh variable $z$, \m{\mapply{\phi}{z}} is a first-order formula without modalities or quantifiers.
  Let $\iget[state]{\It}$ be the (unique) state with
  \m{\relateds{\iaccess[{\pupdate{\lforall[C]{i}{\umod{f(\vec{s})}{\theta}}}}]{\I}}{\iget[state]{\I}}{\iget[state]{\It}} = \iaccess[\jupd]{\I}}.
  By renaming, we can assume the quantified variable $i$ not to occur anywhere else than in $\jupd$.
  We write this occurrence constraint as \m{i\not\in\vec{u}} and \m{i\not\in\mapply{\phi}{z}}.
  \begin{iteMize}{$\bullet$}
  \item Suppose \m{\imodels{\I}{\lexists[C]{i}{\vec{s}=\ddiamond{\jupd}{\vec{u}}}}}, then
   \m{\imodels{\I}{\lexists[C]{i}{(\vec{s}=\ddiamond{\jupd}{\vec{u}}\land\mapply{\phi}{\theta})}}} by premise.
   That is equivalent to: there is an \m{e\in\idomain{\I}{C}} with
   \m{\imodels{\Ie}{\vec{s}=\ddiamond{\jupd}{\vec{u}}\land\mapply{\phi}{\theta}}}.
   That means
   \m{\imodels{\Ied}{\mapply{\phi}{z}}}
   for \m{d \eqdef \ivaluation{\Ie}{\theta}}
   by the substitution lemma.
   This is equivalent to
   \m{\imodels{\Id}{\mapply{\phi}{z}}}, because \m{i\not\in\mapply{\phi}{z}}, i.e., $i$ does not occur in \m{\mapply{\phi}{z}}, so that its value is irrelevant.
   We want to show that
   \m{\imodels{\Id}{\mapply{\phi}{z}}} also holds for \m{d=\ivaluation{\I}{\ddiamond{\jupd}{f(\vec{u})}}}, because this implies
   \m{\imodels{\I}{\mapply{\phi}{\ddiamond{\jupd}{f(\vec{u})}}}}
   by the substitution lemma.
   Now
   \[
   \ivaluation{\I}{\ddiamond{\jupd}{f(\vec{u})}}
   =
   \ivaluation{\It}{f(\vec{u})}
   =
   \iget[state]{\It}(f)\big(\ivaluation{\It}{\vec{u}}\big)
   =
   \iget[state]{\It}(f)\big(\ivaluation{\I}{\ddiamond{\jupd}{\vec{u}}}\big)
   \stackrel{*}{=}
   \iget[state]{\It}(f)\big(\ivaluation{\Ie}{\vec{s}}\big)
   \stackrel{\iaccess[\jupd]{\I}}{=}
   \ivaluation{\Ie}{\theta}
   =
   d
   \]
   Thus
   \m{\imodels{\I}{\mapply{\phi}{\ddiamond{\jupd}{f(\vec{u})}}}}.
   The equality marked $*$ holds, because the premise
   implies
   \m{\imodels{\Ie}{\vec{s}=\ddiamond{\jupd}{\vec{u}}}},
   which yields
   \[
   \ivaluation{\Ie}{\vec{s}}
   =
   \ivaluation{\Ie}{\ddiamond{\jupd}{\vec{u}}}
   \stackrel{i\not\in\vec{u}}{=}
   \ivaluation{\I}{\ddiamond{\jupd}{\vec{u}}}
   \]
   
  \item Suppose \m{\imodels{\I}{\lnot\lexists[C]{i}{\vec{s}=\ddiamond{\jupd}{\vec{u}}}}}, then
   \m{\imodels{\I}{\mapply{\phi}{f(\ddiamond{\jupd}{\vec{u}})}}}
   by the premise.
   Consequently
   \m{\imodels{\Id}{\mapply{\phi}{z}}}
   for \m{d \eqdef \ivaluation{\I}{f(\ddiamond{\jupd}{\vec{u}})}}
   by the substitution lemma.
   We show that
   \m{\imodels{\Id}{\mapply{\phi}{z}}} also holds for \m{d=\ivaluation{\I}{\ddiamond{\jupd}{f(\vec{u})}}}, because this implies
   \m{\imodels{\I}{\mapply{\phi}{\ddiamond{\jupd}{f(\vec{u})}}}}
   by the substitution lemma.
   This time we have
   \[
   \ivaluation{\I}{\ddiamond{\jupd}{f(\vec{u})}}
   =
   \ivaluation{\It}{f(\vec{u})}
   =
   \iget[state]{\It}(f)\big(\ivaluation{\It}{\vec{u}}\big)
   \stackrel{*}{=}
   \iget[state]{\I}(f)\big(\ivaluation{\It}{\vec{u}}\big)
   =
   \iget[state]{\I}(f)\big(\ivaluation{\I}{\ddiamond{\jupd}{\vec{u}}}\big)
   =
   \ivaluation{\I}{f(\ddiamond{\jupd}{\vec{u}})}
   =
   d
   \]
   The equality marked $*$ holds, because---by assumption
   \m{\imodels{\I}{\lnot\lexists[C]{i}{\vec{s}=\ddiamond{\jupd}{\vec{u}}}}}---%
   we know that for position
   \m{\ivaluation{\It}{\vec{u}} = \ivaluation{\I}{\ddiamond{\jupd}{\vec{u}}}}
   there is no \m{e\in\idomain{\I}{C}}
   such that
   \[
   \ivaluation{\Ie}{\vec{s}}
   =
   \ivaluation{\It}{\vec{u}} = \ivaluation{\I}{\ddiamond{\jupd}{\vec{u}}}
   \stackrel{i\not\in\vec{u}}{=}
   \ivaluation{\Ie}{\ddiamond{\jupd}{\vec{u}}}
   \]
   Thus $\jupd$ has no effect on the interpretation of $f$ at position \m{\ivaluation{\It}{\vec{u}}} and $\iget[state]{\I}$ and $\iget[state]{\It}$ agree at that position.
  \end{iteMize}
  In both cases, equivalence of premise and conclusion can be established by following the equations and equivalences backwards, which also gives a proof for the dual rule \irref{assignb}.
  For the case where \m{\mapply{\phi}{z}} contains modalities or quantifiers, the proof is accordingly using the substitution lemma and the fact that the interpretation of the symbols occurring in \m{\ddiamond{\jupd}{f(\vec{u})}}
  is not affected by the modalities and quantifiers in \m{\mapply{\phi}{z}}
  (since all substitutions need to be admissible for \QdL rules to be applicable).
  }%

 \item\noindent{\hskip-12 pt\irref{upskip}\ }\ Local soundness of rule \irref{upskip} for injective quantified assignments \m{\pupdate{\lforall[C]{i}{\umod{f(\vec{s})}{\theta}}}} is a simple consequence of the fact that a quantified assignment to $f$ cannot affect the evaluation of another operator $\mascriptor\neq f$, but only its arguments (assuming admissible substitutions).
 
 \item\noindent{\hskip-12 pt\irref{newex}\ }\
 {\newcommand{\Id}{\imodif[state]{\I}{i}{e}}%
  The soundness of axiom \irref{newex} (i.e., validity of the conclusion) is a simple consequence of the fact that we have assumed finite support for the createdness flag $\laexisting{\cdot}$ and that domains are infinite.
  That is, there are only finitely many \m{e\in\idomain{\I}{C}} with \m{\imodels{\Id}{\laexisting{i}=1}}, yet domain \m{\idomain{\I}{C}} is infinite.
  Consequently, in every state $\iget[state]{\I}$, there always is a choice $e$ for $i$ that has not been created yet (\m{\imodels{\Id}{\laexisting{i}\neq1}}).
 }%

 \item\noindent{\hskip-12 pt\irref{evolved}\ }\
  {\newcommand{\Ir}[1][r]{\imodif[state]{\I}{t}{#1}}%
  \newcommand{\Ifz}{\iconcat[state=\varphi(\zeta)]{\Ir}}%
  Rule \irref{evolved} is locally sound.
  Let~\m{\solutionfor[\vec{s}]{}(t)} be simultaneous solutions for the
  respective differential
  equations\ignore{\m{\hevolve{\lforall[C]{i}{\D{f(\vec{s})}=\theta}}}}
  with symbolic initial values~\m{f(\vec{s})} and
  let~\m{\ddiamond{\pupdate{\lforall[C]{i}{\solutionupdate{t}}}}{}} denote the quantified assignment
  \m{\ddiamond{\pupdate{\lforall[C]{i}{\umod{f(\vec{s})}{\solutionfor[\vec{s}]{}(t)}}}}{}}.
  Assume~$\iportray{\I}$ satisfies the premise:
  \m{\imodels{\I}{\lexists{t{\geq}0}{(\bar\ivr
                \land
                \ddiamond{\pupdate{\lforall[C]{i}{\solutionupdate{t}}}}{\phi}
              \big)}}},
  with \m{\lforall{0{\leq}\tilde{t}{\leq}t}{\ddiamond{\pupdate{\lforall[C]{i}{\solutionupdate{\tilde{t}}}}}{\ivr}}}
  abbreviated as $\bar\ivr$.
  By premise, there is a real~$r\geq0$ such that
  \m{\imodels{\Ir}{\bar\ivr \land \ddiamond{\pupdate{\lforall[C]{i}{\solutionupdate{t}}}}{\phi}}}.
  Abbreviate \m{\hevolvein{\lforall[C]{i}{\D{f(\vec{s})}=\theta}}{\ivr}} by~$\mathcal{D}$.
  We have to show that \m{\imodels{\I}{\ddiamond{\hevolve{\mathcal{D}}}{\phi}}}.
  Equivalently, we show~\m{\imodels{\Ir}{\ddiamond{\hevolve{\mathcal{D}}}{\phi}}}, because~$t$ is a fresh variable that does not occur in~$\mathcal{D}$ or~$\phi$.
  Let function \m{\hastype{\varphi}{\interval{[0,r]}\to\linterpretations{\Sigma}{V}}} be defined such that
  \m{\related{\iaccess[\solutionupdate{t}]{\Ir[\zeta]}}{\iget[state]{\I}}{\varphi(\zeta)}} for all \m{\zeta\in\interval{[0,r]}}.
  By premise,~$\varphi(0)$ is identical to~$\iget[state]{\I}$ and~$\phi$ holds at~$\varphi(r)$. Thus it only remains to be shown that~$\varphi$ respects the constraints for the flow function $\varphi$ in the definition of the semantics of \m{\iaccess[\mathcal{D}]{\I}} in \rref{sec:QdL-semantics}.
  In fact,~$\varphi$ obeys the continuity and differentiability properties required for well-definedness of time-derivatives by the corresponding properties of the solution~$\solutionfor[\vec{s}]{}(t)$.
  Moreover, for any \m{e\in\idomain{\I}{C}},
  \m{\ivaluation{\imodif[state]{\iconcat[state=\varphi(\zeta)]{\Ir}}{i}{e}}{f(\vec{s})}
  = \ivaluation{\imodif[state]{\Ir[\zeta]}{i}{e}}{\solutionfor[\vec{s}]{}(t)}}
  has a derivative of value
  \m{\ivaluation{\imodif[state]{\iconcat[state=\varphi(\zeta)]{\Ir}}{i}{e}}{\theta}},
  because~$\solutionfor[\vec{s}]{}$ is a solution of the quantified differential equation \m{\hevolve{\lforall[C]{i}{\D{f(\vec{s})}=\theta}}} with corresponding initial values~\m{\iget[state]{\I}(f(\vec{s}))}.
  Further, it can be shown that the evolution invariant region~$\ivr$ is respected along~$\varphi$ as follows:
  By premise, \m{\imodels{\Ir}{\bar\ivr}} holds for the initial state~$\iget[state]{\Ir}$, thus
  \m{\imodels{\Ifz}{\ivr}} for all \m{\zeta\in\interval{[0,r]}}.
  Combining these results, we can conclude that~$\varphi$ is a witness for~\m{\imodels{\I}{\ddiamond{\hevolve{\mathcal{D}}}{\phi}}}.
  \\
  The converse direction can be shown accordingly to prove equivalence and the dual rule \irref{evolveb} for quantified differential equations with unique solutions (see end of \rref{sec:QdL-semantics}).
  Without unique solutions, the rule is more complicated, but still works: all parameters of all parametric solutions will need to be quantified over in addition to time \m{t{\geq}0}.
  }%
  
 \item\noindent{\hskip-12 pt\irref{assignrb}\ }\
  Rules \irref{assignrb+assignrd} are locally sound by a simple consequence of the fact that arbitrary nondeterministic assignment of $\theta$ for any $j$ of type $C$ to $\onew{}$ is the same as corresponding quantification over $C$.
  The semantics of \m{\dbox{\pupdate{\lforall[C]{j}{\pumod{\onew{}}{\theta}}}}{}} then is equivalent to universal quantification, that of \m{\ddiamond{\pupdate{\lforall[C]{j}{\pumod{\onew{}}{\theta}}}}{}} is equivalent to existential quantification.

 \item\noindent{\hskip-12 pt\irref{gend}\ }\ 
  Rules \irref{genb+gend+invind+con} are sound (but not locally sound) by a variation of the usual proofs~\cite{Harel_et_al_2000,Platzer10}.
  For \irref{gend}, let premise \m{\lsequent{\phi}{\psi}} be valid.
  Let the antecedent be true in a state: \m{\imodels{\I}{\ddiamond{\alpha}{\phi}}}, i.e., let~\m{\relateds{\iaccess[\alpha]{\I}}{\iget[state]{\I}}{\iget[state]{\It}}} with \m{\imodels{\It}{\phi}}.
  Hence, the premise implies
  \m{\imodels{\It}{\phi\limply\psi}}, thus
  \m{\imodels{\It}{\psi}}, which implies
  \m{\imodels{\I}{\ddiamond{\alpha}{\psi}}}.
  The proof for \irref{genb} is similar.
  
 \item\noindent{\hskip-12 pt\irref{invind}\ }\
  Let premise \m{\lsequent{\inv}{\dbox{\alpha}{\inv}}} be valid and let
  the antecedent of the conclusion be true in $\iportray{\I}$, that is \m{\imodels{\I}{\inv}}.
  By premise, \m{\imodels{\It}{\inv}} for all states $\iget[state]{\It}$ with \m{\relateds{\iaccess[\alpha]{\I}}{\iget[state]{\I}}{\iget[state]{\It}}}.
  We thus conclude
  \m{\imodels{\I}{\inv\limply\dbox{\prepeat{\alpha}}{\inv}}}
  by induction along the series of states reached from~$\iget[state]{\I}$ by repeating~$\alpha$.
  
 \item\noindent{\hskip-12 pt\irref{con}\ }\ 
  {\newcommand{\Id}{\imodif[state]{\I}{v}{d}}%
  \newcommand{\Is}{\imodif[state]{\It}{v}{d-1}}%
  Assume that the antecedent is valid and that the premise holds in~$\iportray{\I}$.
  By premise, we have that
  \m{\imodels{\It}{v>0\land\var(v)\limply\ddiamond{\alpha}{\var(v-1)}}}
  for all states~$\iget[state]{\It}$.
  By antecedent, there is a~\m{d\in\reals} such that \m{\imodels{\Id}{\var(v)}}.
  Now, the proof is a well-founded induction on~$d$.
  If~\m{d\leq0}, we have \m{\imodels{\I}{\ddiamond{\prepeat{\alpha}}{\lexists{v{\leq}0}{\var(v)}}}} directly for zero repetitions.
  Otherwise, if~\m{d>0}, we have, by premise, that \[\imodels{\Id}{v>0\land\var(v)\limply\ddiamond{\alpha}{\var(v-1)}}\]
  As~\m{v>0\land\var(v)} holds true at $\iportray{\Id}$,
  we have  \m{\imodels{\It}{\var(v-1)}} for some~$\iget[state]{\It}$ with~\m{\related{\iaccess[\alpha]{\Id}}{\iget[state]{\Id}}{\iget[state]{\It}}}.
  Thus, \m{\imodels{\Is}{\var(v)}} satisfies the induction hypothesis for a smaller~$d$ and a reachable~$\iget[state]{\It}$, because~\m{\related{\iaccess[\alpha]{\I}}{\iget[state]{\I}}{\iget[state]{\It}}} as~$v$ does not occur in~$\alpha$.
  The induction is well-founded, because~$d$ decreases by~1 up to the base case~\m{d\leq0}.\qed
  }%
\end{desCription}

\section{Completeness} \label{sec:ProofTheory}

The verification problem for distributed hybrid systems is extremely challenging. It has \emph{three independent sources} of undecidability.
Thus, no verification technique can be effective. Hence, \QdL cannot be effectively axiomatizable.
The discrete fragment of \QdL is not effectively axiomatizable and the discrete fragment of \QHPs is a computationally complete sublanguage.
The continuous fragment of \QdL is also not effectively axiomatizable.
The fragment with only structural and dimension-changing dynamics is not effective either, because it can encode two-counter machines in link data structures.
As a stronger result, we give a simple proof showing that each of those fragments of \QdL can define first-order integer arithmetic and are, thus, affected by G\"odel's incompleteness theorem \cite{Goedel_1931}.

\begin{theorem}[Incompleteness of {\QdL}] \label{thm:QdL-incomplete}
  \index{incomplete!QdL@\QdL}
  The discrete fragment of \QdL, the continuous fragment of \QdL, and the fragment of \QdL with structural and dimension-changing dynamics are \dfn[axiomatize]{not effectively axiomatizable}, i.e., they have no sound and complete effective calculus, because natural numbers are definable\index{definable} in each of those fragments.
  \index{integer!arithmetic}
  \index{natural!number!definable}
\end{theorem}
\begin{proof}
  We prove that natural numbers are definable among the real numbers of \QdL interpretations in all three fragments.
  Then these fragments extend first-order \emph{integer} arithmetic such that the incompleteness theorem of G\"odel~\cite{Goedel_1931} applies.
  G\"odel's incompleteness theorem shows that no logic extending first-order integer arithmetic can have a sound and complete effective calculus.
  Natural numbers are definable in the discrete fragment using repetitive additions without continuous evolutions, quantified state change, or first-order function symbols:
  \[
  \textit{nat}(n) ~\lbisubjunct~ \ddiamond{\pupdate{\umod{x}{0}};\prepeat{(\pupdate{\umod{x}{x+1}})}}{~x=n}
  .
  \index{_nat_@$\textit{nat}$}
  \]
  In the continuous fragment, an isomorphic copy of the natural numbers is definable using linear ordinary (non-quantified) differential equations without first-order function symbols:
  \[
  \textit{nat}(n) ~\lbisubjunct~
  \lexists{s}{\lexists{c}{\lexists{\tau}{(s=0\land c=1\land\tau=0 \land \ddiamond{\hevolve{\D{s}={c}\syssep\D{c}={-s}\syssep\D{\htime}={1}}}{(s=0 \land \htime=n)})}}}
  .
  \]
  These differential equations characterise~$\sin$ and~$\cos$ as unique solutions for~$s$ and~$c$, respectively. Their zeros, as detected by~$\tau$, correspond to an isomorphic copy of natural numbers, scaled by~$\pi$, i.e., $\textit{nat}(n)$ holds iff~$n$ is of the form~$k\pi$ for a~$k\in\naturals$; see \rref{fig:QdL-incomplete-sin}.
  The initial values for~$s$ and~$c$ prevent the trivial solution identical to~\m{0}.
  \begin{figure}[tbh]
    \centering
    \begin{tikzpicture}
      \begin{scope}
        \draw[->] (-0.1,0) -- (8.3,0) node[right] {$\tau$} coordinate(t axis);
        \draw[->] (0,-1.2) -- (0,1.2) node[above] {$s$} coordinate(x axis);
      \end{scope}
      \fill[draw=blue,fill=blue!20,domain=0:7.8539815,smooth] plot[id=sin2x] function{sin(2*x)} |- (0,0);
      \foreach \A in {0,1.5707963,3.1415926,...,9}
      {
        \fill[vred] (\A,0) circle (2pt);
      }
      \draw (1.5707963-0.15,0) node[below] {$\phantom{1}\pi$} -- ++(0,0);
      \foreach \A in {3,5}
      {
        \draw (\A*1.5707963-0.15,0) node[below] {$\A\pi$} -- ++(0,0);
      }
      \foreach \A in {2,4}
      {
        \draw (\A*1.5707963+0.15,0) node[below] {$\A\pi$} -- ++(0,0);
      }
    \end{tikzpicture}
    \caption{Characterisation of~$\naturals$ as zeros of solutions of differential equations.}
    \label{fig:QdL-incomplete-sin}
  \end{figure}
  
  Integer arithmetic for natural numbers is also definable in the fragment with only structural and dimensional dynamics.
   The proof is somewhat more involved, because we do not consider data arithmetic to be part of that fragment.
  Instead, we characterize natural numbers by chains of links along the values of a function $p$, where we encode zero by a constant symbol $z$:
  \[
  \textit{nat}(n) ~\lbisubjunct~ \ddiamond{\prepeat{(\ptest{n\neq z};~\pupdate{\umod{n}{p(n)}})}}{~n=z}
  .
  \index{_nat_@$\textit{nat}$}
  \]
  We characterize addition by a \QHP \m{\textit{plus}(s,n,m)} to express that the result of adding the natural numbers represented by $n$ and $m$ yields the number represented by $s$:
  \begin{align*}
  \textit{plus}(s,n,m) \mequiv \pupdate{\pumod{s}{z}};~
  &
  \prepeat{\big( \ptest{n\neq z};~\pupdate{\pumod{n}{p(n)}};~\pupdate{\pumod{\nu}{\pnew{}}};~\pupdate{\pumod{p(\nu)}{s}};~\pupdate{\pumod{s}{\nu}} \big)};\\
  &
  \prepeat{\big( \ptest{m\neq z};~\pupdate{\pumod{m}{p(m)}};~\pupdate{\pumod{\nu}{\pnew{}}};~\pupdate{\pumod{p(\nu)}{s}};~\pupdate{\pumod{s}{\nu}} \big)};~
  \ptest{(n=z\land m=z)}
  \end{align*}
  The idea behind this characterization is to create a new chain of links along the values of $p$ by first creating exactly as many links as we can follow along $p$ when starting from $n$, and then continue creating exactly as many links as we can follow along $p$ when starting from $m$, instead; see \rref{fig:QdL-incomplete-dim}.
  The number of links of the result $s$ then is the sum of the respective numbers of links of $n$ and $m$.
  \begin{figure}[tbh]
    \centering
    \begin{tikzpicture}[every join/.style={trans}]
      \tikzstyle{node}=[draw,circle]
      \tikzstyle{trans}=[draw,<-,blue,thick,>=latex]
      \node[node,label=180:$z$] (z) {};
      \begin{scope}[start chain=n,xshift=1.2cm,yshift=1cm]
        \foreach \i in {1,...,3} {
          \node[node,on chain=n,join] {};
        }
        \node[node,on chain=n,join,label=0:$n$] {};
      \end{scope}
      \draw[trans] (z) -- (n-begin);
      \begin{scope}[start chain=m,xshift=1.2cm,yshift=0cm]
        \tikzstyle{node}+=[fill=blue!20]
        \foreach \i in {1,...,2}
          \node[node,on chain=m,join] {};
        \node[node,on chain=m,join,label=0:$m$] {};
      \end{scope}
      \draw[trans] (z) -- (m-begin) node[pos=0.7,above] {$p$} node[pos=0.7,below] {$p$};
      \tikzstyle{node}+=[cloud,cloud puffs=8]
      \tikzstyle{trans}+=[dashed]
      \begin{scope}[start chain=s,xshift=1.2cm,yshift=-1cm]
        \foreach \i in {1,...,4}
          \node[node,on chain=s,join] {};
       \tikzstyle{node}+=[fill=blue!20]
        \foreach \i in {1,...,2}
          \node[node,on chain=s,join] {};
        \node[node,on chain=s,join,label=0:${s=n+m}$] {};
      \end{scope}
      \draw[trans] (z) -- (s-begin);
      \node (desc) at (0,-1.3) {};
      \draw[decorate,decoration={brace,amplitude=8pt},yshift=1cm] (desc -| s-4) -- node[below=6pt] {new copy of $n$} (desc -| z);
      \draw[decorate,decoration={brace,amplitude=8pt},yshift=1cm] (desc -| s-end) -- node[below=6pt] {append new copy of $m$} (desc -| s-4);
    \end{tikzpicture}
    \caption{Characterization of $\naturals$ addition with $p$ links in dimensional dynamics.}
    \label{fig:QdL-incomplete-dim}
  \end{figure}

  We characterize multiplication by a \QHP \m{\textit{times}(s,n,m)} to express that the result of multiplying the natural numbers represented by $n$ and $m$ yields the number represented by~$s$:
  \begin{align*}
  \textit{times}(s,n,m) \mequiv \pupdate{\pumod{s}{z}};~
  &
  \prepeat{\big( \ptest{n\neq z};~\pupdate{\pumod{n}{p(n)}};~\textit{plus}(t,m,s);~\pupdate{\pumod{s}{t}} \big)};~
  \ptest{n=z}
  \end{align*}
  The idea behind this characterization is to compute multiplication by a corresponding number of additions characterized by $\textit{plus}(t,m,s)$.
  That is, the product of $n$ and $m$ can be computed by adding $m$ to an accumulator $s$, $n$ times.
  \qedhere
\end{proof}

The standard way to show adequacy of proof calculi for problems that are not effective is to prove completeness relative to an oracle for handling a fragment of the logic.
Unlike in Cook/Harel relative completeness for discrete programs \cite{DBLP:journals/siamcomp/Cook78,Harel_et_al_2000}, however, \QdL cannot be complete relative to the fragment of the data logic (many-sorted first-order logic with reals), because first-order real arithmetic is decidable and many-sorted first-order logic is semidecidable.
If the \QdL calculus would be complete relative to its data of many-sorted first-order logic with real arithmetic, then, since this is a semidecidable logic, the \QdL calculus would be complete altogether, which would contradict \rref{thm:QdL-incomplete}.
Thus, we need a different basis for a relative completeness argument.
Unlike in conventional discrete programs, the complexity of distributed hybrid systems truly originates from the actual dynamics, not the data.

\rref{thm:QdL-incomplete} shows that the discrete fragment, the continuous fragment, and also the structural/dimensional fragment of \QdL each cause non-axiomatizability of \QdL.
The combination of these fragments and their repeated interaction in the \QHP dynamics of \QdL cannot be any easier.
We prove that, nevertheless, our \QdL calculus is a complete axiomatization relative to the fragment of \QdL that has only quantified differential equations in modalities.
We call this sublogic \FOQD, the \dfn[\FOQD]{first-order logic of quantified differential equations}, i.e., (many-sorted) first-order logic with real arithmetic augmented with formulas expressing properties of quantified differential equations, that is, \QdL formulas of the form \m{\dbox{\hevolvein{\lforall[C]{i}{\D{f(\vec{s})}=\theta}}{\ivr}}{F}}.
The dual formula \m{\ddiamond{\hevolvein{\lforall[C]{i}{\D{f(\vec{s})}=\theta}}{\ivr}}{F}} is expressible as \m{\lnot\dbox{\hevolvein{\lforall[C]{i}{\D{f(\vec{s})}=\theta}}{\ivr}}{\lnot F}}.
Note that the inclusion of $\ivr$ in \FOQD is not essential \cite{DBLP:conf/lics/Platzer12b}.

\begin{theorem}[Axiomatization] \label{thm:QdL-complete}
  The calculus in \rref{calculus:QdL} is a sound and complete axiomatization of \QdL relative to quantified differential equations, i.e., every valid \QdL formula can be derived from valid \FOQD tautologies.
\end{theorem}
\begin{proof}[Proof Outline]
  The (constructive) proof, which, in full, is contained in the remainder of this section,
  generalizes our earlier proof for static, unquantified hybrid systems \cite{DBLP:journals/jar/Platzer08} to \QdL and distributed hybrid systems.
  We prove that every valid \QdL formula can be proven in the \QdL calculus from elementary properties of quantified differential equations (valid oracle instances).
  The crucial step is to show that every valid property of a repetition~$\prepeat{\alpha}$ of a \QHP~$\alpha$ for a distributed hybrid system can be proven by \irref{invind} or \irref{con} with a sufficiently strong invariant or variant that is expressible in \QdL.
  For this, we show that \QHP transitions can be characterized in \QdL.
  One decisive difference to our previous proof \cite{DBLP:journals/jar/Platzer08} is the need to show that states can be characterized by a fixed-size vector of real numbers, and can thus be quantified over.
  This is easy in static finite-dimensional systems, but a fairly tricky challenge in unbounded varying-dimensional systems with first-order functions.
  \qedhere
\end{proof}
This central result shows that properties of distributed hybrid systems can be proven to exactly the same extent to which properties of quantified differential equations can be proven.
Proof-theoretically, the \QdL calculus completely lifts verification techniques for quantified continuous dynamics to distributed hybrid dynamics.
Even though distributed hybrid systems have numerous independent sources of undecidability, we have shown that all true \QdL formulas can be proven in our \QdL calculus, if only we manage to tame the complexity of the continuous dynamics.
Despite these new independent sources of undecidability, we have shown that \QdL can still be axiomatized completely relative to differential equations, only now they are quantified differential equations.

Another important consequence of this result is that decomposition is successful in taming the complexity of distributed hybrid systems.
The \QdL proof calculus is strictly compositional.
All proof rules prove logical formulas or properties of \QHPs by reducing them to structurally simpler \QdL formulas.
As soon as we understand that the distributed hybrid systems complexity comes from a combination of several simpler aspects, we can, hence, tame the system complexity by reducing it to analyzing the dynamical effects of simpler parts.
This decomposition principle is exactly how \QdL proofs can scale to interesting systems in practice.
The relative completeness theorem~\ref{thm:QdL-complete} gives the theoretical evidence why this principle works in general.

\newcommand{\PTS}[3]{\mathcal{R}_{#1}(#3)}%
\newcommand{\ati}[3]{#1^{(#2)}_{#3}}%
\newcommand{\reduct}[1]{#1^{\flat}}%
\newcommand{\Oracle}{\ensuremath{\mathcal{D}}\xspace}%
\newcommand{\linfersequent}[3][]
{\infers[#1] \lsequent{#2}{#3}}%

\newcommand{\FOQDivr}[2][]{#1}%

In the remainder of this section, we present a fully constructive proof of \rref{thm:QdL-complete}.
We have already shown that the \QdL calculus is a sound axiomatization of \QdL in \rref{thm:QdL-sound}.
We need to prove that the \QdL calculus is a complete axiomatization relative to quantified differential equations: every valid \QdL formula can be derived in the \QdL calculus from elementary properties of quantified differential equations.
We need to prove that every valid \QdL formula can be derived in the \QdL calculus from a finite set of valid \FOQD tautologies.
A road map of the proof of \rref{thm:QdL-complete} that we present here is above.

The basic structure follows that of our relative completeness proof for unquantified differential dynamic logic for fixed-dimensional static hybrid systems in previous work \cite{DBLP:journals/jar/Platzer08}.
Here we generalize the proof to \QdL.
A fundamental difference to previous work is that states can be characterized trivially in fixed-dimensional static hybrid systems, but it is not obvious why a finite formula would be sufficient in varying dimensions.
In (dynamic) distributed hybrid systems, we have to prove that there is a finite formula that can characterize and identify all states (see \rref{sec:identification}).
In fixed-dimensional static hybrid systems, states can be characterized and identified trivially by a fixed vector of real numbers for each system variable.
In \QdL, instead, states are full first-order structures with interpretations of functions for all function symbols and the ability to characterize semantic states in logic is no longer obvious.
States are no longer assignments of real numbers to a finite number of variables.
In \QdL, states are full first-order interpretations of function symbols.

Natural numbers are definable in \FOD by \rref{thm:QdL-incomplete}.
Thus, we allow quantifiers over natural numbers like \m{\lforall[\naturals]{x}{\phi}}  and \m{\lexists[\naturals]{x}{\phi}} and over integers \m{\lforall[\integers]{x}{\phi}} as abbreviations.

\subsection{Characterizing Real G{\"o}del Encodings}
As the central device for constructing a \FOQD formula that captures the effect of unboundedly many repetitive hybrid transitions and just uses finitely many real variables, we show that a real version of G{\"o}del encoding is definable in \FOD.
That is, we show that there is a \FOD formula that reversibly packs finite sequences of real values into a single real number.

Observe that a single differential equation system is \emph{not} sufficient for defining these pairing functions as their solutions are differentiable, yet, as a consequence of Morayne's theorem \cite{Morayne87}, there is no differentiable surjection \m{\reals\to\reals^2}, nor to any part of~$\reals^2$ of positive measure\ignore{, nor differentiable surjections to a part \m{L\subseteq\reals^2} of positive measure (only continuous surjections which are almost everywhere differentiable)}.
We show that real sequences can be encoded nevertheless by chaining the effects of solutions of multiple differential equations and quantifiers.

\begin{lemma}[$\reals$-G{\"o}del encoding] \label{lem:realGodel}
  The formula~$\text{at}(Z,n,j,z)$, which holds iff~$Z$ is a real number that represents a G{\"o}del encoding of a sequence of~$n$ real numbers with real value~$z$ at position~$j$ (for a position~$j$ with~\m{1\leq j\leq n}), is definable in \FOD.
  For a formula~\m{\mapply{\phi}{z}} we abbreviate~\m{\lexists{z}{(\text{\normalfont{at}}(Z,n,j,z)\land\mapply{\phi}{z})}} by~\m{\mapply{\phi}{\ati{Z}{n}{j}}}.
\end{lemma}%
\begin{proof}
The proof is an immediate corollary to a result from previous work \cite[Lemma~4]{DBLP:journals/jar/Platzer08}.
\qedhere
\end{proof}
\ignore{
\begin{figure}[bht]%
  \newcommand{\mtime}{t}%
  \centering
  \begin{tabular}{@{}c@{}}%
      \begin{minipage}{9.5cm}%
      \begin{tikzpicture}%
        \node[anchor=east] (a) at (0,+0.63)
          {\begin{minipage}{2.8cm}%
            \[\sum_{i=0}^\infty \frac{a_i}{2^{i}} = a_0.a_1a_2\dots\]%
          \end{minipage}};
        \node[anchor=east] (b) at (0,-0.63)
          {\begin{minipage}{2.9cm}%
            \[\sum_{i=0}^\infty \frac{b_i}{2^{i}} = b_0.b_1b_2\dots\]%
          \end{minipage}};
        \node[anchor=west] (p) at (0.8,0)
          {\begin{minipage}{5.3cm}%
            \[\sum_{i=0}^\infty \left(\frac{a_i}{2^{2i-1}}+\frac{b_i}{2^{2i}}\right) = a_0b_0.a_1b_1a_2b_2\dots\]%
          \end{minipage}};
        \draw[<->,bend right=20] (a.-10) to (p.182);
        \draw[<->,bend left=20]  (b.01) to (p.182);
      \end{tikzpicture}%
      \end{minipage}%
    \vspace{0.4cm}
    \\
    a.~~ \footnotesize{Fractional encoding principle by bit interleaving}
    \label{fig:Godelprinciple}%
    \\
      \begin{minipage}{\textwidth}%
      \begin{eqnarray*}%
        \text{at}(Z,n,j,z)
        &\,\lbisubjunct\,&
        \lforall[\integers]{i}{\digit(z,i)=\digit(Z,n(i-1)+j)}
        \land\textit{nat}(n)\land\textit{nat}(j)\land n>0
        \\
        \digit(a,i)
        &=&
        \intpart(2\fraction(2^{i-1}a))
        \\
        \intpart(a) &=& a-\fraction(a)
        \\
        \fraction(a)=z
        &\lbisubjunct&
        \lexists[\integers]{i}{z=a-i} \land -1<z \land z<1 
        \land az\geq0
        \\
        2^i=z
        &\lbisubjunct&
        i\geq0\land
        \lexists{x}{\lexists{\mtime}{(x=1\land\mtime=0\land
        \ddiamond{\hevolve{\D{x}=x\ln2\land\D{\mtime}=1}}{(\mtime=i\land x=z)})}}
        \\
        &&\lor~
        i<0\land
        \lexists{x}{\lexists{\mtime}{(x=1\land\mtime=0\land
        \ddiamond{\hevolve{\D{x}=-x\ln2\land\D{\mtime}=-1}}{(\mtime=i\land x=z)})}}
        \\
        \ln 2=z
        &\lbisubjunct&
        \lexists{x}{\lexists{\mtime}{(x=1\land\mtime=0\land
        \ddiamond{\hevolve{\D{x}=x\land\D{\mtime}=1}}{(x=2\land\mtime=z)})}}
      \end{eqnarray*}%
      \end{minipage}%
    \vspace{0.4cm}
    \\
    b.~~ \footnotesize{Definition of $\reals$-G{\"o}del encoding in \FOD}
    \label{fig:GodelencodingFOD}%
    \end{tabular}
    \caption{Characterizing G{\"o}del encoding of $\reals$-sequences in one real number}
    \label{fig:Godelencoding}
\end{figure}
\begin{proof}
  The basic idea of the $\reals$-G{\"o}del encoding is to interleave the bits of real numbers as depicted in \rref{fig:Godelencoding}a (for a pairing of~$n=2$ numbers~$a$ and~$b$).
  For defining~\m{\text{at}(Z,n,j,z)}, we use several auxiliary functions to improve readability, see \rref{fig:Godelencoding}b.
  Note that these definitions need no recursion, hence, like in the notation~\m{\mapply{\phi}{\ati{Z}{n}{j}}}, we can consider occurrences of the function symbols as syntactic abbreviations for quantified variables satisfying the respective definitions.
  
  The function symbol~\m{\digit(a,i)} gives the $i$-th bit of~$a\in\reals$ when represented with basis~2.
  For~\m{i>0},~\m{\digit(a,i)} yields fractional bits, and, for~\m{i\leq0}, it yields bits of the integer part.
  For instance,~\m{\digit(a,1)} yields the first fractional bit,~\m{\digit(a,0)} is the least-significant bit of the integer part of~$a$.
  The function~$\intpart(a)$ represents the integer part of~$a\in\reals$.
  The function~$\fraction(a)$ represents the fractional part of~$a\in\reals$, which drops all integer bits.
  The last constraint in its definition implies that~$\fraction(a)$ keeps the sign of~$a$ (or~0).
  Consequently,~$\intpart(a)$ and~$\digit(a,i)$ also keep the sign of~$a$ (or~0).
  Exponentiation~$2^i$ is definable using differential equations, using an auxiliary characterization of the natural logarithm~$\ln2$.
  The definition of~$2^i$ splits into the case of exponential growth when~\m{i\geq0} and a symmetric case of exponential decay when~\m{i<0}.
  \qedhere
\end{proof}
}

\subsection{First-order State Identification} \label{sec:identification}
The crucial step in the proof of \rref{thm:QdL-complete} is the construction of \QdL (in)variants that are strong enough to characterize properties of repetition.
In order to be able to characterize \QHP state transitions in \QdL (in)variants for the completeness proof, we first need to find formulas that characterize/identify states.
For finite-dimensional systems of a fixed dimension $n$, states can simply be characterized completely by the values of all $n$ real state variables.
A particular state could be characterized uniquely by the formula \m{x=2\land y=0.5 \land z=-0.382}, for example.
As a trivial corollary to \rref{lem:realGodel}, states can then even be characterized uniquely by one real number when using the $\reals$-G\"odel encoding.
For infinite-dimensional systems, systems with changing dimension, or systems with a dynamics that depends on evolving interpretations of function symbols $f(\vec{s})$, the situation is more difficult.
After all, a state of \QdL is a full first-order structure with functions as interpretations of function symbols, and these interpretations can change from state to state.
Furthermore, in order to navigate among states during the completeness proof, we need to be able to characterize the current first-order state, but also to recall a previously identified first-order state and express what holds true at this state.

We show that the first-order states reachable with \QHP $\alpha$ from an initial state can, nevertheless, be characterized uniquely by real numbers, which can thus be quantified over.
Furthermore, we show that this correspondence can be axiomatized in \FOQD.
One key observation is that the first-order interpretations can change from state to state, but only according to the dynamics of the \QHP.
Intuitively, the difference of any reachable first-order state to the initial state can be characterized by a finite list of differences to the initial state.
Clearly this difference concerns only finitely many symbols occurring in $\alpha$. It also concerns only finitely many positions of their interpreted functions, because actualist quantified assignments and actualist quantified differential equations only change the interpretation of finitely many function symbols at finitely many positions (actual quantified domains $\laetype{C}$ occurring in actualist quantifiers of \QHPs are finite).
Note that it is crucial for this argument that we have assumed the actual existence predicate $\laexisting{i}$ to have finite support.

\newcommand{\sortof}[1]{S_{#1}}%
\newcommand{\lstateid}[1]{\mathfrak{#1}}%
\renewcommand{\lnow}[1]{\mathop{\downarrow}#1}%
\renewcommand{\lthen}[2]{\mathop{@}#1\,#2}%
\newcommand{\isval}[4]{\ifthenelse{\equal{#4}{}}{}{#4=}\textit{is}_{#1}(#2,#3)}
\newcommand{\stateid}{\lstateid{I}}%
\newcommand{\stateidz}{\lstateid{B}}%
\begin{lemma}[State identification] \label{lem:identification}
  \newcommand{\Iu}{\iconcat[state=\iota]{\stdI}}%
  \renewcommand{\It}{\iconcat[state=\tau]{\stdI}}%
  Let $\Sigma_b$ be a finite set of function symbols containing $\laexisting{\cdot}$.
  The operators $\lnow{}$ and $\lthen{}{}$, which identify and recall states reachable by \QHPs, are definable in \FOQD such that:
  \begin{enumerate}[\em(1)]
  \item \label{case:identification-now}
    For every \QHP~$\alpha$ with \m{\bvar{\alpha}\subseteq\Sigma_b}, every variable \m{\stateid \not\in \Sigma_b} of sort $\reals$, and every $\iname[state]{\I}~\iget[state]{\I}$,
    the formula $\lnow{\stateid}$ is true in at most one of the states reachable by $\alpha$ from $\iget[state]{\I}$.
    That is, there is at most one $\iname[state]{\Iu}~\iget[state]{\Iu}$ such that
    \m{\relateds{\iaccess[\alpha]{\I}}{\iget[state]{\I}}{\iget[state]{\Iu}}} and \m{\imodels{\Iu}{\lnow{\stateid}}}.
  \item \label{case:identification-then}
    For every \QHP~$\alpha$ with \m{\bvar{\alpha}\subseteq\Sigma_b}, every variable \m{\stateid \not\in \Sigma_b} of sort $\reals$, every formula $\phi$, and every $\iname[state]{\I}~\iget[state]{\I}$,
    the formula $\lthen{\stateid}{\phi}$ is true in any state reachable by $\alpha$ from $\iget[state]{\I}$ if and only if $\phi$ is true in the (unique) state that is reachable by $\alpha$ from $\iget[state]{\I}$ in which $\lnow{\stateid}$ holds (provided such a state is reachable at all, otherwise the truth-value of $\lthen{\stateid}{\phi}$ is arbitrary).
    That is, suppose there is a $\iname[state]{\Iu}~\iget[state]{\Iu}$ such that
    \m{\relateds{\iaccess[\alpha]{\I}}{\iget[state]{\I}}{\iget[state]{\Iu}}} and \m{\imodels{\Iu}{\lnow{\stateid}}} (thus, by \rref{case:identification-now}, $\iget[state]{\Iu}$ is unique with that property).
    Then for any $\iname[state]{\It}~\iget[state]{\It}$ with
    \m{\relateds{\iaccess[\alpha]{\I}}{\iget[state]{\I}}{\iget[state]{\It}}}, it is the case that \m{\imodels{\It}{\lthen{\stateid}{\phi}}} if and only if \m{\imodels{\Iu}{\phi}}.
    If, on the contrary, there is no $\iname[state]{\Iu}~\iget[state]{\Iu}$ with
    \m{\relateds{\iaccess[\alpha]{\I}}{\iget[state]{\I}}{\iget[state]{\Iu}}} and \m{\imodels{\Iu}{\lnow{\stateid}}}, then this lemma makes no statement concerning the truth of formula $\lthen{\stateid}{\phi}$ at any  $\iname[state]{\I}~\iget[state]{\It}$.
  \end{enumerate}
\end{lemma}
\begin{proof}
  The formulas $\lnow{\stateid}$ and $\lthen{\stateid}{\phi}$ are like the \textit{here} and \textit{at} operators of hybrid-nominal logic. We show that they can be characterized by \FOQD formulas.
  For defining $\lnow{\stateid}$ and $\lthen{\stateid}{\phi}$, we use an auxiliary function \m{\isval{f}{\stateid}{\vec{o}}{}} to improve readability.
  The formula \m{\isval{f}{\stateid}{\vec{o}}{\theta}} is true if the value of $\theta$ coincides with the value of $f$ at position $\vec{o}$ according to the state characterized by $\stateid$ (i.e., where $\lnow{\stateid}$ is true).
  We characterize \m{\isval{f}{\stateid}{\vec{o}}{\theta}} by the following \FOQD formula:
  \begin{align*}
    &
    \piif{\lexists[\naturals]{s}{(s<m\land \ati{X}{m}{s}=\vec{o})}}
    {\lexists[\naturals]{s}{(s<m\land \ati{X}{m}{s}=\vec{o} \land \theta=\ati{Y}{m}{s})}}
    {\theta=f(\vec{o})}\\
    &\quad\text{where}~\stateid ~\text{is split into the following abbreviations}~
    m\eqdef\ati{\ati{\stateid}{d}{i}\,}{3}{1}, X\eqdef\ati{\ati{\stateid}{d}{i}\,}{3}{2}, Y\eqdef\ati{\ati{\stateid}{d}{i}\,}{3}{3}
    \\&\quad\text{further~$d$ is the number of symbols in $\Sigma_b$ and $i$ is the index of $f$ in $\Sigma_b$}
  \end{align*}
  The function symbol $\isval{f}{\stateid}{\vec{o}}{}$ gives the value ($\theta$) of function $f$ at position $\vec{o}$ at the state characterized by the real number denoted by $\stateid$.
  It can be defined easily using the real pairing function from \rref{lem:realGodel}.
  The basic idea is to understand $\stateid$ via the real pairing function as a list of length $m$ of position/value pairs ($\ati{X}{m}{s}/\ati{Y}{m}{s}$), which characterize changes to the value $f(\vec{o})$ for each of the finitely many function symbols $f\in\Sigma_b$.
  Using an arbitrary but fixed ordering, these function symbols $f$ are identified with their index $d$ in $\Sigma_b$.
  The most important insight for the proof is that, for every state reachable by $\alpha$ from $\iget[state]{\I}$, the list of changes of $f$ compared to $f(\vec{o})$ at $\iget[state]{\I}$ is always finite after finitely many transitions of quantified state change with finite support (see end of \rref{sec:objectcreation}).
  Consequently, the list of changes can always be encoded by one (finite) real number according to \rref{lem:realGodel}.
  
  Using the auxiliary definition \m{\isval{f}{\stateid}{\vec{o}}{\theta}}, we characterize cases~\ref{case:identification-now} and \ref{case:identification-then}, that is $\lnow{\stateid}$ and $\lthen{\stateid}{\phi}$ by the following \FOQD formulas:
  \begin{align*}
    \lnow{\stateid} &\mequiv \landfold_{f\in\Sigma_b} \lforall[\sortof{f}]{\vec{o}} \isval{f}{\stateid}{\vec{o}}{f(\vec{o})}
    \qquad\text{where~$\sortof{f}$ is the sort of the arguments of $f$}
    \\
    \lthen{\stateid}{\phi} &\mequiv 
    \ddiamond{\hevolve{\lforall[C]{i}{\lforall[\reals]{u}{\D{f(i)}=u}}}}{(\phi \land \lnow{\stateid})}
  \end{align*}
  The definitions do not need recursion, so that we can consider occurrences of the defined notations as syntactic abbreviations for quantified variables satisfying the respective definitions (like for \rref{lem:realGodel}).

  Case~\ref{case:identification-now}:
  The characterization for $\lnow{\stateid}$ is defined as a conjunction over all relevant function symbols $f\in\Sigma_b$ asserting that the value $f(\vec{o})$ of $f$ at each position $\vec{o}$ of the sort $\sortof{f}$ of $f$ is identical to the corresponding value $\isval{f}{\stateid}{\vec{o}}{}$ characterized by $\stateid$.
  
  Case~\ref{case:identification-then}:
  The characterization for $\lthen{\stateid}{\phi}$ uses a quantified differential equation with a variable $u$ that only occurs on the right hand side and thus changes $f$ at all positions $i$ with an arbitrary slope $u$.
  The $\lthen{\stateid}{\phi}$ characterization then checks if the appropriate state characterized by $\stateid$ has been reached using $\lnow{\stateid}$ and further expresses that $\phi$ holds at this state.
  By \rref{case:identification-now}, we know that $\lnow{\stateid}$ holds in at most one of the states reachable by $\alpha$ from $\iget[state]{\I}$.
  In the quantified differential equation system for $\lthen{\stateid}{\phi}$, the second quantified variable $u$ amounts to nondeterministically specifying a slope $u$ for each $f(i)$.
  Unlike $i$, quantified variable $u$ only occurs on the right hand side of the quantified differential equation.
  Consequently, the semantics (\rref{case:QdL-QHP-transition-evolve} of the transition relation $\iaccess[\alpha]{\I}$ defined in \rref{sec:QdL-QHP-transition}) defines the states corresponding to \emph{all choices} for $u$ to be reachable.
  These respective choices for $u$ include the choice that leads to the state characterized by $\lnow{\stateid}$, e.g., by choosing slope
  \m{u\eqdef\isval{f}{\stateid}{i}{}-f(i)} for each $i$ and evolving for 1 time unit.
  To simplify notation, we define \m{\lthen{\stateid}{\phi}} only for \m{\Sigma_b=\{f\}}.
  The construction is repeated accordingly (by nesting modalities) for each \m{f\in\Sigma_b}, which are finitely many.
  The createdness flag $\laexisting{\cdot}$ needs to be part of $\Sigma_b$ so that object creation is taken care of on the fly.
  \qedhere
\end{proof}

\subsection{Expressibility and Rendition of Quantified Hybrid Program Semantics} \label{sec:expressive}

In order to show that \QdL is sufficiently expressive to state the invariants and variants that are needed for proving valid statements about \QHP loops with \irref{invind} and \irref{con}, we prove an expressibility result.
We give a constructive proof that the state transition relation of \QHPs is definable in \FOQD, i.e., there is a \FOQD-formula~$\PTS{\alpha}{\vec{x}}{\stateid}$ characterizing the state transitions of quantified hybrid program~$\alpha$ from the current state to the state characterized by~$\stateid$ (a real variable that characterizes a state by way of \rref{lem:identification}).
For this, we need to characterize the dynamics of \QHPs, which are dynamic distributed hybrid processes with repetitively evolving discrete, continuous, structural, and dimension-changing dynamics, equivalently by quantified differential equations in \FOQD.

\begin{lemma}[Program rendition] \label{lem:programrendition}
  For every \QHP~$\alpha$ with symbols among a finite set $\Sigma_b\supseteq\{\laexisting{\cdot}\}$ there is a \FOQD-formula
  \m{\PTS{\alpha}{\vec{x}}{\stateid}}
  with one additional free variable $\stateid$ of sort $\reals$
  such that
  \begin{equation*}
    \entails \PTS{\alpha}{\vec{x}}{\stateid} \lbisubjunct
    \ddiamond{\alpha}{\lnow{\stateid}}
    \label{eq:transitioncharacteriztic}
  \end{equation*}
\end{lemma}
  \newcommand{\mtime}{t}%
  \begin{figure}[hbt]
    \vspace*{-\baselineskip}
    \begin{align*}
      \PTS{\pupdate{\lforall[C]{i}{\umod{f(\vec{s})}{\theta}}}}{\vec{x}}{\stateid}
      &\mequiv
      \lforall[\sortof{f}]{\vec{o}}{}
      \\&\quad
      {\big(\piif{\lexists[C]{i}{\vec{o}=\vec{s}}}
      {\lexists[C]{i}{(\vec{o}=\vec{s} \land \isval{f}{\stateid}{\vec{o}}{\theta})}}
      {\isval{f}{\stateid}{\vec{o}}{f(\vec{o})}}
      \big)}
      \\&\phantom{\mequiv}~\land \landfold_{g\in\Sigma_b\setminus\{f\}} \lforall[\sortof{g}]{\vec{o}}{\isval{g}{\stateid}{\vec{o}}{g(\vec{o})}}
     \\
      \PTS{\hevolve{\lforall[C]{i}{\D{f(\vec{s})}=\genDE{}}}}{\vec{x}}{\stateid}
      &\mequiv
      \ddiamond{\hevolve{\lforall[C]{i}{\D{f(\vec{s})}=\genDE{}}}}{\lnow{\stateid}}
      \\
\FOQDivr[{
      \PTS{\hevolvein{\lforall[C]{i}{\D{f(\vec{s})}=\genDE{}}}{\chi}}{\vec{x}}{\stateid}
      &\mequiv
      \ddiamond{\hevolvein{\lforall[C]{i}{\D{f(\vec{s})}=\genDE{}}}{\chi}}{\lnow{\stateid}}
      \\
}]{
      \PTS{\hevolvein{\lforall[C]{i}{\D{f(\vec{s})}=\genDE{}}}{\chi}}{\vec{x}}{\stateid}
      &\mequiv
      \lexists{\mtime}{\big(\mtime=0\land
        \ddiamond{\hevolve{\D{\mtime}=1\syssep\lforall[C]{i}{\D{f(\vec{s})}=\genDE{}}}}
        {\big(
        \lnow{\stateid}
        \land %
        \dbox{\hevolve{\D{\mtime}=-1\syssep\lforall[C]{i}{\D{f(\vec{s})}=-\genDE{}}}}
        {(\mtime\geq0\limply\chi)}\big)}\big)}
      \\
}
      \PTS{\ptest{\chi}}{\vec{x}}{\stateid}
      &\mequiv
      \chi \land \lnow{\stateid}
      \\
      \PTS{\pchoice{\beta}{\gamma}}{\vec{x}}{\stateid}
      &\mequiv
      \PTS\beta{\vec{x}}{\stateid} \lor \PTS\gamma{\vec{x}}{\stateid}
      \\
      \PTS{\beta;\,\gamma}{\vec{x}}{\stateid}
      &\mequiv
      \lexists{\stateidz}{(
      \PTS{\beta}{\vec{x}}{\stateidz} \land \lthen{\stateidz}{\PTS{\gamma}{\stateidz}{\stateid}})}
      \\
      \PTS{\prepeat{\beta}}{\vec{x}}{\stateid}
      &\mequiv
      \lexists{\stateidz}{\lexists[\naturals]{n}{\big(
        \lnow{\ati{\stateidz}{n}{1}} \land \ati{\stateidz}{n}{n}=\stateid
        \land
        \lforall[\naturals]{i}{(1\leq i<n \limply
          \internal{i<n is required here as some loops may not be able to be continued indefinitely}
          \lthen{\ati{\stateidz}{n}{i}}{\PTS{\beta}{\ati{\stateidz}{n}{i}}{\ati{\stateidz}{n}{i+1}}}
          \internal{Use the computably bijective
            $\mathbf{N} \cong \mathbf{N}^n$, here}
          )}\big)}}
    \end{align*}
    \caption{Explicit rendition of \QHP transition semantics in \FOQD}
    \label{fig:programrendition}
  \end{figure}
\begin{proof}
  The program rendition is defined inductively in \rref{fig:programrendition}.
  The characterization of quantified assignments is a variation of the characterization of $\lnow{\stateid}$ from the proof of \rref{lem:identification}.
  The only difference is that the value $\theta$ is used instead of $f(\vec{o})$ for positions $\vec{o}$ that are affected by the quantified state change, i.e., $\vec{o}$ is of the form $\vec{s}$ for some $i$ (where the quantified assignment matches as expressed by \m{\lexists[C]{i}{\vec{o}=\vec{s}}}).
  Quantified differential equations give \FOQD-formulas already, because $\lnow{\stateid}$ is a \FOQD-formula, hence no further reduction is necessary.
\FOQDivr{
  Evolution along quantified differential equations with additional invariant region $\ivr$ is not part of \FOQD.
  It is definable, however, by following the unique flow (\rref{lem:uniqueflow}) backwards.
  Continuous evolution is reversible, i.e., the transitions of~\m{\hevolve{\D{x_i}=-\theta}} are inverse to those of~\m{\hevolve{\D{x_i}=\theta}}.
  \internal{assuming appropriate domain restrictions for definedness of differential equations to carry over in the form of \m{\dbox{\hevolve{\D{x_i}=\theta_i}}{(defined(\theta)\limply\phi)}}}%
  Consequently, when using auxiliary variable~$t$, all evolutions of
  \m{\dbox{\hevolve{\D{x_1}={-\theta_1}\syssep\sdots\syssep\D{x_k}={-\theta_k}\land\D{\mtime}=-1}}{}}
  follow the same flow as
  \m{\ddiamond{\hevolve{\D{x_1}={\theta_1}\syssep\sdots\syssep\D{x_k}={\theta_k}\land\D{\mtime}=1}}{}}
  but backwards.
  By also reverting clock~$\mtime$, we ensure that, along the reverse flow,~$\chi$ has been true at all times (because of the box modality) until starting time~$t=0$, see \rref{fig:backflow}.
\begin{figure}[htb]
  \newcommand{\ws}{\nu}\newcommand{\wt}{\vec{v}}%
  \renewcommand{\I}{\iconcat[state=\ws]{\stdI}}%
  \renewcommand{\It}{\iconcat[state=\wt]{\stdI}}%
  \tikzstyle{axes}=[]
  \tikzstyle{mode switch}=[black!70,thin,dotted]
  \centering%
    \begin{tikzpicture}[scale=1.5]
      \begin{scope}[style=axes]
        \draw[->] (-0.1,0) -- (2.4,0) node[right] {$\mtime$} coordinate(t axis);
        \draw[->] (0,-0.1) -- (0,1.2) node[above] {$\vec{x}$} coordinate(x axis);
      \end{scope}
      {
        \draw[draw=vgreen,fill=vgreen!10] (1.1,0.8) ellipse (0.9cm and 0.4cm);
        \node[vgreen] at (1.6,0.6) {$\chi$};
      }
      \newcommand{\breakp}{1.8}
      \begin{scope}[xshift=0.7cm,yshift=-0.1cm]
        {
          \draw[thick,domain=-0.7:0.6,smooth,xshift=.5cm]
            plot
            (\x,{exp(-1.5*\x)+1.2*(1-exp(-1.5*\x))})
          node (flowend) {}
          node[above] {$\iget[state]{\It}$};
        }
        \node (flowstart) at (-0.7,0.62847) {};
        \ignore{
          \draw[thick,domain=-1:-0.4,smooth,xshift=1cm] plot (\x,{exp(-1.5*\x)+1.2*(1-exp(-1.5*\x))})
          node[above] {$\iget[state]{\It}$};
          \draw[thick,dotted,color=vblue,domain=-0.4:0.6,smooth,xshift=1cm] plot (\x,{exp(-1.5*\x)+1.2*(1-exp(-1.5*\x))});
        }
        \ignore{
        \draw[thick,domain=-1:-1.01,smooth,xshift=1cm]
        plot (\x,{exp(-1.5*\x)+1.2*(1-exp(-1.5*\x))})
        node[right] {$\iget[state]{\I}$};
        }
        \draw[thick,vred,domain=-0.7:0.6,smooth,xshift=0.5cm,yshift=-0.11cm]
          plot[mark=triangle*,mark options={vred},mark phase=5,mark repeat=8]
          (\x,{exp(-1.5*\x)+1.2*(1-exp(-1.5*\x))})
          node (backstart) {};
        \draw[thick,dotted,vred,domain=-1:-0.7,smooth,xshift=0.5cm,yshift=-0.11cm] plot (\x,{exp(-1.5*\x)+1.2*(1-exp(-1.5*\x))});
        \tikzstyle{my loop}=[->,to path={
          .. controls +(10:0.5) and +(-10:0.5) .. (\tikztotarget) \tikztonodes}]
        \draw[thick] (flowend) to[my loop] (backstart)
          node[midway,right,text width=4cm] {revert flow and time\\and check~$\chi$ backwards};
      \end{scope}
      \node[above=-1pt,rotate=5] at (1,0.9) {$\hevolve{\D{x}={\genDE{x}}}$};
      \draw[mode switch] (0.5,0) node[below,black] {$0$} -- ++(0,1.1);
      \draw[mode switch] (\breakp,0) node[below,black] {$r$} -- ++(0,1.1);
      \path (0.7,0.4) -- node[below,vred] {$\hevolve{\D{x}={-\genDE{x}}}$} (\breakp,0.4);
    \end{tikzpicture}
  \caption{Invariant region checks along backwards flow over time~$\mtime$}
  \label{fig:backflow}
\end{figure}

  {\newcommand{\ws}{\nu}\newcommand{\wt}{\omega}%
  \newcommand{\wf}{f}\newcommand{\wg}{g}%
  \renewcommand{\I}{\iconcat[state=\ws]{\stdI}}%
  \renewcommand{\It}{\iconcat[state=\wt]{\stdI}}%
  \newcommand{\Ifz}{\iconcat[state=f(\zeta)]{\I}}%
  To show reversibility, let~\m{\related{\iaccess[\hevolve{\D{x_1}={\theta_1}\syssep\sdots\syssep\D{x_k}={\theta_k}}]{\I}}{\ws}{\wt}}, that is, 
  let~$\wf:\interval{[0,r]}\to\linterpretations{\Sigma}{V}$ be a solution of~\m{\D{x_1}={\theta_1},\sdots,\D{x_k}={\theta_k}} starting in~$\ws$ and ending in~$\wt$.
  Then~\m{\wg:\interval{[0,r]}\to\linterpretations{\Sigma}{V}}, defined as~\m{\wg(\zeta)=\wf(r-\zeta)}, starts in~$\wt$ and ends in~$\ws$.
  Thus, it only remains to show that~$\wg$ is a solution of~\m{\D{x_1}={-\theta_1},\sdots,\D{x_k}={-\theta_k}}, which can be seen for~\m{1\leq i\leq k} as follows:
  \begin{equation*}
    \begin{split}
      \D[t]{\wg(t)(x_i)}(\zeta)
      =& \D[t]{\wf(r\tweak{{-}}t)(x_i)}(\zeta)
      = \D[u]{\wf(u)(x_i)}\D[t]{(r\tweak{{-}}t)}(\zeta)
      = -\D[u]{\wf(u)(x_i)}(\zeta)
      \\
      =& -\ivaluation{\Ifz}{\theta_i}
      = \ivaluation{\Ifz}{-\theta_i}
      \enspace.
    \end{split}
  \end{equation*}
  }%
  Unlike all other cases, case
  \m{\PTS{\hevolvein{\D{x_1}=\genDE{x_1}_1\syssep\sdots\syssep\D{x_k}=\genDE{x_k}_k}{\chi}}{\vec{x}}{\vec{v}}}
  in \rref{fig:programrendition} uses nested \FOD modalities.
  Yet nested modalities can be avoided in~\m{\PTS{\alpha}{\vec{x}}{\vec{v}}} using an equivalent \FOD formula without them, see \rref{fig:backflow}:
  \begin{align*}
    &
    \lexists{\mtime}{\lexists{r}{\big(\mtime=0\land
      \ddiamond{\hevolve{\D{x_1}=\genDE{x_1}_1\syssep\sdots\syssep\D{x_k}=\genDE{x_k}_k\syssep\D{\mtime}=1}}
      {(\vec{v}=\vec{x}\land r=\mtime)}
      \land
      \\&\phantom{\lexists{\mtime}{\lexists{r}{\big(}}}
      \lforall{\vec{x}}{\lforall{\mtime}{(\vec{x}=\vec{v}\land \mtime=r \limply
      \dbox{\hevolve{\D{x_1}=-\genDE{x_1}_1\syssep\sdots\syssep\D{x_k}=-\genDE{x_k}_k\land\D{\mtime}=-1}}
      {(\mtime\geq0\limply\chi)})}}
      \big)}}
    \enspace.
  \end{align*}
}%
  
  With a finite formula, the characterization of repetition~\m{\PTS{\prepeat{\beta}}{\vec{x}}{\vec{v}}} in \FOQD needs to capture arbitrarily long sequences of intermediate first-order states and the correct transition between successive states of such a sequence.
  To achieve this with first-order quantifiers, we use the real G{\"o}del encoding from \rref{lem:realGodel} in \rref{fig:programrendition} along with the first-order state identification from \rref{lem:identification} to map unbounded sequences of real first-order states reversibly to a single real variable~$\stateidz$, which can be quantified over in first-order logic and identify a first-order state with it by \rref{lem:identification}.
 \qedhere
\end{proof}

Using the \QHP rendition from \rref{lem:programrendition} to characterize modalities, we prove that every \QdL formula can be expressed equivalently in \FOQD by structural induction.

\begin{lemma}[Expressibility] \label{lem:expressive}
  \QdL is \dfn{expressible} in \FOQD:
  for all \QdL formulas~\m{\phi \in \lformulas{\Sigma}{V}}
  there is a \FOQD-formula~\m{\reduct{\phi} \in \lformulas[\FOQD]{\Sigma}{V}}
  that is equivalent, i.e., \m{\entails{\phi \lbisubjunct \reduct{\phi}}}.
  The converse holds trivially.
  \internal{It seems
    that only the direction $\imodels{\I}{\phi \limply \reduct{\phi}}$ is
    required for the relative completeness proof.}
\end{lemma}
\proof
  The proof follows an induction on the structure of formula~$\phi$ for which it is
  imperative to find an equivalent~$\reduct{\phi}$ in \FOQD.
  Observe that the construction of~$\reduct{\phi}$ from~$\phi$ is effective.
  \begin{enumerate}[(1)]
  \item[0.] If $\phi$ is a first-order formula, then \m{\reduct{\phi} \eqdef \phi} already is a \FOQD-formula such that nothing has to be shown.
  \item If $\phi$ is of the form~\m{\varphi \lor \psi}, then by induction hypothesis there are
    \FOQD-formulas $\reduct{\varphi},\reduct{\psi}$ such that \m{\entails \varphi \lbisubjunct \reduct{\varphi}} and
    \m{\entails \psi \lbisubjunct \reduct{\psi}}, from which we can conclude by congruence\internal{of $\lbisubjunct$} that
    \m{\entails (\varphi \lor \psi) \lbisubjunct (\reduct{\varphi} \lor \reduct{\psi})}
    giving \m{\entails \phi \lbisubjunct \reduct{\phi}}
    by choosing \m{\reduct{\varphi} \lor \reduct{\psi}} for~\m{\reduct{\phi}}.
    Likewise reasoning concludes the other propositional connectives
    or quantifiers.
  \item The case where~$\phi$ is of the form~$\ddiamond{\alpha}{\psi}$ is a consequence of the characterization of the semantics of \QHPs in \FOQD.
    The expressibility conjecture holds by induction hypothesis using the equivalence of explicit \QHP renditions from \rref{lem:programrendition}:
    \[
    \entails ~~ \ddiamond{\alpha}{\psi}
    \lbisubjunct
    \lexists{\stateid}{(\PTS\alpha{\vec{x}}{\stateid} \land \lthen{\stateid}{\reduct{\psi}})}
    \enspace.
    \]
  \item The case where~$\phi$ is~$\dbox{\alpha}{\psi}$ is again a consequence of \rref{lem:programrendition}:
    \[
    \entails ~~ \dbox{\alpha}{\psi}
    \lbisubjunct
    \lforall{\stateid}{(\PTS\alpha{\vec{x}}{\stateid} \limply \lthen{\stateid}{\reduct{\psi}})}
    \enspace.\eqno{\qEd}
    \]
  \end{enumerate}

\subsection{Relative Completeness of First-order Assertions} \label{sec:ccHoare}

As special cases of \rref{thm:QdL-complete}, we first prove relative completeness for first-order assertions about \QHPs.
These first-order cases constitute the basis for the general completeness proof for arbitrary \QdL formulas.

In the sequel, we use the notation \m{\infers[\Oracle] \phi} to indicate that a \QdL formula~$\phi$ is derivable from a set of \FOQD-tautologies,
which is equivalent to saying that~$\phi$ is derivable in the \QdL calculus augmented with a single \dfn[oracle]{oracle axiom}~\dfnn{\Oracle}, that gives all valid \FOQD-instances.
The \QdL calculus contains a complete calculus for propositional logic and for many-sorted first-order logic.
We implicitly use simple propositional reasoning (using the \irref{cut}-rule) to glue together subproofs propositionally.

\begin{proposition}[Relative completeness of first-order safety\untweak{assertions}] \label{prop:CAbcomplete}
  For every \QHP \m{\alpha \ignore{\in \lprograms{\Sigma}{V}}} and all \FOQD formulas \m{F,G}
  \[
  \entails F \limply \dbox{\alpha}{G}
  ~\text{implies}~
  \infers[\Oracle] F \limply \dbox{\alpha}{G}
  \enspace.
  \]
\end{proposition}
\proof
  We generalize the relative completeness proof by Cook \cite{DBLP:journals/siamcomp/Cook78} to \QdL
  and follow an induction on the structure of program~$\alpha$.
  In the following, \emph{IH} is short for the induction hypothesis.
  \begin{enumerate}[(1)]
  \item The cases where~$\alpha$ is of the form
    \m{\pupdate{\umod{f(\vec{s})}{\theta}}}, \m{\ptest{\chi}}, \m{\pchoice{\beta}{\gamma}}, or \m{\beta;\gamma}
    are consequences of the soundness of the rules \irref{composeb}, \irref{choiceb}, \irref{testb}, and \irref{assignb}, which are equivalence rules.
    Consequently, whenever their conclusion is valid, their premise is valid and of smaller complexity (the programs get simpler), hence the premise is derivable by IH.
    Thus, we can derive \m{F\limply\dbox{\alpha}{G}} by applying the respective rule.
    For \irref{assignb} and \irref{upskip}, respectively, the premise is simpler because the quantified assignment is only applied to structurally simpler expressions ($\vec{u}$) in the premise than in the conclusion ($f(\vec{u})$) while the program stays the same.
    For nondeterministic assignments, the reasoning is similar using equivalence rule \irref{assignrb} instead of \irref{assignb}.
    Again, the premise is valid, and already a \FOQD formula, hence derivable as an $\Oracle$ axiom directly.
    A formal rewrite proof along these lines is a simple modification of prior work \cite{DBLP:conf/cade/BeckertP06}.
    We explicitly show the proof for~\m{\beta;\gamma} as it contains an extra twist.
  \item
    \m{\entails F \limply \dbox{\beta;\gamma}{G}}
    implies \m{\entails F \limply \dbox{\beta}{\dbox{\gamma}{G}}}.
    By \rref{lem:expressive}, there is a \FOQD-formula~$\reduct{G}$ such that
    \m{\entails \reduct{G} \lbisubjunct \dbox{\gamma}{G}}.
    From the validity of
    \m{\entails F \limply \dbox{\beta}{\reduct{G}}},
    we can conclude by IH that
    \m{\linfersequent[\Oracle]{F}{\dbox{\beta}{\reduct{G}}}} is derivable.
    Similarly, 
    \m{\entails \reduct{G} \limply \dbox{\gamma}{G}}
    yields
    \m{\infers[\Oracle]{\reduct{G}}\limply{\dbox{\gamma}{G}}} by IH.
    With an application of \irref{genb}, the latter derivation can be extended to a derivation of
    \m{\linfersequent[\Oracle] {\dbox{\beta}{\reduct{G}}} {\dbox{\beta}{\dbox{\gamma}{G}}}}.
    Combining the above derivations propositionally by a cut with \m{\dbox{\beta}{\reduct{G}}}, we can derive
    \m{\linfersequent[\Oracle]{F} {\dbox{\beta}{\dbox{\gamma}{G}}}},
    from which \irref{composeb} yields
    \m{\linfersequent[\Oracle]{F} {\dbox{\beta;\gamma}{G}}} as desired.
  \item \m{\entails F \limply \dbox{\hevolvein{\lforall[C]{i}{\D{f(\vec{s})}=\theta}}{\ivr}}{G}}
    is a \FOQD-formula and hence derivable as a \Oracle axiom directly.
  \item \m{\entails F \limply \dbox{\prepeat{\beta}}{G}}
    can be derived by induction.
    For this, we define the invariant as a \FOQD encoding of the statement that all potential poststates of~$\prepeat{\beta}$ satisfy~$G$ according to \rref{lem:expressive}:
    \[
    \inv \mequiv
    \reduct{(\dbox{\prepeat{\beta}}{G})}
    \mequiv
    \lforall{\stateid}{(\PTS{\prepeat{\beta}}{\vec{x}}{\stateid} \limply \lthen{\stateid}{G})}
    \enspace.
    \]
    Since \m{F \limply \inv} and \m{\inv \limply G} are
    valid \FOQD-formulas according to the semantics, they are derivable by \Oracle. %
    By %
    \irref{genb}, \m{\linfersequent[\Oracle] {\dbox{\prepeat{\beta}}{\inv}} {\dbox{\prepeat{\beta}}{G}}}
    is derivable from the latter.
    Likewise, \m{\inv \limply \dbox{\beta}{\inv}} is valid according to the semantics of repetition, thus derivable by IH, since~$\beta$ is less complex.
    Now \irref{invind} yields
    \m{\linfersequent[\Oracle]{\inv}{\dbox{\prepeat{\beta}}{\inv}}}.
    Combining the above derivations propositionally by a cut with \m{\dbox{\prepeat{\beta}}{\inv}} and~\m{\inv} yields
    \m{\linfersequent[\Oracle]{F}{\dbox{\prepeat{\beta}}{G}}}.\qed
  \end{enumerate}

\begin{proposition}[Relative completeness of first-order liveness\untweak{assertions}] \label{prop:CAdcomplete}
  For each \QHP \m{\alpha \ignore{\in \lprograms{\Sigma}{V}}} and all \FOQD-formulas \m{F,G}
  \[
  \entails F \limply \ddiamond{\alpha}{G}
  ~\text{implies}~
  \infers[\Oracle] F \limply \ddiamond{\alpha}{G}
  \enspace.
  \]
\end{proposition}
\proof
  We generalize the integer arithmetic completeness proof by Harel \cite{Harel_1979} to the hybrid case.
  Most cases of the proof are simple adaptations of the corresponding cases in \rref{prop:CAbcomplete}.
  What remains to be shown is the case of repetitions.
  Assume that \m{\entails F \limply \ddiamond{\prepeat{\beta}}{G}}.
  To derive this formula by \irref{con}, we use a \FOQD-formula~$\var(n)$ as a variant expressing that, after~$n$ iterations,~$\beta$ can lead to  a state satisfying~$G$.
  This formula is obtained from \rref{lem:programrendition}-\ref{lem:expressive} as
  \m{\reduct{(\ddiamond{\prepeat{\beta}}{G})}
  \mequiv
  \lexists{\stateid}{(\PTS{\prepeat{\beta}}{\vec{x}}{\stateid} \land \lthen{\stateid}{G})}},
  \emph{except} that the quantifier on the repetition count~$n$ is removed such that~$n$ becomes a free variable
  (plus index shifting to count repetitions):
  \begin{equation*}
    \var(n-1)
    \,\mequiv\,
    \lexists{\stateidz}{\big(
      \lnow{\ati{\stateidz}{n}{1}} \land \ati{\stateidz}{n}{n}=\stateid
      \land
      \lforall[\naturals]{i}{(1\leq i<n \limply
        \internal{i<n is required here as some loops may not be able to be continued indefinitely}
        \lthen{\ati{\stateidz}{n}{i}}{\PTS{\beta}{\ati{\stateidz}{n}{i}}{\ati{\stateidz}{n}{i+1}}}
        \internal{Use the computably bijective
          $\mathbf{N} \cong \mathbf{N}^n$, here}
        )}
        \land\lthen{\stateid}{G}
        \big)}
    \enspace.
  \end{equation*}
  By \rref{lem:realGodel},~$\var(n)$ can only hold true if~$n$ is a natural number.

  According to the loop semantics,
  \m{\entails {n>0 \land \var(n) \limply \ddiamond{\beta}{\var(n-1)}}}
  is valid by construction:
  If~\m{n>0} is a natural number then so is~\m{n-1}, and if~$\beta$ reaches~$G$ after~$n$ repetitions, then, after executing~$\beta$ once,~\m{n-1} repetitions of~$\beta$ reach~$G$.
  By IH, this formula is derivable, since~$\beta$ contains less loops.
  We have derived
  \m{\infers[\Oracle] n>0 \land \var(n) \limply \ddiamond{\beta}{\var(n-1)}}.
  Thus
  \m{\linfersequent[\Oracle]{\lexists{v}{\var(v)}} {\ddiamond{\prepeat{\beta}}{\lexists{v{\leq}0}{\var(v)}}}}
  by \irref{con}.
  It only remains to show that the antecedent is derivable from~$F$ and that~\m{\ddiamond{\prepeat{\beta}}{G}} is derivable from the succedent.
  From our assumption, we conclude that the following are valid \FOQD-formulas, hence \Oracle-axioms:
  \begin{iteMize}{$\bullet$}
   \item \m{\entails F \limply \lexists{v}{\var(v)}}, 
    because~\m{\entails F\limply\ddiamond{\prepeat{\beta}}{G}}, and
   \item \m{\entails (\lexists{v{\leq}0}{\var(v)}) \limply G},
    because~\m{v{\leq}0} and the fact, that, by \rref{lem:realGodel},~\m{\var(v)} only holds true for natural numbers, imply~\m{\var(0)}.
    Further,~\m{\var(0)} entails~$G$, because zero repetitions of~$\beta$ have no effect.
  \end{iteMize}
  We extend the latter
  derivation to
  \m{\linfersequent[\Oracle] {\ddiamond{\prepeat{\beta}}{\lexists{v{\leq}0}{\var(v)}}} {\ddiamond{\prepeat{\beta}}{G}}} by \irref{gend}.
  Now, the above derivations can be combined propositionally by a cut with \m{\ddiamond{\prepeat{\beta}}{\lexists{v{\leq}0}{\var(v)}}} and with \m{\lexists{v}{\var(v)}}
  to yield \m{\linfersequent[\Oracle]{F}{\ddiamond{\prepeat{\beta}}{G}}}.\qed

\subsection{Relative Completeness of the \texorpdfstring{\QdL}{QdL} Calculus} \label{sec:relativeCompletenessProof}
Having succeeded with the proofs of the above statements about parts of the completeness proof, we can finish the proof of \rref{thm:QdL-complete}.

\proof[Proof of \rref{thm:QdL-complete}]
  \newcommand{\measure}[1]{|#1|}%
  The proof follows a basic structure similar to that of Harel's proof for the discrete case \cite[Theorem~3.1]{Harel_1979}.
  We have to show that every valid \QdL formula~$\phi$
  can be proven from \FOQD axioms within the \QdL calculus:
  from \m{\entails\phi} we have to prove \m{\infers[\Oracle] \phi}.
  The proof proceeds as follows: By propositional
  recombination, we inductively identify fragments of~$\phi$ that correspond to
  \m{\phi_1 \limply \dbox{\alpha}{\phi_2}}
  or
  \m{\phi_1 \limply \ddiamond{\alpha}{\phi_2}} logically.
  Next, we express subformulas $\phi_i$ equivalently in \FOQD by \rref[lemma]{lem:expressive},
  and use \rref{prop:CAbcomplete} and~\ref{prop:CAdcomplete} to resolve these first-order safety or liveness assertions.
  Finally, we prove that the original \QdL formula can be re-derived from the subproofs.

  We can assume~$\phi$ to be given in conjunctive normal form by appropriate propositional reasoning.
  In particular, we assume that negations are pushed inside over modalities using the dualities
  \m{\lnot\dbox{\alpha}{\phi} \mequiv \ddiamond{\alpha}{\lnot\phi}}
  and
  \m{\lnot\ddiamond{\alpha}{\phi} \mequiv \dbox{\alpha}{\lnot\phi}}.
  The remainder of the proof follows an induction on a measure~$\measure{\phi}$ defined as the number of modalities in~$\phi$.
  For a uniform proof, we assume real quantifiers to be abbreviations for modal formulas by
  \m{\lexists[\reals]{x}{\phi} \mequiv \ddiamond{\hevolve{\D{x}=1}}{\phi} \lor \ddiamond{\hevolve{\D{x}=-1}}{\phi}} and
  \m{\lforall[\reals]{x}{\phi} \mequiv \dbox{\hevolve{\D{x}=1}}{\phi} \land \dbox{\hevolve{\D{x}=-1}}{\phi}}.
  Following either \m{\hevolve{\D{x}=1}} or \m{\hevolve{\D{x}=-1}}, we can reach any real number as a value for $x$.
  Similarly, we assume quantifiers for sort $C\neq\reals$ to be abbreviations for modal formulas by
  \m{\lexists[C]{x}{\phi} \mequiv \ddiamond{\pupdate{\lforall[C]{j}{\pumod{x}{j}}}}{\phi}} and
  \m{\lforall[C]{x}{\phi} \mequiv \dbox{\pupdate{\lforall[C]{j}{\pumod{x}{j}}}}{\phi}}.
  We can obtain any object of sort $C$ by an appropriate choice of $j$.
  Now the proof is by induction on the measure $\measure{\phi}$ of $\phi$.
  \begin{enumerate}[(1)]
  \item[0.] \m{\measure{\phi}=0} then~$\phi$ is a first-order formula, hence derivable by \Oracle.
  
  \item $\phi$ is of the form \m{\lnot\phi_1}, then~$\phi_1$ is first-order, as we assumed negations to be pushed inside.
   Hence, case~0 applies: \m{\measure{\phi}=0}.
  
  \item $\phi$ is of the form \m{\phi_1 \land \phi_2}, then individually deduce the simpler proofs for
   \m{\infers[\Oracle] \phi_1} and \m{\infers[\Oracle] \phi_2}
    by IH, which can be combined by \irref{andr}.
  
  \item $\phi$ is a disjunction and---without loss of
    generality---has one of the following forms
    (otherwise use associativity and commutativity to select a different order for the disjunction):
    \begin{displaymath}
    \begin{array}{r@{~}c@{~}l}
      \phi_1 &\lor& \dbox{\alpha}{\phi_2}\\
      \phi_1 &\lor& \ddiamond{\alpha}{\phi_2}
      \internal{subsumed by modal cases using the  encoding of quantifiers as modalities
      \phi_1 &\lor& \lexists{x}{\phi_2}\\
      \phi_1 &\lor& \lforall{x}{\phi_2}
      \enspace.
      }
    \end{array}
    \end{displaymath}
    As a unified notation for those cases we use~\m{\phi_1 \lor \dmodality{\alpha}{\phi_2}}.
    Then, \m{\measure{\phi_2}<\measure{\phi}},
    since~$\phi_2$ has less modalities.
    Likewise, \m{\measure{\phi_1}<\measure{\phi}} because
    \m{\dmodality{\alpha}{\phi_2}} contributes one modality to
    \m{\measure{\phi}} that is not part of $\phi_1$.
    
    According to \rref{lem:expressive} there are \FOQD-formulas \m{\reduct{\phi_1},\reduct{\phi_2} \untweak{\in \lformulas[\FOQD]{\Sigma}{V}}} that satisfy
    \m{\entails \phi_i \lbisubjunct \reduct{\phi_i}} for \m{i=1,2}.
    By congruence, the validity \m{\entails \phi} yields that
    \m{\entails \reduct{\phi_1} \lor \dmodality{\alpha}{\reduct{\phi_2}}},
    which directly implies
    \m{\entails \lnot\reduct{\phi_1} \limply
    \dmodality{\alpha}{\reduct{\phi_2}}}.
    Then by \rref{prop:CAbcomplete} or~\ref{prop:CAdcomplete}, respectively, we can derive
    \begin{equation} \label{star:11}
      \linfersequent[\Oracle]{\lnot\reduct{\phi_1}} {\dmodality{\alpha}{\reduct{\phi_2}}}
      \enspace.
    \end{equation}
    Further \m{\entails \phi_1 \lbisubjunct \reduct{\phi_1}} implies
    \m{\entails \lnot\phi_1 \limply \lnot\reduct{\phi_1}}, 
    which is derivable by IH, because
    \m{\measure{\phi_1}<\measure{\phi}}.
    We combine the resulting derivation \m{\linfersequent[\Oracle]{\lnot\phi_1} {\lnot\reduct{\phi_1}}},
    with \eqref{star:11} by a cut with \m{\lnot\reduct{\phi_1}} to obtain
    \begin{equation} \label{star:12}
      \linfersequent[\Oracle]{\lnot\phi_1} {\dmodality{\alpha}{\reduct{\phi_2}}}
      \enspace.
    \end{equation}
    Likewise~\m{\entails \phi_2 \lbisubjunct \reduct{\phi_2}} implies
    \m{\entails \reduct{\phi_2} \limply \phi_2}, which is derivable by IH,
    as \m{\measure{\phi_2}<\measure{\phi}}.
    We can extend the derivation of
    \m{\infers[\Oracle] \reduct{\phi_2} \limply \phi_2}
    \m{\linfersequent[\Oracle] {\dmodality{\alpha}{\reduct{\phi_2}}}
    {\dmodality{\alpha}{\phi_2}}}
    by \irref{genb}--\irref{gend}.
    Finally we combine
    the latter propositionally with~\eqref{star:12} by a cut with \m{\dmodality{\alpha}{\reduct{\phi_2}}} to derive
    \m{\linfersequent[\Oracle]{\lnot\phi_1} {\dmodality{\alpha}{\phi_2}}},
    from which
    \m{\infers[\Oracle] \phi_1 \lor \dmodality{\alpha}{\phi_2}}
    can be obtained, again using \irref{cut}, to complete the proof.\qed
  \end{enumerate}

\section{Distributed Car Control Verification} \label{sec:distributed-car-control-verification}
{%
With the \QdL calculus and the compatibility condition $\dcseparate{i}{j}$ from eqn.\,\rref{eq:distributed-car-control-separate}, we can easily prove collision freedom, i.e., formula \rref{eq:distributed-car-control-new-ae}, in the distributed car control system \rref{eq:distributed-car-control-new-model}:
\begin{multline}
({\dcinv}) \limply\\ {\dbox{\prepeat{(\dcsys)}}{~\laforall[C]{i{\neq}j}{\oa{x}{i}{\neq}\oa{x}{j}}}}
  \label{eq:distributed-car-control-simple-new}
\end{multline}
The biggest challenge in the proof of this \QdL formula is that it involves continuous dynamics, discrete dynamics, and dimensional dynamics, and that all parts of the system need to interact safely for the system to stay collision-free.
In particular, formula \rref{eq:distributed-car-control-simple-new} states a safety property of unboundedly many cars driving on a road, where an unbounded number of new cars may additionally appear dynamically during the evolution of the system.
See \rref{fig:new-car-proof} for a formal \QdL proof of this \QdL formula, which proves collision freedom despite dynamic appearance of new cars.

{%
\let\dcnuo\dcnu%
\let\paexistingorg\paexisting
\renewcommand*{\paexisting}[1]{\laexisting{#1}}%
\newcommand{\dcnuu}{\paexisting{\dcnu}}%
\newcommand*{\dcnuud}[1]{[\dcnuu]{#1}}%
\newcommand{\dcnuosep}{\laforall[C]{i}{\dcseparate{i}{\dcnuo}}}%
\newcommand*{\dcsolution}[1]{\oa{\mathcal{S}_t}{#1}}%
\newcommand*{\ddiadcsolnew}[1]{\dmodality{\pupdate{\dcsolvenew}}{#1}}%
  \newcommand*{\oasol}[2]{%
    \ifthenelse{\equal{#1}{x}}{x(#2)+v(#2)t+\frac{a(#2)}{2}t^2}
    {\ifthenelse{\equal{#1}{v}}{v(#2)+a(#2)t}
    {#1(#2)}}}%
  \newcommand*{\oaskol}[2]{\textsc{#1}_{#2}}%
  \newcommand*{\oasolskol}[2]{%
    \ifthenelse{\equal{#1}{x}}{\oaskol{x}{#2}+\oaskol{v}{#2}t+\frac{\oaskol{a}{#2}}{2}t^2}
    {\ifthenelse{\equal{#1}{v}}{\oaskol{v}{#2}+\oaskol{a}{#2}t}
    {\oaskol{#1}{#2}}}}%
\newcommand{\dcsolve}{\pupdate{\laforall[C]{i}{\dcsolution{i}}}}%
\newcommand{\dcsolvenew}{\pupdate{\laforallplus[C]{\dcnuo}{i}{\dcsolution{i}}}}%
\newcommand*{\dcsolseparate}[2]{\mathcal{S}_t\mathcal{M}(#1,#2)}%
\newcommand*{\dcseparateisk}[2]{\mathcal{M}_{#1,#2}}%
\newcommand*{\dcsolseparateisk}[2]{\mathcal{S}_t\mathcal{M}_{#1,#2}}%
\let\dcsys\DCCS%

\begin{figure}[tbh]
  \footnotesize
  \renewcommand{\linferSequentSeparation}{}
  \newcommand*{\dboxorg}[2]{[#1]#2}%
  \let\dmodality\dbox
  \let\weakaway\ignore  
  \let\weaken\ignore  
  \advance\leftskip-1cm
  \begin{minipage}{\textwidth}
  \begin{sequentdeduction}[array]
    \linfer[invind]
    {\linfer[composeb]
      {\linfer[new]%
        {\linfer[testb]
          {\linfer[implyr]
            {\linfer[upnewall+andl]
              {\linfer[evolveb]
                {\linfer[allr+implyr] %
                  {\linfer[upnewu]
                    {\linfer[upnewall]
                        {\linfer[allr]
                          {\linfer[upapply]
                            {\linfer[andr] %
                              {\linfer[alllinst]
                                {\linfer[iallr]
                                    {\lclose}
                                  {\lsequent{\weakaway{\laforall[C]{i,j}{\dcseparate{i}{j}}}, \dcseparate{i}{\dcnuo}, t{\geq}0} {\dcsolseparate{i}{\dcnuo}}}
                                } %
                                {\lsequent{\weakaway{\laforall[C]{i,j}{\dcseparate{i}{j}}}\dots, \dcnuosep, t{\geq}0} {\dcsolseparate{i}{\dcnuo}}}
                              ! %
                               \linfer[alllinst]
                                 {\linfer[iallr]
                                     {\lclose}
                                   {\lsequent{\dcseparate{i}{j}, \weaken{\dcnuosep}, t{\geq}0} {\dcsolseparate{i}{j}}}
                                 } %
                                 {\lsequent{\laforall[C]{i,j}{\dcseparate{i}{j}}, \weaken{\dcnuosep}\dots, t{\geq}0} {\dcsolseparate{i}{j}}}
                              } %
                              {\lsequent{\laforall[C]{i,j}{\dcseparate{i}{j}}, \dcnuosep, t{\geq}0} {\dcsolseparate{i}{\dcnuo}\land\dcsolseparate{i}{j}}}
                            } %
                            {\lsequent{\laforall[C]{i,j}{\dcseparate{i}{j}}, \dcnuosep, t{\geq}0} {
                              \ddiadcsolnew{(\dcseparate{i}{\dcnuo}\land\dcseparate{i}{j})}
                            }}
                          } %
                          {\lsequent{\laforall[C]{i,j}{\dcseparate{i}{j}}, \dcnuosep, t{\geq}0} {
                            \ddiadcsolnew{\laforall[C]{i,j}{(\dcseparate{i}{\dcnuo}}\land\dcseparate{i}{j})}
                          }}
                        } %
                      {\lsequent{\laforall[C]{i,j}{\dcseparate{i}{j}}, \dcnuosep, t{\geq}0} {\ddiadcsolnew{\dcnuud{\dcinv}}}}
                    } %
                    {\lsequent{\laforall[C]{i,j}{\dcseparate{i}{j}}, \dcnuosep, t{\geq}0} {\dcnuud{\dmodality{\dcsolve}{\dcinv}}}}
                  } %
                  {\lsequent{\laforall[C]{i,j}{\dcseparate{i}{j}}, \dcnuosep} {\dcnuud{\lforall{t{\geq}0}{\dmodality{\dcsolve}{\dcinv}}}}}
                } %
                {\lsequent{\laforall[C]{i,j}{\dcseparate{i}{j}}, \weaken{\dcseparate{\dcnuo}{\dcnuo}}, \dcnuosep} {\dcnuud{\dbox{\dcevo}{\dcinv}}}}
              } %
              {\lsequent{\laforall[C]{i,j}{\dcseparate{i}{j}}, \dcnuud{\dcnusep}} {\dcnuud{\dbox{\dcevo}{\dcinv}}}}
            } %
            {\lsequent{\laforall[C]{i,j}{\dcseparate{i}{j}}} {\dcnuud{(\dcnusep\limply\dbox{\dcevo}{\dcinv})}}}
          } %
          {\lsequent{\textcolor{gray}{\lnaexisting{\dcnu}},\laforall[C]{i,j}{\dcseparate{i}{j}}} {\dcnuud{\dbox{\ptest{\dcnusep};\dcevo}{\dcinv}}}}
        } %
        {\lsequent{\laforall[C]{i,j}{\dcseparate{i}{j}}} {\dboxorg{\dcnup}{\dbox{\ptest{\dcnusep};\dcevo}{\dcinv}}}}
      } %
      {\lsequent{\laforall[C]{i,j}{\dcseparate{i}{j}}} {\dbox{\dcsys}{\dcinv}}}
    } %
    {\lsequent{\dcinv} {\dbox{\prepeat{(\dcsys)}}{\laforall[C]{i{\neq}j}{\oa{x}{i}{\neq}\oa{x}{j}}}}}
  \end{sequentdeduction}
  \end{minipage}

  \caption{\QdL proof for collision freedom in distributed car control with dynamic appearance.}
  \label{fig:new-car-proof}
\end{figure}
The proof in \rref{fig:new-car-proof} uses induction (rule \irref{invind}) with invariant \m{\dcinv}.
Figure~\ref{fig:new-car-proof} does not show the branch proving that the invariant \m{\dcinv} implies the postcondition \m{\laforall[C]{i{\neq}j}{\oa{x}{i}{\neq}\oa{x}{j}}}, which is easy to prove.

\irlabel{new|$new$}%
The proof step marked by \irref{new} uses the definition of $\pnew{C}$ from eqn.~\rref{eq:new}.
To save space, we abbreviate \m{\dbox{\paexistingorg{\dcnu}}{}} by \m{\dbox{\paexisting{\dcnu}}{}} in \rref{fig:new-car-proof}.
The proof uses the derived rules \irref{upnewall} and \irref{upnewu} from \rref{sec:QdL-calculus} to propagate the effect of object creation on actualist quantifiers and actualist quantified assignments respectively.
In rule \irref{upnewu}, the shorthand notation \m{\dcsolvenew} in the resulting formula indicates that the new object $\dcnu$ is also updated according to the solution $\dcsolution{\dcnu}$, not just the previously existing objects (\m{\dcsolve}).
Here, we abbreviate by \m{\dcsolution{i}} the solution
\m{\umod{\oa{x}{i}}{\oa{x}{i}+\oa{v}{i}t+\frac{\oa{a}{i}}{2}t^2} \syssep \umod{\oa{v}{i}}{\oa{v}{i}+\oa{a}{i}t}} of the quantified differential equation \m{\hevolve{\lforall[C]{i}{\D[2]{x(i)}=a(i)}}}, which rule \irref{evolveb} introduces.
For the top-most application of rule \irref{assignb}, we denote by \m{\dcsolseparate{i}{j}} the result of substituting \m{\laforall[C]{i}{\dcsolution{i}}} into \m{\dcseparate{i}{j}} according to rule \irref{assignb}.
In \rref{fig:new-car-proof}, we leave out some irrelevant formulas, indicated by ellipsis ($\dots$) or gray print.
The proof closes (indicated by $\ast$) by \qelim{} with rule \irref{iallr}.
Hence, \QdL formula \rref{eq:distributed-car-control-simple-new} is valid by \rref{thm:QdL-sound}.
}%

In a similar way, the \QdL proof rules can prove collision freedom in an advanced distributed car control system that has both dynamic appearance of cars on the road as in \rref{eq:distributed-car-control-new-model} and more flexibility in acceleration and braking choices of the individual cars as in \rref{eq:distributed-car-control-accel-ae}.
For this, we choose a weaker constraint for \m{\dcseparate{i}{j}} that allows cars that move with quite different accelerations, if only the respective safety distances are compatible with the different velocities:
\begin{align*}
  i\neq j \,\limply\,& \big((\dcaccelseparatetf{i}{j})
  \\&
  \lor(\dcaccelseparatets{i}{j})\big)%
\end{align*}
{%
With this choice for \m{\dcseparate{i}{j}}, the \QdL proof calculus can be used to prove the following \QdL formula with a proof very similar to that in \rref{fig:new-car-proof}:
\begin{align}
  &\hspace*{-0.6cm}\dcinv \limply
  \notag\\
  &\big[\big( \dcnup;~ \ptest{\dcnusep};
  \notag\\&~~\dcaccelallt;
  \notag\\&~~\dcaccelevot )\prepeat{}
  \notag\\&\big]{~\laforall[C]{i{\neq}j}{\oa{x}{i}{\neq}\oa{x}{j}}}%
  \label{eq:distributed-car-control-accel-new}
\end{align}
The \QHP in \QdL formula~\rref{eq:distributed-car-control-accel-new} allows all cars to change their respective acceleration freely when all other cars are sufficiently far away like in \rref{eq:distributed-car-control-accel-ae}.
For this, we choose a condition characterizing that the distributed car control system stays controllable for at least $\cyct$ time units (which is the maximum reaction time of the controller):
\[
\SBform{i}{j} ~\mequiv~ \SBformt{i}{j}
\]
The continuous dynamics in \rref{eq:distributed-car-control-accel-new} is bounded by the evolution domain constraint \m{\tau\leq\cyct} to evolve for at most $\cyct$ time units, at which point, at the latest, the discrete controllers will have a chance to react to situation changes again (i.e., the control loop repeats).
The \QdL proof of \rref{eq:distributed-car-control-accel-new} has the same structure as that in \rref{fig:new-car-proof} except that the arithmetic is more involved to handle the resulting nonlinear and nonmonotonic arithmetic constraints, see \cite{DBLP:conf/csl/Platzer10:TR}.

For a \QdL proof extending the above ideas to a proof of collision-freedom for a more realistic distributed car control system having arbitrarily many cars switching between arbitrarily many lanes with dynamic appearance and disappearance of arbitrarily many cars, we refer to follow-up work \cite{DBLP:conf/fm/LoosPN11}.
Unlike our simplified system model, this follow-up work does not assume that all cars use the same braking power.

}%
}%

\section{Conclusions}

We have introduced a formal system model and semantics for dynamic distributed hybrid systems together with a compositional verification logic and proof calculus.
We believe this is the \emph{first formal verification approach for distributed hybrid dynamics}, where structure and dimension of the system can evolve jointly with the discrete and continuous dynamics.
Our approach handles \emph{distributed hybrid systems} with interacting discrete dynamics, continuous dynamics, structural dynamics, and dimensional dynamics.
We have proven our calculus to be a \emph{sound and complete axiomatization} relative to quantified differential equations.
Our calculus proves collision avoidance in distributed car control with dynamic appearance of new cars on the road, which is out of scope for other approaches.

Future work includes full modular concurrency in distributed hybrid systems, which is already challenging in discrete programs. %

\section*{Acknowledgement}
I thank Frank Pfenning for his helpful comments and the reviewers for their feedback.

\bibliographystyle{alpha}
\bibliography{QdLax}

\end{document}